\documentclass{elsarticle}
\title{
	Programming from Metaphorisms % [JOURNAL VERSION]
}

\author[uminho]{Jos\'{e} Nuno Oliveira} %corref{cor1}}
\ead{jno@di.uminho.pt}

\address[uminho]{
          High Assurance Software Laboratory\\
	 (http://haslab.uminho.pt) \\
          INESC TEC and University of Minho\\
	  Braga, Portugal
}
\date\today
\newtheorem{theorem}{Theorem} % [section]
\newtheorem{lemma}[theorem]{Lemma} % [section]
\def\junc#1#2{\mathopen{[} #1,#2 \mathclose{]}}

\newtheorem{todobody}{To-do} % [section]
\usepackage{xcolor,colortbl}
\newcommand{\one}{\cellcolor{gray!20}1}
\makeatletter
\@ifundefined{lhs2tex.lhs2tex.sty.read}%
  {\@namedef{lhs2tex.lhs2tex.sty.read}{}%
   \newcommand\SkipToFmtEnd{}%
   \newcommand\EndFmtInput{}%
   \long\def\SkipToFmtEnd#1\EndFmtInput{}%
  }\SkipToFmtEnd

\newcommand\ReadOnlyOnce[1]{\@ifundefined{#1}{\@namedef{#1}{}}\SkipToFmtEnd}
\usepackage{amstext}
\usepackage{amssymb}
\usepackage{stmaryrd}
\DeclareFontFamily{OT1}{cmtex}{}
\DeclareFontShape{OT1}{cmtex}{m}{n}
  {<5><6><7><8>cmtex8
   <9>cmtex9
   <10><10.95><12><14.4><17.28><20.74><24.88>cmtex10}{}
\DeclareFontShape{OT1}{cmtex}{m}{it}
  {<-> ssub * cmtt/m/it}{}

\DeclareFontShape{OT1}{cmtt}{bx}{n}
  {<5><6><7><8>cmtt8
   <9>cmbtt9
   <10><10.95><12><14.4><17.28><20.74><24.88>cmbtt10}{}
\DeclareFontShape{OT1}{cmtex}{bx}{n}
  {<-> ssub * cmtt/bx/n}{}
\newcommand{\Conid}[1]{\mathit{#1}}
\newcommand{\Varid}[1]{\mathit{#1}}
\newcommand{\anonymous}{\kern0.06em \vbox{\hrule\@width.5em}}

\newcommand{\bind}{\mathbin{>\!\!\!>\mkern-6.7mu=}}
\renewcommand{\leq}{\leqslant}
\renewcommand{\geq}{\geqslant}
\usepackage{polytable}

\@ifundefined{mathindent}%
  {\newdimen\mathindent\mathindent\leftmargini}%
  {}%

\def\resethooks{%
  \global\let\SaveRestoreHook\empty
  \global\let\ColumnHook\empty}
\newcommand*{\savecolumns}[1][default]%
  {\g@addto@macro\SaveRestoreHook{\savecolumns[#1]}}
\newcommand*{\restorecolumns}[1][default]%
  {\g@addto@macro\SaveRestoreHook{\restorecolumns[#1]}}
\newcommand*{\aligncolumn}[2]%
  {\g@addto@macro\ColumnHook{\column{#1}{#2}}}

\resethooks

\newcommand{\onelinecommentchars}{\quad-{}- }
\newcommand{\commentbeginchars}{\enskip\{-}
\newcommand{\commentendchars}{-\}\enskip}

\newcommand{\visiblecomments}{%
  \let\onelinecomment=\onelinecommentchars
  \let\commentbegin=\commentbeginchars
  \let\commentend=\commentendchars}

\newcommand{\invisiblecomments}{%
  \let\onelinecomment=\empty
  \let\commentbegin=\empty
  \let\commentend=\empty}

\visiblecomments

\newlength{\blanklineskip}
\newcommand{\hsindent}[1]{\quad}% default is fixed indentation
\let\hspre\empty
\let\hspost\empty

\EndFmtInput
\makeatother
\ReadOnlyOnce{polycode.fmt}%
\makeatletter

\newcommand{\hsnewpar}[1]%
  {{\parskip=0pt\parindent=0pt\par\vskip #1\noindent}}

\newcommand{\hscodestyle}{}

\newcommand{\sethscode}[1]%
  {\expandafter\let\expandafter\hscode\csname #1\endcsname
   \expandafter\let\expandafter\endhscode\csname end#1\endcsname}

  {\par\noindent
   \advance\leftskip\mathindent
   \hscodestyle
   \let\\=\@normalcr
   \let\hspre\(\let\hspost\)%
   \pboxed}%
  {\endpboxed\)%
   \par\noindent
   \ignorespacesafterend}

  {\hsnewpar\abovedisplayskip
   \advance\leftskip\mathindent
   \hscodestyle
   \let\hspre\(\let\hspost\)%
   \pboxed}%
  {\endpboxed%
   \hsnewpar\belowdisplayskip
   \ignorespacesafterend}

  {\hsnewpar\abovedisplayskip
   \advance\leftskip\mathindent
   \hscodestyle
   \let\\=\@normalcr
   \(\pboxed}%
  {\endpboxed\)%
   \hsnewpar\belowdisplayskip
   \ignorespacesafterend}

\newcommand{\plainhs}{\sethscode{plainhscode}}

\plainhs

  {\hsnewpar\abovedisplayskip
   \advance\leftskip\mathindent
   \hscodestyle
   \let\\=\@normalcr
   \(\parray}%
  {\endparray\)%
   \hsnewpar\belowdisplayskip
   \ignorespacesafterend}

  {\parray}{\endparray}

  {\(\parray}{\endparray\)}

\def\codeframewidth{\arrayrulewidth}
\RequirePackage{calc}

  {\parskip=\abovedisplayskip\par\noindent
   \hscodestyle
   \arrayrulewidth=\codeframewidth
   \tabular{@{}|p{\linewidth-2\arraycolsep-2\arrayrulewidth-2pt}|@{}}%
   \hline\framedhslinecorrect\\{-1.5ex}%
   \let\endoflinesave=\\
   \let\\=\@normalcr
   \(\pboxed}%
  {\endpboxed\)%
   \framedhslinecorrect\endoflinesave{.5ex}\hline
   \endtabular
   \parskip=\belowdisplayskip\par\noindent
   \ignorespacesafterend}

\newcommand{\framedhslinecorrect}[2]%
  {#1[#2]}

  {\(\def\column##1##2{}%
   \let\>\undefined\let\<\undefined\let\\\undefined
   \newcommand\>[1][]{}\newcommand\<[1][]{}\newcommand\\[1][]{}%
   \def\fromto##1##2##3{##3}%
   }{\) }%

  {\let\orighscode=\hscode
   \let\origendhscode=\endhscode
   \def\endhscode{\def\hscode{\endgroup\def\@currenvir{hscode}\\}\begingroup}
   \orighscode\def\hscode{\endgroup\def\@currenvir{hscode}}}%
  {\origendhscode
   \global\let\hscode=\orighscode
   \global\let\endhscode=\origendhscode}%

\makeatother
\EndFmtInput
\newenvironment{beqnarray*}{\blue\vspace*{-1ex}\begin{eqnarray*}}{\end{eqnarray*}\black\vspace*{-1ex}\hskip-1pt\vskip-0.9em\relax}
\def\myxym#1{\vcenter{\xymatrix{#1}}}
\def\arrayin#1{\begin{array}{rcl}#1\end{array}}

\def\sse{\subseteq}
\newenvironment{lcbr}{\left\{\begin{array}{l}}{\end{array}\right.}

\def\unary#1#2{\def\arg{#2}\def\omisso{}\ifx\arg\omisso{\mathit{#1}}\else\ap{\mathit{#1}}{#2}\fi}

\def\LNCS{LNCS\index{LNCS!\uk{Lecture Notes in Computer Science}}}

\def\alt#1#2{\mathopen{[}#1\hskip 1pt,#2\mathclose{]}}
\def\ap#1#2{#1\,#2}
\def\arrayin#1{\begin{array}{rcl}#1\end{array}}
\def\bang{{!}}

\def\comp{\mathbin{\cdot}}
\def\conv#1{#1^\circ}
\def\eqnnewpage{\end{eqnarray*}\newpage~\vskip-2em\begin{eqnarray*}&& }

\def\ff#1{\ap\f{#1}}
\def\fun#1{{\sf #1}}
\def\f{\fun F}
\def\from{\mathbin{\leftarrow}}
\def\implied{\mathbin\Leftarrow}
\def\just#1#2{\\ &#1& \rule{2em}{0pt} \{ \mbox{\rule[-.7em]{0pt}{1.8em} \small #2 \/} \} \nonumber\\ && }

\def\p#1{\pi_{#1}}
\def\qed{\\&\Box&}
\def\rcb#1#2#3#4{\def\nothing{}\def\range{#3}\mathopen{\langle}#1 \ #2 \ \ifx\range\nothing::\else: \ #3 :\fi \ #4\mathclose{\rangle}}
\def\scata#1{\mathopen{(\!|}#1\mathclose{|\!)}}
\def\shrunkby{\mathbin{\upharpoonright}}
\def\wider#1{~ #1 ~}

\def\muK{\mu_{\fun K}}
\def\muF{\mu_{\fun F}}
\def\muG{\mu_{\fun G}}
\def\muH{\mu_{\fun H}}

\let\kons=\underline

\def\rdiv{\mathbin{\setminus}}

\def\kr{\mathbin{\hbox{\tiny${}^\triangledown$}}}

\def\aspas#1{``#1''}
\def\EXIT{ \bibliographystyle{plain} \bibliography{/Users/jno/share/texinputs/jno} \end{document}}

\def\start{&&}
\def\more{\\&&}

\def\rcbnb#1#2#3#4{\def\nothing{}\def\range{#3}#1 \ #2 \ \ifx\range\nothing::\else: \ #3 :\fi \ #4}
\def\kr{\mathbin{\hbox{\tiny${}^\vartriangle$}}}

\def\thinnedby{\mathbin{\downharpoonleft}}
\usepackage{fleqn}
\usepackage[utf8x]{inputenc}
\usepackage{amsmath}
\usepackage{wrapfig}
\usepackage{amssymb,amsmath}
\usepackage{hyperref}
\usepackage[all]{xy}
\def\larrow#1#2#3{\xymatrix{ #3 & #1 \ar[l]_-{#2} }}
\def\rarrow#1#2#3{\xymatrix{ #1 \ar[r]^-{#2} & #3 }}
\def\bdarrow#1#2#3#4{\xymatrix{#1\ar@<.5ex>[r]^{#2}&#4\ar@<.5ex>[l]^{#3}}}

\def\arLaw#1#2#3#4#5{
\xymatrix{
        #1      \ar@/^1pc/[rr]^-{#4} &
        #5 &
        #2      \ar@/^1pc/[ll]^-{#3}
}}

\def\eqnnewpagex{\end{eqnarray}\newpage~\vskip-2em\begin{eqnarray}\nonumber}
\def\secref#1{Sect.~\ref{#1}}
\def\implies{\mathbin{\Rightarrow}}

\def\htmladdnormallink#1#2{#1}

\def\implied{\mathbin\Leftarrow}

\def\bang{{!}}

\def\muF{\mu_{\fun F}}

\def\bibliometrics#1{\relax} % #2 to swallow full stop :-)

\def\equiv{\Leftrightarrow}
\def\B{\mathbb{B}}
\begin{document}

\begin{abstract}
This paper presents a study of the \emph{metaphorism} pattern of relational specification,
showing how it can be refined into recursive programs.

Metaphorisms express input-output relationships which preserve relevant
information while at the same time some intended optimization takes place.
Text processing, sorting, representation changers, etc., are examples of
metaphorisms.

The kind of metaphorism refinement studied in this paper is a strategy known
as \emph{change of virtual data structure}. By framing metaphorisms in the class
of (inductive) \emph{regular} relations, sufficient conditions are given
for such implementations to be calculated using relation algebra.

The strategy is illustrated with examples including the derivation of the
\emph{quicksort} and \emph{mergesort} algorithms, showing what they have in common
and what makes them different from the very start of development.
\end{abstract}

\begin{keyword}
	Programming from specifications
\sep
	Algebra of programming
\sep
	Weakest precondition calculus.
\end{keyword}

\maketitle

%epigraph{I recall seeing a package to make quotes}{Snowball}
\vskip2em ~ \hfill
\begin{minipage}{.55\textwidth}\footnotesize\em
% Politicians and diapers should be changed frequently and all for the same reason.
% Read more at http://izquotes.com/quote/290079
Politicians and diapers should be changed often and for the same reason.%
\vskip 1.5ex \em
\hfill (attributed to Mark Twain) % E\c{c}a de Queir\'os (1845-1900)
\end{minipage}

\section{Context}

The witty quote by 19th century author Mark Twain % diplomat E\c{c}a de Queir\'os
that provided inspiration for the title of this paper embodies a \emph{metaphor}
which the reader will surely
appreciate. But, what do {metaphors} of this kind have to do with computer programming?

A synergy between metaphors in cognitive linguistics and some relational
patterns common in the field of formal specification, termed \emph{metaphorisms},
was suggested in our earlier conference paper \cite{Ol15a}, which the current
paper extends by framing the approach into the study of the wider class of
inductive \emph{regular relations} \cite{JMBD91}. In particular, an algebra
useful for reasoning about such specification patterns is developed, whose
ubiquity is already observed by \citet{JMBD91}:
\begin{quote}\em
\def\omit#1{[...]}
We have found regular relations to be very general; in particular, \omit{all
but the} most \omit{pathological} specifications we encounter in practice
are regular.
\end{quote}

Metaphorisms are regular relations represented by symmetric divisions of
inductive functions (aka.\ \emph{folds} or \emph{catamorphisms}) restricted
by other regular relations expressing some kind of optimization. After
introducing the metaphor and metaphorism concepts and their underlying algebra,
this paper presents a generic process of implementing metaphorisms towards
\emph{divide \& conquer} program strategies based on implicit,
\emph{virtual} data structures.

\paragraph{Related work}
This paper follows our previous line of research \cite{MO12a} in investigating
relational specification patterns which involve the \emph{shrinking} combinator
for controlling vagueness and non-determinism. It also relates to the 
work on representation changers \cite{HM93b} and on the relational algebra
of programming in general \cite{BM97,MDM94}. Our calculation of sufficient
conditions for implementing metaphorisms via change of virtual data-structure,
illustrated with the \emph{quicksort} and \emph{mergesort} algorithms, can be regarded
as a generalization and expansion of the derivation of \emph{quicksort}
by \citet{BM97}, where it is given in a rather brief and terse style.

Interest in so-called \emph{regular}
% (\emph{difunctional}, \emph{rational})
relations dates back to at least the work by Riguet in the late
1940’s \cite{Ri48}. Their use as specification devices was pioneered by \citet{JMBD91} and \citet{MDM94} in the early 1990's.
Shortly afterwards, Hutton's PhD thesis \cite{Hu92} presents a number of
program derivations % (in Ruby)
in which such % difunctional
relations are in evidence. Rectangular relations, a special case
of regular relations, have also been studied in \cite{Ri48,Sc08}.
Interestingly, \citet{JMBD91} already acknowledge that \emph{it
is common for specifications to be written as the intersection of an equivalence
relation with a rectangular relation}, which is precisely the specification pattern
at focus in the current paper.

Metaphorisms can also be regarded as relational generalizations
of so-called \emph{metamorphisms} \cite{Er98,Gi07}. \emph{Virtual
data structures} have been studied mainly from the perspective
of \emph{deforestation} \cite{SM93,TM95}. Their role in structuring
\emph{divide and conquer} algorithms is commonly accepted but less
worked out in a formal context.\footnote{\label{fn:170701a} In the words of \citet{SM93}
\emph{``virtual data structures (...) play the role of a catalyst in the
development of programs, in the sense that in the final program they have
been transformed away"}.}
Sorting is addressed from this perspective in Bird \& de Moor's textbook \cite{BM97}, which also
stresses algorithm classification through synthesis in the spirit
of \cite{Da78}. In the same vein, 
the role of intermediate, virtual types in classifying and cataloguing
specifications in software repositories has also been emphasized \cite{Ol02}.

\section{Introduction}
Programming theory has been structured around concepts such as \emph{syntax},
\emph{semantics}, \emph{generative grammar} and so on, that have been imported from
Chomskian linguistics. The basis is that syntax provides the \emph{shape} of information
and that semantics express information \emph{contents} in a syntax-driven way
(e.g.\ the meaning of the whole dependent on the meaning of the parts).

Cognitive linguistics breaks with such a \emph{generative} tradition in its belief
that semantics are conveyed in a different way, just by juxtaposing concepts
in the form of \emph{metaphors} which let meanings permeate each other by an
innate capacity of our brain to function metaphor-wise.
Thus we are led to the {\em metaphors we live by}, quoting the classic textbook
by Lakoff and Johnson \cite{LJ80}. If in a public discussion
one of the opponents is said to have \emph{counterattacked} with a \emph{winning} argument, the
underlying metaphor is \emph{argument is war}; metaphor \emph{time is money}
underlies everyday phrases such as \emph{wasting time}, \emph{investing time} and so on;
Twain's % Queir\'os'
quote lives in the metaphor \emph{politics is dirty}, the same that would enable
one to say that somebody might need to \emph{clean his/her reputation}, for instance.

In his \emph{Philosophy of Rhetoric} \cite{Ri36},
Richards finds three kernel ingredients in a metaphor, namely
a \emph{tenor} (e.g.\ \emph{politicians}), a \emph{vehicle} (e.g.\ \emph{diapers})
and a shared \emph{attribute} (e.g.\ soiling). % ... left for the reader to guess
The \emph{flow of meaning} is from vehicle to tenor, through the (as a rule
left unspecified) common attribute.
A sketchy characterization of this construction in the form of a ``cospan''
\begin{eqnarray}
\xymatrix{
	\ensuremath{\fun T }
		\ar[dr]_{g}
&&
	\ensuremath{\mathsf{V}}
		\ar[dl]^{f}
\\
&
	A
}
	\label{eq:150328a}
\end{eqnarray}
is given in \cite{Ol13c}.
Functions \ensuremath{\Varid{f}\mathbin{:}\mathsf{V}\to \Conid{A}} and \ensuremath{\Varid{g}\mathbin{:}\fun T \to \Conid{A}}, the ``witnesses''
of the metaphor, extract a common attribute (\ensuremath{\Conid{A}})
from both tenor (\ensuremath{\fun T }) and vehicle (\ensuremath{\mathsf{V}}). The cognitive, \ae sthetic, or witty power
of a metaphor is obtained by \emph{hiding} \ensuremath{\Conid{A}}, thereby establishing a
\emph{composite}, binary relationship\footnote{%
Given a binary relation \ensuremath{\Conid{R}}, writing \ensuremath{\Varid{b}\;\Conid{R}\;\Varid{a}} (read: ``\ensuremath{\Varid{b}} is related to \ensuremath{\Varid{a}} by \ensuremath{\Conid{R}}'')
means the same as \ensuremath{\Varid{a}\;\conv{\Conid{R}}\;\Varid{b}}, where \ensuremath{\conv{\Conid{R}}} is said to be the \emph{converse} of \ensuremath{\Conid{R}}.
So \ensuremath{\conv{\Conid{R}}} corresponds to the \emph{passive voice}; compare e.g.\
% p3, footnote 1: "passive voice; compare eg 'John loves Mary' with 'Mary is
\ensuremath{\Conid{John}\;\Varid{loves}\;\Conid{Mary}} with \ensuremath{\Conid{Mary}\;\Varid{is}\;\Varid{loved}\;\Varid{by}\;\Conid{John}}: \ensuremath{\conv{(\Varid{loves})}} = \ensuremath{(\Varid{is}\;\Varid{loved}\;\Varid{by})}.}
	\ensuremath{\larrow{\mathsf{V}}{\conv{\Varid{g}} \comp \Varid{f}}{\fun T }}
between tenor and vehicle --- the ``\ensuremath{\fun T } is \ensuremath{\mathsf{V}}'' metaphor --- which leaves \ensuremath{\Conid{A}} implicit.

It turns out that, in the field of program specification, many problem statements
are \emph{metaphorical} in the same (formal) sense: they are characterized
as input-output relationships in which the \emph{preservation} of some kernel
information is kept implicit, possibly subject to some form of optimization.

% Interestingly, both metaphors capturing information preservation (\ref{eq:150328a}) and a
A wide class of optimization criteria can be characterized
by so called \emph{regular}, or \emph{rational} relations.\footnote{Also called
\emph{difunctional} or \emph{uniform} --- see e.g.\ \cite{Ri48,JMBD91,Hu92,BM97}.}
First, some intuition about what \emph{regularity} means in this context: a regular relation
is such that, wherever two inputs have a common image, then they have \emph{exactly the same} set of images.
In other words, the image sets of two different inputs are either disjoint or the same.
As a counterexample, take following relation, represented as matrix with inputs taken from set \ensuremath{\{\mskip1.5mu a_1 ,\mathinner{\ldotp\ldotp},a_5 \mskip1.5mu\}}
and outputs delivered into set \ensuremath{\{\mskip1.5mu b_1,\mathinner{\ldotp\ldotp},b_5 \mskip1.5mu\}}:
\begin{eqnarray}
\begin{tabular}{r|p{1mm}p{1mm}p{1mm}p{1mm}p{1mm}}\ensuremath{\Conid{R}}&\ensuremath{a_1 }&\ensuremath{a_2 }&\ensuremath{a_3 }&\ensuremath{a_4 }&\ensuremath{a_5 }
\\\hline
   \ensuremath{b_1}&0&0&\one&0&\one
\\ \ensuremath{b_2}&0&0&0&0&0
\\ \ensuremath{b_3 }&0&{\one}&0&0&0
\\ \ensuremath{b_4 }&0&{\one}&0&{\one}&0
\\ \ensuremath{b_5 }&0&0&0&{\one}&0
\\ \end{tabular}
	\label{eq:170813a}
\end{eqnarray}
Concerning inputs \ensuremath{a_3 } and \ensuremath{a_5 }, regularity holds; but sets \ensuremath{\{\mskip1.5mu b_3 ,b_4 \mskip1.5mu\}} and \ensuremath{\{\mskip1.5mu b_4 ,b_5 \mskip1.5mu\}} --- the images of \ensuremath{a_2 }
and \ensuremath{a_4 }, respectively --- are neither disjoint nor the same: so \ensuremath{\Conid{R}} isn't regular. (It will become so
if e.g.\ \ensuremath{b_4 } is dropped from both image sets or one of \ensuremath{b_3 } or \ensuremath{b_5 } is replaced for the other in the 
corresponding image set.)

These relations are also called \emph{rational} because they can be represented by ``fractions''
of the form \ensuremath{\frac{\Varid{f}}{\Varid{g}}}, where \ensuremath{\Varid{f}} and \ensuremath{\Varid{g}} are functions and notation \ensuremath{\frac{\Conid{R}}{\Conid{S}}} expresses
the so-called \emph{symmetric division} \cite{BSZ89,FS90} of two relations \ensuremath{\Conid{R}} and \ensuremath{\Conid{S}}.
As detailed in the sequel, it can be easily shown that \ensuremath{\conv{\Varid{g}} \comp \Varid{f}\mathrel{=}\frac{\Varid{f}}{\Varid{g}}},
meaning that metaphors (\ref{eq:150328a}) are rational relations.

This paper is organized in two main parts. In the first part we develop an
\emph{algebra of metaphors} expressed as rational relations and address the
combination of metaphors with another class of rational relations
(called \emph{rectangular}) used to express requirements on the ``tenor'' (``output'')
side of metaphors.
The second part focusses on metaphors \ensuremath{\frac{\Varid{f}}{\Varid{g}}} where \ensuremath{\mathsf{V}} and \ensuremath{\fun T } are inductive
(recursive) types and \ensuremath{\Varid{f}} and \ensuremath{\Varid{g}} are morphisms which extract a common view
of such types. That is, \ensuremath{\Varid{f}} and \ensuremath{\Varid{g}} become \emph{catamorphisms} \cite{BM97},
also known as \emph{folds} \cite{Gi16}.

We use the word \emph{metaphorism} \cite{Ol15a} to refer to the specification pattern just
described.
An example of this is \emph{text formatting}, a relationship between
formatted and unformatted text whose metaphor consists in
preserving the sequence of words of both, while the output text is
optimized wrt.\ some visual criteria.\footnote{%
It is the privilege of those who don't work with \textsc{wysiwyg} text processors
to feel the rewarding (if not \ae sthetic) contrast between the window
where source text is edited %, subject to formatting tags and the like,
and that showing the corresponding, nice-looking PDF output.
%Well, such is the \emph{metaphor} at work!
}
Other examples could have been given:
%Examples are abundant:
\begin{itemize}
\item	Change of base of numeric representation --- the number represented in the
	source is the same represented by the result, cf.\ the `representation changers' studied by \citet{HM93b}.
%item	Conversion of finite lists into balanced search trees --- the information preserved is the set of elements of the source list; the optimization is the invariant induced on the output tree, making it adequate for searching, etc.
\item	Source code refactoring --- the meaning of the source program is preserved,
	the target code being better styled wrt.\ coding conventions and best practices.
\item	Gaussian elimination --- it transforms a system of linear equations 
	into a triangular system that has the same set of roots.% (0, infinity, or 1 root).
\item	Sorting --- the bag (multiset) of elements of the source list is preserved, the
	optimization consisting in obtaining an ordered output. % \cite{He09}.
\end{itemize}

The \emph{optimization} implicit in all these examples can be expressed by reducing
the \emph{vagueness} of relation \ensuremath{\conv{\Varid{g}} \comp \Varid{f}} in (\ref{eq:150328a}) according to some
criterion telling which outputs are better than others. This can be achieved by
adding such criteria in the form of a relation \ensuremath{\Conid{R}} that ``shrinks'' \ensuremath{\conv{\Varid{g}} \comp \Varid{f}},
\begin{eqnarray}
	\ensuremath{\Conid{M}\mathrel{=}{(\conv{\Varid{g}} \comp \Varid{f})}\shrunkby{\Conid{R}}}
& \rule{20ex}{0pt}&
\myxym{
&
	\ensuremath{\fun T }
\\
	\ensuremath{\fun T }
		\ar[dr]_{g}
		\ar[ur]^{\ensuremath{\Conid{R}}}
&&
	\ensuremath{\mathsf{V}}
		\ar[dl]^{f}
		\ar[ul]_{\ensuremath{\Conid{M}}}
		\ar[ll]_{\ensuremath{\conv{\Varid{g}} \comp \Varid{f}}}
\\
&
	A
}
	\label{eq:150205e}
\end{eqnarray}
using the ``shrinking'' operator proposed by \citet{MO12a} for reducing non-determinism.
By unfolding the meaning of this relational operator, the relationship
(\ref{eq:150205e}) established by \ensuremath{\Conid{M}} is the following:
\begin{eqnarray*}
	\ensuremath{\Varid{t}\;\Conid{M}\;\Varid{v}} &\equiv& \ensuremath{(\Varid{g}\;\Varid{t}\mathrel{=}\Varid{f}\;\Varid{v})\mathrel{\wedge}\rcb{\forall}{\Varid{t'}}{\Varid{g}\;\Varid{t'}\mathrel{=}\Varid{f}\;\Varid{v}}{\Varid{t}\;\Conid{R}\;\Varid{t'}}}
\end{eqnarray*}
In words: for each vehicle \ensuremath{\Varid{v}}, choose among all tenors \ensuremath{\Varid{t'}} with the same (hidden) attribute
of \ensuremath{\Varid{v}} those that are better than any other with respect to \ensuremath{\Conid{R}}, if any.

A \emph{metaphorism} \ensuremath{\Conid{M}\mathrel{=}{(\conv{\Varid{g}} \comp \Varid{f})}\shrunkby{\Conid{R}}} therefore
involves two functions and an optimization criterion.
In the text formatting metaphorism, for instance,
\begin{eqnarray*}
\myxym{
	\ensuremath{[\mskip1.5mu \Conid{String}\mskip1.5mu]}
		\ar[dr]_{\ensuremath{(\bind \Varid{words})}~~~~~}
&&
	\ensuremath{\Conid{String}}
		\ar[dl]^{\ensuremath{\Varid{words}}}
		\ar[ll]_{\ensuremath{\Conid{Format}}}
\\
&
	\ensuremath{[\mskip1.5mu \Conid{String}\mskip1.5mu]}
}
\end{eqnarray*}
arrow \ensuremath{\Conid{Format}} relates a string (source text) to a list of strings
(output text lines) such that the original sequence of words is preserved
when white space is discarded. (Monadic function \ensuremath{\bind \Varid{words}}
promotes \ensuremath{\Varid{words}} from strings to lists of strings.\footnote{Technically,
\ensuremath{(\bind \Varid{words})} is termed the \emph{Kleisli lifting} (or \emph{extension})  of function \ensuremath{\Varid{words}} \cite{Mog91}.})
Formatting consists in (re)introducing white space evenly throughout
the output text lines. For economy of presentation, the diagram omits the
optimization part in
\begin{eqnarray}
	\ensuremath{\Conid{Format}} = \ensuremath{{(\conv{(\bind \Varid{words})} \comp \Varid{words})}\shrunkby{\Conid{R}}}
	\label{eq:150318z}
\end{eqnarray}
where relation \ensuremath{\Conid{R}\mathbin{:}[\mskip1.5mu \Conid{String}\mskip1.5mu]\leftarrow [\mskip1.5mu \Conid{String}\mskip1.5mu]} should capture the intended formatting criterion
on lines of text, e.g.\
evenly spaced lines better than unevenly spaced ones, and so on.

Formally, nothing precludes \ensuremath{\Varid{f}} and \ensuremath{\Varid{g}} from being the same attribute function,
in which case types \ensuremath{\mathsf{V}} and \ensuremath{\fun T } are also the same. Although less interesting
from a strictly (cognitive) metaphorical perspective, metaphorisms of this
instance of (\ref{eq:150205e}) are very common in programming
--- take \emph{sorting} as example, where \ensuremath{\mathsf{V}} and \ensuremath{\fun T } are inhabited by finite
sequences of the same (ordered) type. Interestingly, some sorting algorithms
actually involve \emph{another} data-type, but this is hidden and kept implicit
in the whole algorithmic process. Quicksort, for instance, unfolds recursively
in a binary fashion which makes its use of the run-time heap look like a
binary search tree.%
\footnote{Similar patterns can be found in other \emph{divide \& conquer} algorithms.}
Because such a tree is not visible from outside, some authors refer to it
as a \emph{virtual} data structure \cite{SM93}.

\paragraph{Contribution}
This paper addresses a generic process of implementing metaphorisms that
introduces \emph{divide \& conquer} strategies through implicit,
virtual data structures.
%Conditions for the semantics of (\ref{eq:150205e})
%to be preserved along the calculation process are discussed. Altogether, reasoning
%shows how the ``outer metaphor'' of the specification (\ref{eq:150205e})
%disappears and is replaced by a more implicit but more interesting
%``inner metaphor'' which is at the heart of each implementation.
%
In particular, it
% the paper contributes to the discipline of program calculation by:
\begin{itemize}
\item	introduces the relational notions of \emph{metaphor} and \emph{metaphorism}
	and develops their algebra based on rational relations,
	including \emph{divide \& conquer} factorization laws
	(Sections \ref{sec:160110a} and \ref{sec:160125c});
\item	gives results for implementing metaphorisms as hylomorphisms \cite{BM97}
	% as implementations of metaphorisms by \emph{changing the virtual structure} 
	(Sections \ref{sec:160125e} and \ref{sec:150406g}), of which two examples are given:
	\emph{quicksort} (\secref{sec:150406i}) and \emph{mergesort} (\secref{sec:160121b}).
\end{itemize}
The paper also includes Sections \ref{sec:150406h} and \ref{sec:160122b},
which conclude and discuss future work, respectively.
Proofs of some auxiliary results are given in
	 %appendix
	\ref{sec:150329b}.

\section{Relation algebra preliminaries} \label{sec:170319a}
\paragraph{Functions}
A function \ensuremath{\Varid{f}\mathbin{:}\Conid{X}\to \Conid{Y}} is a special case of a relation, such that \ensuremath{\Varid{y}\;\Varid{f}\;\Varid{x}~\Leftrightarrow~\Varid{y}\mathrel{=}\Varid{f}\;\Varid{x}}.%
	\footnote{%
	Following a widespread convention, functions (i.e.\ total and deterministic relations)
	will be denoted by lowercase characters (e.g.\ $f$, $g$)
	or identifiers starting with a lowercase characters,
	while uppercase letters are reserved to arbitrary relations.

	In order to save parentheses in relational expressions,
	we adopt the following precedence rules:
	(a) unary operators take precedence over binary ones ;
	(b) composition binds tighter than any other binary operator;
	(c) intersection binds tighter than union;
	(d) division binds tighter than intersection.
}
The equality sign forces \ensuremath{\Varid{f}} to be totally defined and deterministic.
(We read \ensuremath{\Varid{y}\mathrel{=}\Varid{f}\;\Varid{x}} saying \aspas{\ensuremath{\Varid{y}} is \emph{the} result --- not \emph a result --- of applying \ensuremath{\Varid{f}} to \ensuremath{\Varid{x}}}.)
This makes (total) functions quite rich in relational algebra.
For instance, any function \ensuremath{\Varid{f}} satisfies not only the \emph{shunting rules}
\begin{eqnarray}
	\ensuremath{\Varid{f} \comp \Conid{R}} \subseteq S & \equiv & R \subseteq \ensuremath{\conv{\Varid{f}} \comp \Conid{S}}
		\label{eq:020617e}
\\
	\ensuremath{\Conid{R} \comp \conv{\Varid{f}}} \subseteq S & \equiv & R \subseteq \ensuremath{\Conid{S} \comp \Varid{f}}
		\label{eq:020617f}
\end{eqnarray}
where \ensuremath{\Conid{R}}, \ensuremath{\Conid{S}} are arbitrary (suitably typed) binary relations, but also
\begin{eqnarray}
	b(\conv g\comp R\comp f)a
&
	\wider\equiv
&
	(g\ b) R (f\ a)
	\label{eq:040120c}
\end{eqnarray}
a rule which helps moving variables outwards in expressions.
For \ensuremath{\Conid{R}} the identity function \ensuremath{{id}\;\Varid{x}\mathrel{=}\Varid{x}}, (\ref{eq:040120c}) instantiates to
$b(\conv g\comp f)a \wider\equiv \ensuremath{\Varid{g}\;\Varid{b}\mathrel{=}\Varid{f}\;\Varid{a}}$, that is, to metaphor (\ref{eq:150328a}).

Given \ensuremath{\Varid{k}\;{\in}\;\Conid{K}}, \ensuremath{\kons{\Varid{k}}\mathbin{:}\Conid{X}\to \Conid{K}} denotes the polymorphic \emph{constant}
function which always yields \ensuremath{\Varid{k}} as result: \ensuremath{\kons{\Varid{k}} \comp \Varid{f}\mathrel{=}\kons{\Varid{k}}}, for every \ensuremath{\Varid{f}}.
Predicates are functions of type \ensuremath{\Conid{X}\to \B}, where \ensuremath{\B\mathrel{=}\{\mskip1.5mu \textsc{t},\textsc{f}\mskip1.5mu\}} is the set
of truth values. The constant predicates
\begin{eqnarray}
	\ensuremath{\Varid{true}\mathrel{=}\kons{\textsc{t}}} & ~,~ & \ensuremath{\Varid{false}\mathrel{=}\kons{\textsc{f}}}
	\label{eq:170519a}
\end{eqnarray}
are used in the sequel. Notation
\begin{eqnarray}
	\ensuremath{\mathop{!}\mathbin{:}\Conid{X}\to \mathrm{1}}
	\label{eq:170519c}
\end{eqnarray}
is chosen to describe the unique (constant) function of its type, where \ensuremath{\mathrm{1}} denotes the singleton type.

\paragraph{Symmetric division}
Given two arbitrary relations \ensuremath{\Conid{R}} and \ensuremath{\Conid{S}} typed as in the diagram below,
define the \emph{symmetric division} \ensuremath{\frac{\Conid{S}}{\Conid{R}}} \cite{FS90} of \ensuremath{\Conid{S}} by \ensuremath{\Conid{R}} by:
\begin{eqnarray}
	\ensuremath{\Varid{b}\;\frac{\Conid{S}}{\Conid{R}}\;\Varid{c}} \wider\equiv
%	|lcbr
%		(rcb forall a (a R b) (a S c))
%		(rcb forall a (a S b) (a R c))
%	|
%		 \equiv
	\ensuremath{\rcb{\forall}{\Varid{a}}{}{\Varid{a}\;\Conid{R}\;\Varid{b}~\Leftrightarrow~\Varid{a}\;\Conid{S}\;\Varid{c}}}
&&
\xymatrix@R=1.2em{
	B
		\ar[dr]_R
&
&
	C
		\ar[dl]^S
		\ar[ll]_{\mbox{\large\ensuremath{\frac{\Conid{S}}{\Conid{R}}}}}
\\
&
	A
}
	\label{eq:160107a}
\end{eqnarray}
That is, \ensuremath{\Varid{b}\;\frac{\Conid{S}}{\Conid{R}}\;\Varid{c}} means that \ensuremath{\Varid{b}} and \ensuremath{\Varid{c}} are related to exactly the same outputs (in \ensuremath{\Conid{A}}) by \ensuremath{\Conid{R}} and by
\ensuremath{\Conid{S}}. Another way of writing (\ref{eq:160107a}) is
\(
	\ensuremath{\Varid{b}\;\frac{\Conid{S}}{\Conid{R}}\;\Varid{c}} \equiv
	\ensuremath{\{\mskip1.5mu \Varid{a}\mid \Varid{a}\;\Conid{R}\;\Varid{b}\mskip1.5mu\}\mathrel{=}\{\mskip1.5mu \Varid{a}\mid \Varid{a}\;\Conid{S}\;\Varid{c}\mskip1.5mu\}}
\) which is the same as
\begin{eqnarray}
	\ensuremath{\Varid{b}\;\frac{\Conid{S}}{\Conid{R}}\;\Varid{c}} & \wider\equiv &
	\ensuremath{\Lambda{\Conid{R}}\;\Varid{b}\mathrel{=}\Lambda{\Conid{S}}\;\Varid{c}}
	\label{eq:160110b}
\end{eqnarray}
where \ensuremath{\Lambda } is the \emph{power transpose} \cite{BM97} operator which maps
a relation \ensuremath{\Conid{Q}\mathbin{:}\Conid{Y}\leftarrow \Conid{X}} to the set valued function \ensuremath{\Lambda{\Conid{Q}}\mathbin{:}\Conid{X}\to \fun P \;\Conid{Y}} such
that \ensuremath{\Lambda{\Conid{Q}}\;\Varid{x}\mathrel{=}\{\mskip1.5mu \Varid{y}\mid \Varid{y}\;\Conid{Q}\;\Varid{x}\mskip1.5mu\}}.
Another way to define \ensuremath{\frac{\Conid{S}}{\Conid{R}}} is \cite{FS90}
\begin{eqnarray}
	\ensuremath{\frac{\Conid{S}}{\Conid{R}}} & = & \ensuremath{{\Conid{R}\rdiv \Conid{S}}\mathbin\cap{\conv{\Conid{R}}\mathbin{/}\conv{\Conid{S}}}}
	\label{eq:160122a}
\end{eqnarray}
which factors symmetric division into the two asymmetric divisions
\ensuremath{\Conid{R}\setminus \Conid{S}} and \ensuremath{\Conid{R}\mathbin{/}\Conid{S}} which can be defined by Galois connections:
\begin{eqnarray}
	\ensuremath{{\Conid{R} \comp \Conid{X}}\subseteq{\Conid{S}}~\Leftrightarrow~{\Conid{X}}\subseteq{\Conid{R}\setminus \Conid{S}}}
	\label{eq:020614b}
\\
	\ensuremath{{\Conid{X} \comp \Conid{R}}\subseteq{\Conid{S}}~\Leftrightarrow~{\Conid{X}}\subseteq{\Conid{S}\mathbin{/}\Conid{R}}}
	\label{eq:100707a}
\end{eqnarray}
Pointwise,
\ensuremath{\Varid{b}\;(\Conid{P}\mathbin{/}\Conid{Q})\;\Varid{a}} means \ensuremath{\rcbnb{\forall}{\Varid{x}}{\Varid{a}\;\Conid{Q}\;\Varid{x}}{\Varid{b}\;\Conid{P}\;\Varid{x}}}
(right division)
and 
\ensuremath{\Varid{b}\;(\Conid{P}\rdiv \Conid{Q})\;\Varid{a}} means \ensuremath{\rcbnb{\forall}{\Varid{x}}{\Varid{x}\;\Conid{P}\;\Varid{b}}{\Varid{x}\;\Conid{Q}\;\Varid{a}}}
(left division).
%\footnote{Cf.\ (\ref{eq:100707a}) in
%appendix
%ref{sec:150329b}.}
Note that, by (\ref{eq:020614b}, \ref{eq:100707a}), (\ref{eq:160122a}) is equivalent to
the universal property:
\begin{eqnarray}
	\ensuremath{{\Conid{X}}\subseteq{\frac{\Conid{S}}{\Conid{R}}}} & \equiv & \ensuremath{{\Conid{R} \comp \Conid{X}}\subseteq{\Conid{S}}\mathrel{\wedge}{\Conid{S} \comp \conv{\Conid{X}}}\subseteq{\Conid{R}}}
	\label{eq:151118b}
\end{eqnarray}
From the definitions above a number of standard properties arise \cite{FS90}:
\begin{eqnarray}
	\ensuremath{\left(\frac{\Conid{S}}{\Conid{R}}\right)^{\hskip-3pt\circ}} &=& \ensuremath{\frac{\Conid{R}}{\Conid{S}}}
	\label{eq:151119a}
\\
	\ensuremath{\frac{\Conid{S}}{\Conid{R}} \comp \frac{\Conid{Q}}{\Conid{S}}} &\ensuremath{ \subseteq }& \ensuremath{\frac{\Conid{Q}}{\Conid{R}}}
	\label{eq:151118d}
\\
	\ensuremath{\conv{\Varid{f}} \comp \frac{\Conid{S}}{\Conid{R}} \comp \Varid{g}} & = & \ensuremath{\frac{\Conid{S} \comp \Varid{g}}{\Conid{R} \comp \Varid{f}}}
	\label{eq:151118c}
\\
	\ensuremath{{id}} &\ensuremath{ \subseteq }& \ensuremath{\frac{\Conid{R}}{\Conid{R}}}
	\label{eq:170104a}
\end{eqnarray}
Thus \ensuremath{\frac{\Conid{R}}{\Conid{R}}} is always an \emph{equivalence relation}, for any given \ensuremath{\Conid{R}}. Furthermore, 
\begin{eqnarray}
	\ensuremath{\Conid{R}\mathrel{=}\frac{\Conid{R}}{\Conid{R}}} & \equiv & \mbox{\ensuremath{\Conid{R}} is an equivalence relation}
	\label{eq:160115b}
\end{eqnarray}
holds.\footnote{\label{fn:170320a}This is proved by Riguet on page 134 of \cite{Ri48}, where
the symmetric division \ensuremath{\frac{\Conid{R}}{\Conid{R}}} is denoted by $noy.(R)$, for ``noyaux" of \ensuremath{\Conid{R}} (``noyaux" means ``kernel").
For those readers not wishing to delve into the notation of \citet{Ri48} we
give a simple proof of (\ref{eq:160115b}) in \ref{sec:150329b} based on the laws of relation division.}
Finally note that, even in the case of functions, (\ref{eq:151118d}) remains an inclusion:
\begin{eqnarray}
	\ensuremath{{\frac{\Varid{f}}{\Varid{g}} \comp \frac{\Varid{h}}{\Varid{f}}}\subseteq{\frac{\Varid{h}}{\Varid{g}}}}
	\label{eq:160112d}
\end{eqnarray}
%since:
%\begin{eqnarray*}
%\start
%	|sse ((syd f g) . (syd h f)) (syd h g) |
%%
%\just\implied{ factor |syd id g| out }
%%
%	|sse (f . (syd h f)) h |
%%
%\just\implied{ factor |h| out }
%%
%	|sse (f . (syd id f)) id |
%%
%\just\equiv{ shunting rule (\ref{eq:020617f}) }
%%
%	|sse f f|
%%
%\just\equiv{trivial}
%%
%	true
%\qed
%\end{eqnarray*}
%From (\ref{eq:160112d}) it follows that |syd f f| is always transitive.
%By (\ref{eq:160112c}) it is symmetric and by (\ref{eq:160112e}) it is reflexive.
%Thus every kernel metaphor |syd f f| is an \emph{equivalence relation}.

\paragraph{Relation shrinking \cite{MO12a}}
Given relations $S : A \from B$ and $R : A \from A$, define $S \shrunkby R : A \from B$, pronounced ``$S$
shrunk by $R$'', by
\begin{eqnarray}
  X \sse S \shrunkby R ~~\equiv~~ X \sse S ~\wedge~ X \comp \conv{S} \sse R
&
\mbox{cf.\ diagram: }
& 
\myxym{
&
	B
		\ar[d]^-{S}
		\ar[dl]_-{S \shrunkby R}
\\
	A
&
	A
		\ar[l]^-{R}
}
	\label{eq:100116d}
\end{eqnarray}
This states that $S \shrunkby R$ is the largest part of \ensuremath{\Conid{S}} such that, if it
yields an output for an input $x$, it must be a maximum, with respect
to $R$, among all possible outputs of $x$ by $S$.
By indirect equality, (\ref{eq:100116d}) is equivalent to the closed definition:
\begin{eqnarray}
	S \shrunkby R ~=~ S \cap R/\conv{S}
	\label{eq:100211c}
\end{eqnarray}
Among the properties of shrinking \cite{MO12a} we single out two \emph{fusion} rules 
\begin{eqnarray}
	\ensuremath{{(\Conid{S} \comp \Varid{f})}\shrunkby{\Conid{R}}} & = & \ensuremath{({\Conid{S}}\shrunkby{\Conid{R}}) \comp \Varid{f}}
	\label{eq:fn-shrink-r}
\\
  	(f\comp S)\shrunkby R &=&  f\comp (S\shrunkby (\conv{f}\comp R \comp f))
	\label{eq:fn-shrink-l}
\end{eqnarray}
that will prove useful in the sequel.
Putting universal properties (\ref{eq:151118b},\ref{eq:100116d}) together we get,
by indirect equality,
\begin{eqnarray}
\start
	\ensuremath{\frac{\Conid{R}}{\Varid{g}}\mathrel{=}\conv{\Varid{g}} \comp ({\Conid{R}}\shrunkby{{id}})}
	\label{eq:170413a}
\more
	\ensuremath{\frac{\Varid{f}}{\Conid{R}}\mathrel{=}\conv{({\Conid{R}}\shrunkby{{id}})} \comp \Varid{f}}
	\label{eq:170413b}
\end{eqnarray}
capturing a relationship between shrinking and symmetric division:
knowing that \ensuremath{{\Conid{R}}\shrunkby{{id}}} is nothing but the deterministic fragment of \ensuremath{\Conid{R}},
we see how the \emph{vagueness} of arbitrary \ensuremath{\Conid{R}} replacing either \ensuremath{\Varid{f}} or \ensuremath{\Varid{g}}
in \ensuremath{\frac{\Varid{f}}{\Varid{g}}} is forced to shrink.

\paragraph{Recursive relations}
Later in the paper we shall need a number of standard constructions in relation algebra
that are briefly introduced next. (For the many details omitted please see e.g.\ the textbook by \citet{BM97}.)

Let \ensuremath{\fun F } be a \emph{relator} \cite{R*92}, that is, a mathematical construction such as, for any type \ensuremath{\Conid{A}}, type \ensuremath{\fun F \;\Conid{A}} is defined
and for any relation \ensuremath{\Conid{R}\mathbin{:}\Conid{B}\leftarrow \Conid{A}}, relation
\ensuremath{\fun F \;\Conid{R}\mathbin{:}\fun F \;\Conid{B}\leftarrow \fun F \;\Conid{A}} is defined such that \ensuremath{\fun F \;{id}\mathrel{=}{id}}, \ensuremath{\fun F \;\conv{\Conid{R}}\mathrel{=}\conv{(\fun F \;\Conid{R})}} and 
\ensuremath{\fun F \;(\Conid{R} \comp \Conid{S})\mathrel{=}(\fun F \;\Conid{R}) \comp (\fun F \;\Conid{S})}.

Any relation \ensuremath{\Conid{R}\mathbin{:}\Conid{A}\leftarrow \fun F \;\Conid{A}} is said to be a (relational) \emph{\ensuremath{\fun F }-algebra}.
Special cases include functional \ensuremath{\fun F }-algebras and, among these, those that
are isomorphisms. Within these, the so-called \emph{initial} \ensuremath{\fun F }-algebras, say
\ensuremath{\mathsf{in}_{\fun F}\mathbin{:}\fun T \leftarrow \fun F \;\fun T }, are such that, given any other \ensuremath{\fun F }-algebra \ensuremath{\Conid{R}\mathbin{:}\Conid{A}\leftarrow \fun F \;\Conid{A}},
there is a unique relation of type \ensuremath{\Conid{A}\leftarrow \fun T }, usually written \ensuremath{\mathopen{(\!|}\Conid{R}\mathclose{|\!)}},
such that \ensuremath{\mathopen{(\!|}\Conid{R}\mathclose{|\!)} \comp \mathsf{in}_{\fun F}\mathrel{=}\Conid{R} \comp \fun F \;\mathopen{(\!|}\Conid{R}\mathclose{|\!)}} holds. Type \ensuremath{\fun T }
(often denoted by \ensuremath{\muF } to express its relationship with the base relator \ensuremath{\fun F })
is also referred to as \emph{initial}. The meaning of such relations \ensuremath{\mathopen{(\!|}\Conid{R}\mathclose{|\!)}}, usually
referred to as \emph{catamorphisms}, or \emph{folds}, is captured by the
\emph{universal property}:
\begin{eqnarray}
	\ensuremath{\Conid{X}\mathrel{=}\mathopen{(\!|}\Conid{R}\mathclose{|\!)}} & \wider\equiv & \ensuremath{\Conid{X} \comp \mathsf{in}_{\fun F}\mathrel{=}\Conid{R} \comp (\fun F \;\Conid{X})}
	\label{eq:cataUniv-rel}
\end{eqnarray}
The base \ensuremath{\fun F } captures the recursive pattern of type \ensuremath{\fun T } (which we write as \ensuremath{\muF }). For instance, for \ensuremath{\fun T } the datatype of finite
lists over a given type \ensuremath{\Conid{A}} one has 
\begin{eqnarray}
	\ensuremath{\begin{lcbr}\fun F \;\Conid{X}\mathrel{=}\mathrm{1}\mathbin{+}\Conid{A} \times \Conid{X}\\\fun F \;\Varid{f}\mathrel{=}{id}\mathbin{+}{id} \times \Varid{f}\end{lcbr}}
	\label{eq:170518a}
\end{eqnarray}
This instance is relevant for the examples that come  later in this paper.

Given \ensuremath{\fun F }-algebras \ensuremath{\Conid{R}\mathbin{:}\Conid{A}\leftarrow \fun F \;\Conid{A}} and \ensuremath{\Conid{S}\mathbin{:}\Conid{B}\leftarrow \fun F \;\Conid{B}}, the composition
\ensuremath{\Conid{H}\mathrel{=}\mathopen{(\!|}\Conid{R}\mathclose{|\!)} \comp \conv{\mathopen{(\!|}\Conid{S}\mathclose{|\!)}}}, of type \ensuremath{\Conid{A}\leftarrow \Conid{B}}, is usually referred to as 
a \emph{hylomorphism} \cite{BM97}. \ensuremath{\Conid{H}} is the least fixpoint of the
relational equation \ensuremath{\Conid{X}\mathrel{=}\Conid{R} \comp (\fun F \;\Conid{X}) \comp \conv{\Conid{S}}}. The intermediate type \ensuremath{\muF }
generated by \ensuremath{\conv{\mathopen{(\!|}\Conid{S}\mathclose{|\!)}}} and consumed by \ensuremath{\mathopen{(\!|}\Conid{R}\mathclose{|\!)}} is known as the
\emph{virtual data structure} \cite{SM93} of the hylomorphism.
The opposite composition \ensuremath{\conv{\mathopen{(\!|}\Conid{S}\mathclose{|\!)}} \comp \mathopen{(\!|}\Conid{R}\mathclose{|\!)}}, for suitably typed \ensuremath{\Conid{S}} and \ensuremath{\Conid{R}},
is sometimes termed a \emph{metamorphism} \cite{Gi07}. 

Two properties stem from (\ref{eq:cataUniv-rel}) that prove particularly useful in calculations about \ensuremath{\mathopen{(\!|}\Conid{R}\mathclose{|\!)}},
namely \emph{fusion}
\begin{eqnarray}
	\ensuremath{\Conid{S} \comp \mathopen{(\!|}\Conid{R}\mathclose{|\!)}\mathrel{=}\mathopen{(\!|}\Conid{Q}\mathclose{|\!)}} & \wider\implied & \ensuremath{\Conid{S} \comp \Conid{R}\mathrel{=}\Conid{Q} \comp \fun F \;\Conid{S}}
	\label{eq:150402a}
\end{eqnarray}
and \emph{cancellation} (cf.\ above):
\begin{eqnarray}
	\ensuremath{\mathopen{(\!|}\Conid{R}\mathclose{|\!)} \comp \mathsf{in}_{\fun F}} &=& \ensuremath{\Conid{R} \comp \fun F \;\mathopen{(\!|}\Conid{R}\mathclose{|\!)}}
	\label{eq:150402b}
\end{eqnarray}
Fusion is particularly helpful in the sense of finding a sufficient condition
on \ensuremath{\Conid{S}}, \ensuremath{\Conid{R}} and \ensuremath{\Conid{Q}} for merging \ensuremath{\Conid{S} \comp \mathopen{(\!|}\Conid{R}\mathclose{|\!)}} into \ensuremath{\mathopen{(\!|}\Conid{Q}\mathclose{|\!)}}. In the words
of \citet{BM97}, law (\ref{eq:150402a}) \emph{is probably the
most useful tool in the arsenal of techniques for program derivation}.
The remainder of this paper will give further evidence of this statement.

\section{On the algebra of metaphors} \label{sec:160110a}
\paragraph{Metaphors as symmetric divisions}
Substituting \ensuremath{\Conid{S},\Conid{R}\mathbin{:=}\Varid{f},\Varid{g}} in (\ref{eq:151118b}) and using the shunting rules (\ref{eq:020617e},\ref{eq:020617f})
we obtain, by indirect equality:
\begin{eqnarray}
	\ensuremath{\frac{\Varid{f}}{\Varid{g}}}& =&\ensuremath{\conv{\Varid{g}} \comp \Varid{f}}
	\label{eq:160117a}
\end{eqnarray}
So, a metaphor \ensuremath{\conv{\Varid{g}} \comp \Varid{f}} (\ref{eq:150205e}) can be expressed as a symmetric division.
% --- just substitute |S,R := f,g| in (\ref{eq:160107a},\ref{eq:160110b}) and simplify.
On the other hand, moving the variables of (\ref{eq:160110b}) outwards
by use of (\ref{eq:040120c}), we obtain the following \emph{power transpose cancellation} rule:\footnote{%
This rule is nothing but another way of stating exercise 4.48 proposed by \citet{BM97}. Note that \ensuremath{\Lambda{\Conid{R}}} is always a (total) function.}
\begin{eqnarray}
	\ensuremath{\frac{\Lambda{\Conid{S}}}{\Lambda{\Conid{R}}}} &=& \ensuremath{\frac{\Conid{S}}{\Conid{R}}}
	\label{eq:160108a}
\end{eqnarray}
Read from right to left, this shows a way of converting arbitrary symmetric divisions into
metaphors.%
%\footnote{Ie.\ rational relations --- recall that |pT R| and |pT S| are functions.}

Hereafter we will adopt \ensuremath{\frac{\Varid{f}}{\Varid{g}}} as our canonical notation for metaphors. This has the advantage
of suggesting an analogy with \emph{rational numbers}
\footnote{This analogy was first noted by \citet{JMBD91} where functions, rational relations and arbitrary relations
are paralleled with integers, rationals and reals, respectively.
} which makes calculation rules easy to understand and memorize.
From (\ref{eq:151119a}) we immediately get that \emph{converses of metaphors} are metaphors:\footnote{%
As (the perception of) time predates money in human evolution, it is reasonable
to guess that metaphor \textsc{time is money} might have started the other
way around, by its \emph{converse} \textsc{money is time}, although this is
highly speculative of course.}
\begin{eqnarray}
	\ensuremath{\left(\frac{\Varid{f}}{\Varid{g}}\right)^{\hskip-3pt\circ}\mathrel{=}\frac{\Varid{g}}{\Varid{f}}}
	\label{eq:160112c}
\end{eqnarray}
% easy to justify thanks to involution (|conv((conv R))=R|) and contravariance (|conv((R .S)) = conv S . (conv R)|).
	%label{eq:020624d}
Moreover, \ensuremath{\frac{\Varid{f}}{{id}}\mathrel{=}\Varid{f}} and \ensuremath{\conv{\Varid{g}}\mathrel{=}\frac{{id}}{\Varid{g}}}, consistent with \ensuremath{{id}} being
the unit of composition, \ensuremath{\Conid{R} \comp {id}\mathrel{=}\Conid{R}\mathrel{=}{id} \comp \Conid{R}}.
%, the multiplicative operation of the monoid of binary relations.
As expected,
\begin{eqnarray*}
	\ensuremath{\frac{{id}}{\Varid{g}} \comp \frac{\Varid{f}}{{id}}\mathrel{=}\frac{\Varid{f}}{\Varid{g}}}
\end{eqnarray*}
holds, a corollary of the more general:\footnote{%
Equality (\ref{eq:160112b}) can be regarded as a generalization of Proposition 4.5 given by \citet{JMBD91}.
%	The right relative product of a regular relation by a function is regular.
%My version: |conv f . R . g| is regular provided |R| is so.
}
\begin{eqnarray}
	\ensuremath{\frac{{id}}{\Varid{g}} \comp \frac{\Varid{h}}{\Varid{k}} \comp \frac{\Varid{f}}{{id}}\mathrel{=}\frac{\Varid{h} \comp \Varid{f}}{\Varid{k} \comp \Varid{g}}}
	\label{eq:160112b}
\end{eqnarray}

If \ensuremath{{id}} plays the role of the multiplicative identity \ensuremath{\mathrm{1}} in the rational number analogy, what is the
counterpart of number \ensuremath{\mathrm{0}}? It is the empty relation \ensuremath{\bot }, which is represented by any \ensuremath{\frac{\Varid{f}}{\Varid{g}}} such that
\ensuremath{\conv{\Varid{g}} \comp \Varid{f}} is empty, that is, \ensuremath{\Varid{g}\;\Varid{y}\not=\Varid{f}\;\Varid{x}} for any choice of \ensuremath{\Varid{x}} and \ensuremath{\Varid{y}}.
(Any two relations \ensuremath{\Conid{R}} and \ensuremath{\Conid{S}} such that \ensuremath{\conv{\Conid{R}} \comp \Conid{S}\; \subseteq \;\bot } are said to be \emph{range disjoint}.)
For instance, \ensuremath{\frac{\Varid{true}}{\Varid{false}}\mathrel{=}\bot }, where \ensuremath{\Varid{true}} and \ensuremath{\Varid{false}} are the constant functions
yielding the corresponding truth values (\ref{eq:170519a}). In general,
\begin{eqnarray*}
	\ensuremath{\Varid{a}\not=\Varid{b}} \wider\equiv \ensuremath{\frac{\wider{\kons{\Varid{a}}}}{\kons{\Varid{b}}}\mathrel{=}\bot }
\end{eqnarray*}
where \ensuremath{\kons{\Varid{a}}} and \ensuremath{\kons{\Varid{b}}} denote constant functions.
% \footnote{Given some |k ins K|, we define polymorphic constant function |const k : A -> K| by |const k a = k|. So |true = const True| and |false = const False|.}
The opposite situation of \ensuremath{\Varid{a}\mathrel{=}\Varid{b}} above leads to \ensuremath{\frac{\wider{\kons{\Varid{a}}}}{\kons{\Varid{a}}}\mathrel{=}\top }, where \ensuremath{\Varid{y}\;\top \;\Varid{x}} holds for
all \ensuremath{\Varid{y},\Varid{x}}. The canonical presentation of this largest possible metaphor is
\begin{eqnarray}
	\ensuremath{\frac{\mathop{!}_{\Conid{A}}}{\mathop{!}_{\Conid{B}}}\mathrel{=}\top _{\Conid{B}\leftarrow \Conid{A}}}
	\label{eq:060124a}
\end{eqnarray}
% the largest relation of its type
recall (\ref{eq:170519c}).

\paragraph{Intersecting metaphors}
In the words of C.S.\ Peirce, \ensuremath{\Varid{y}\;\top \;\Varid{x}} simply means that ``\ensuremath{\Varid{y}} is coexistent with \ensuremath{\Varid{x}}''
\cite{Mad91}. Suggestively, the symbol chosen by Peirce to denote \ensuremath{\top } is $\infty$.
Although semantically poor, this metaphor surely holds about any \ensuremath{\Varid{x}} and \ensuremath{\Varid{y}} related by any other metaphor.
%footnote{Leaving out the nonsensical |bot|.}
In a sense, it can be regarded as the starting point for any metaphorical relationship, obtained by
some form of refinement. Metaphor conjunction is one way of doing such refinement,
%A most useful fact is that the conjunction of two metaphors is a metaphor,
\begin{eqnarray}
	\ensuremath{{\frac{\Varid{f}}{\Varid{g}}}\mathbin\cap{\frac{\Varid{h}}{\Varid{k}}}\mathrel{=}\frac{{\Varid{f}}\kr{\Varid{h}}}{{\Varid{g}}\kr{\Varid{k}}}}
	\label{eq:160111a}
\end{eqnarray}
where the \emph{pairing} of two functions, say \ensuremath{{\Varid{f}}\kr{\Varid{h}}}, is defined by \ensuremath{({\Varid{f}}\kr{\Varid{h}})\;\Varid{x}\mathrel{=}(\Varid{f}\;\Varid{x},\Varid{h}\;\Varid{x})}.%
\footnote{%
The fact that metaphors are preserved by intersection, captured by (\ref{eq:160111a}),
follows immediately from a more general law of relation algebra \cite{BM97}:
\ensuremath{\conv{({\Conid{R}}\kr{\Conid{S}})} \comp ({\Conid{P}}\kr{\Conid{Q}})\mathrel{=}{\conv{\Conid{R}} \comp \Conid{P}}\mathbin\cap{\conv{\Conid{S}}\;\Conid{Q}}}, where pairing is extended to arbitrary
relations in the expected way: \ensuremath{(\Varid{y},\Varid{z})\;({\Conid{R}}\kr{\Conid{S}})\;\Varid{x}~\Leftrightarrow~(\Varid{y}\;\Conid{R}\;\Varid{x})\mathrel{\wedge}(\Varid{z}\;\Conid{S}\;\Varid{x})}.
Note how these laws % (\ref{eq:160111a})
include what are normally regarded as the two key benefits of
the calculus of relations: \emph{converse functions as specifications and intersection of specifications} \cite{BGM02,Na16} .}
As an example of the intersection rule consider
\begin{eqnarray*}
	\ensuremath{{\frac{\Varid{true}}{\Varid{q}}}\mathbin\cap{\frac{\Varid{p}}{\Varid{true}}}} = \ensuremath{\frac{{\Varid{true}}\kr{\Varid{p}}}{{\Varid{q}}\kr{\Varid{true}}}}
\end{eqnarray*}
where \ensuremath{\Varid{p}} and \ensuremath{\Varid{q}} are predicates and \ensuremath{\Varid{true}} is the everywhere true predicate already introduced.
It is easy to show that
\ensuremath{\Varid{y}\;\frac{\Varid{true}}{\Varid{q}}\;\Varid{x}\mathrel{=}\Varid{q}\;\Varid{y}} and \ensuremath{\Varid{y}\;\frac{\Varid{p}}{\Varid{true}}\;\Varid{x}\mathrel{=}\Varid{p}\;\Varid{x}} hold; so the intersection should
mean \ensuremath{\Varid{q}\;\Varid{y}\mathrel{\wedge}\Varid{p}\;\Varid{x}}. In fact:
\begin{eqnarray*}
\start
	\ensuremath{\Varid{y}\;\frac{{\Varid{true}}\kr{\Varid{p}}}{{\Varid{q}}\kr{\Varid{true}}}\;\Varid{x}}
\just\equiv{ pointwise meaning of \ensuremath{\frac{\Varid{f}}{\Varid{g}}} and \ensuremath{{\Varid{f}}\kr{\Varid{g}}} }
	\ensuremath{(\Varid{q}\;\Varid{y},\textsc{t})\mathrel{=}(\textsc{t},\Varid{p}\;\Varid{x})}
\just\equiv{ equality of pairs ; predicate logic }
	\ensuremath{\Varid{q}\;\Varid{y}\mathrel{\wedge}\Varid{p}\;\Varid{x}}
\qed
\end{eqnarray*}
We will focus on the particular metaphoric pattern
\ensuremath{{\frac{\Varid{f}}{\Varid{g}}}\mathbin\cap{\frac{\Varid{true}}{\Varid{p}}}\mathrel{=}\frac{{\Varid{f}}\kr{\Varid{true}}}{{\Varid{g}}\kr{\Varid{p}}}}
later in this paper:
\begin{eqnarray*}
	\ensuremath{\Varid{y}\;({\frac{\Varid{f}}{\Varid{g}}}\mathbin\cap{\frac{\Varid{true}}{\Varid{p}}})\;\Varid{x}} \hskip1ex\ensuremath{~\Leftrightarrow~}\hskip1ex \ensuremath{(\Varid{g}\;\Varid{y}\mathrel{=}\Varid{f}\;\Varid{x})} \hskip1ex\ensuremath{\mathrel{\wedge}}\hskip1ex \ensuremath{\Varid{p}\;\Varid{y}}
\end{eqnarray*}
In this relational specification pattern, outputs \ensuremath{\Varid{y}} preserve some common information
wrt.\ inputs \ensuremath{\Varid{x}} with the additional ingredient of satisfying post-condition \ensuremath{\Varid{p}}.

\paragraph{Rectangular metaphors}
A relation \ensuremath{\Conid{R}} is said to be \emph{rectangular} iff \ensuremath{\Conid{R}\mathrel{=}\Conid{R} \comp \top  \comp \Conid{R}} holds \cite{Ri48,Sc08}.
Note that \ensuremath{{\Conid{R}}\subseteq{\Conid{R} \comp \top  \comp \Conid{R}}} always holds: \ensuremath{\Varid{b}\;\Conid{R}\;\Varid{a}} implies that there exist \ensuremath{\Varid{a'},\Varid{b'}} such that
\ensuremath{\Varid{b}\;\Conid{R}\;\Varid{a'}} and \ensuremath{\Varid{b'}\;\Conid{R}\;\Varid{a}}. 
Metaphors of the form \ensuremath{\frac{\wider{\kons{\Varid{a}}}}{\Varid{f}}}
(meaning \ensuremath{\Varid{f}\;\Varid{x}\mathrel{=}\Varid{a}} for some given \ensuremath{\Varid{a}})
are rectangular, as the following calculation shows:
\begin{eqnarray*}
\start
	\ensuremath{\frac{\wider{\kons{\Varid{a}}}}{\Varid{f}} \comp \top  \comp \frac{\wider{\kons{\Varid{a}}}}{\Varid{f}}\mathrel{=}\frac{\wider{\kons{\Varid{a}}}}{\Varid{f}}}
\just\equiv{ since \ensuremath{{\Conid{R}}\subseteq{\Conid{R} \comp \top  \comp \Conid{R}}} always holds }
	\ensuremath{{\frac{\wider{\kons{\Varid{a}}}}{\Varid{f}} \comp \top  \comp \frac{\wider{\kons{\Varid{a}}}}{\Varid{f}}}\subseteq{\frac{\wider{\kons{\Varid{a}}}}{\Varid{f}}}}
\just\implied{ monotonicity of composition (by \ensuremath{\conv{\Varid{f}}}) }
	\ensuremath{{\kons{\Varid{a}} \comp \top  \comp \frac{\kons{\Varid{a}}}{\Varid{f}}}\subseteq{\kons{\Varid{a}}}}
\just\equiv{ shunting (\ref{eq:020617e}) ; \ensuremath{\frac{\kons{\Varid{a}}}{\kons{\Varid{a}}}\mathrel{=}\top } }
	\ensuremath{{\top  \comp \frac{\kons{\Varid{a}}}{\Varid{f}}}\subseteq{\top }}
\just\equiv{ any relation is at most \ensuremath{\top } }
	\ensuremath{\Varid{true}}
\qed
\end{eqnarray*}
As rectangularity is preserved by converse, \ensuremath{\frac{\Varid{f}}{\wider{\kons{\Varid{a}}}}} is also rectangular.

\paragraph{Kernel metaphors}
In keeping with the analogy between fractions of integers and
\emph{fractions of functions}
one might wish the equality \ensuremath{\frac{\Varid{f}}{\Varid{f}}\mathrel{=}{id}} to hold, but this only happens for \ensuremath{\Varid{f}} injective.%
	\footnote{Morphisms such that \ensuremath{\frac{\Conid{S}}{\Conid{S}}\mathrel{=}{id}} are referred to as \emph{straight} by \citet{FS90} % pag 234
	and generically underlie the proof strategy known as \emph{indirect equality} \cite{BM97}.}
As seen in \secref{sec:170319a}, metaphor \ensuremath{\frac{\Varid{f}}{\Varid{f}}} is an \emph{equivalence relation}
and therefore % symmetric, transitive and
reflexive:
\begin{eqnarray}
	\ensuremath{{{id}}\subseteq{\frac{\Varid{f}}{\Varid{f}}}}
	\label{eq:160112e}
\end{eqnarray}
It is known as the \emph{kernel} of \ensuremath{\Varid{f}} and it ``measures'' the injectivity of \ensuremath{\Varid{f}},
as defined by the preorder
\begin{eqnarray}
	\ensuremath{\Varid{f}\leq \Varid{g}} \wider\equiv \ensuremath{{\frac{\Varid{g}}{\Varid{g}}}\subseteq{\frac{\Varid{f}}{\Varid{f}}}}
	\label{eq:160112f}
\end{eqnarray}
where \ensuremath{\Varid{f}\leq \Varid{g}} means that \ensuremath{\Varid{f}} is \emph{less injective} than
\ensuremath{\Varid{g}}.\footnote{See e.g.\ \cite{Ol14a,Gi16}. This injectivity preorder is the
converse of the \emph{determination order} of \cite{BS81}.}
Clearly, \ensuremath{\mathop{!}\leq \Varid{f}\leq {id}} for any \ensuremath{\Varid{f}} (\ref{eq:060124a},\ref{eq:160112e}).
The following alternative way of stating (\ref{eq:160112f})
\begin{eqnarray}
	\ensuremath{\Varid{f}\leq \Varid{g}} & \ensuremath{~\Leftrightarrow~} & \ensuremath{\rcb{\exists }{\Varid{k}}{}{\Varid{f}\mathrel{=}\Varid{k} \comp \Varid{g}}}
	\label{eq:160118f}
\end{eqnarray}
is given by Gibbons \cite{Gi16}.

Every equivalence relation \ensuremath{\larrow{\Conid{A}}{\Conid{R}}{\Conid{A}}} is representable by a kernel metaphor,
and canonically by
\begin{eqnarray}
	\ensuremath{\Conid{R}\mathrel{=}\frac{\Lambda{\Conid{R}}}{\Lambda{\Conid{R}}}}
	\label{eq:160115a}
\end{eqnarray}
where the power transpose \ensuremath{\Lambda{\Conid{R}}} maps each element of \ensuremath{\Conid{A}} into its \emph{equivalence class}.
(\ref{eq:160115a}) follows immediately from (\ref{eq:160115b}) and (\ref{eq:160108a}).
%proved \ref{sec:150329b}:
%\begin{eqnarray*}
%\start
%	|syd (pT R)(pT R)|
%%
%\just={ (\ref{eq:160108a}) }
%%
%	|syd (R)(R)|
%%
%\just={ (\ref{eq:160115b}) since |R| is an equivalence }
%%
%	|R|
%\qed
%\end{eqnarray*}

\paragraph{Weakest preconditions}
Given a predicate \ensuremath{\Varid{p}}, we define the metaphor \ensuremath{\Varid{p}\hskip-1pt\mathit{?}} by
\begin{eqnarray}
	\ensuremath{\Varid{p}\hskip-1pt\mathit{?}\mathrel{=}{{id}}\mathbin\cap{\frac{\Varid{true}}{\Varid{p}}}}
	\label{eq:160121c}
\end{eqnarray}
This is called the \emph{partial identity}%
\footnote{
%We use uppercase Greek letters (e.g.\ |vPsi|, |vPhi|, ...) to denote
\emph{Partial identities} are also known as \emph{coreflexives},
\emph{monotypes} or \emph{tests} \cite{DBW97,FS90,Ko97a}.
% Every partial identity |vPsi| is such that |vPsi atmost id| and is in one-to-one correspondence with some predicate |q|. As in \cite{MO12a} we write |vPsi=crflx q| wherever we want to indicate that |q| is the predicate captured by |vPsi|.  Thus |vPsi=crflx q| has the pointwise meaning \( |b vPsi a| \equiv |b=a && q a| \).
}
for \ensuremath{\Varid{p}} in the sense that
\begin{quote}
\ensuremath{\Varid{y}\;(\Varid{p}\hskip-1pt\mathit{?})\;\Varid{x}~\Leftrightarrow~(\Varid{p}\;\Varid{y})\mathrel{\wedge}\Varid{y}\mathrel{=}\Varid{x}}
\end{quote}
holds. That is, \ensuremath{\Varid{p}\hskip-1pt\mathit{?}} is the fragment of \ensuremath{{id}} where \ensuremath{\Varid{p}} holds. Note that \ensuremath{\Varid{p}\hskip-1pt\mathit{?}\mathrel{=}{{id}}\mathbin\cap{\frac{\Varid{p}}{\Varid{true}}}} by converses. The rectangular metaphor \ensuremath{\frac{\Varid{true}}{\Varid{p}}}
can be recovered from \ensuremath{\Varid{p}\hskip-1pt\mathit{?}} by
\begin{eqnarray}
	\ensuremath{\Varid{p}\hskip-1pt\mathit{?} \comp \top \mathrel{=}\frac{\Varid{true}}{\Varid{p}}}
\end{eqnarray}
(Conversely, \ensuremath{\top  \comp \Varid{p}\hskip-1pt\mathit{?}\mathrel{=}\frac{\Varid{p}}{\Varid{true}}}.)
It is easy to show that
\begin{eqnarray*}
	\ensuremath{{\Varid{f}}\mathbin\cap{\frac{\Varid{true}}{\Varid{q}}}\mathrel{=}\Varid{q}\hskip-1pt\mathit{?} \comp \Varid{f}}
\\
	\ensuremath{{\Varid{f}}\mathbin\cap{\frac{\Varid{p}}{\Varid{true}}}\mathrel{=}\Varid{f} \comp \Varid{p}\hskip-1pt\mathit{?}}
\end{eqnarray*}
hold.\footnote{Check (\ref{eq:081025a}) and (\ref{eq:071215a}) in %appendix
\ref{sec:150329b}.
}
Thus, \ensuremath{\Varid{q}} (resp.\ \ensuremath{\Varid{p}}) work as \emph{post} (resp.\ \emph{pre}) conditions
for function \ensuremath{\Varid{f}}. The particular situation in which \ensuremath{\Varid{q}\hskip-1pt\mathit{?} \comp \Varid{f}\mathrel{=}\Varid{f} \comp \Varid{p}\hskip-1pt\mathit{?}}
holds captures a \emph{weakest/strongest} pre/post-condition relationship
expressed by the following universal property:
\begin{eqnarray}
	\ensuremath{\Varid{f} \comp \Varid{p}\hskip-1pt\mathit{?}} = \ensuremath{\Varid{q}\hskip-1pt\mathit{?} \comp \Varid{f}} & \wider\equiv & \ensuremath{\Varid{p}\mathrel{=}\Varid{q} \comp \Varid{f}}
	\label{eq:150406c}
\end{eqnarray}
Condition \ensuremath{\Varid{p}\mathrel{=}\Varid{q} \comp \Varid{f}} is equivalent to \ensuremath{\Varid{p}\mathrel{=}\Varid{wp}\;(\Varid{f},\Varid{q})}, the weakest precondition (\textsc{wp}) for the outputs of \ensuremath{\Varid{f}}
to fall within \ensuremath{\Varid{q}}.
Property (\ref{eq:150406c}) enables a ``logic-free'' calculation of weakest preconditions,
as we shall soon see:
given \ensuremath{\Varid{f}} and post-condition \ensuremath{\Varid{q}}, there exists a unique (weakest) precondition \ensuremath{\Varid{p}} such that
\ensuremath{\Varid{q}\hskip-1pt\mathit{?} \comp \Varid{f}} can be replaced by \ensuremath{\Varid{f} \comp \Varid{p}\hskip-1pt\mathit{?}}.
% is always possible (cf.\ \emph{existence}) and \textsc{wp} |p| is \emph{unique}.
Moreover:
\begin{eqnarray}
	\ensuremath{\frac{\Varid{f}}{\Varid{f}} \comp \Varid{p}\hskip-1pt\mathit{?}} = \ensuremath{\Varid{p}\hskip-1pt\mathit{?} \comp \frac{\Varid{f}}{\Varid{f}}} & \wider\implied & \ensuremath{\Varid{p}\leq \Varid{f}} %  |rcb exists q () (p = q.f)|
	\label{eq:150407a-modified}
\end{eqnarray}
where \ensuremath{\leq } denotes the injectivity preorder on functions (\ref{eq:160112f},\ref{eq:160118f}).
Relational proofs for (\ref{eq:150406c}) and (\ref{eq:150407a-modified}) are given in
	% Appendix
	\ref{sec:150329b}.

\paragraph{Products of metaphors}
Metaphors can also be combined pairwise, leading to metaphors on pairs.%
	\footnote{By referring to the \emph{quixotic} plot of a couple of politicians in some particular
situation, one might wish to suggest that one of them behaved like Don Quixote \emph{and}
the other like Sancho Panza.}
This situation is captured by the \emph{product rule},
\begin{eqnarray}
	\ensuremath{\frac{\Varid{f}}{\Varid{g}} \times \frac{\Varid{h}}{\Varid{k}}} = \ensuremath{\frac{\Varid{f} \times \Varid{h}}{\Varid{g} \times \Varid{k}}}
	\label{eq:160112a}
\end{eqnarray}
telling that the product of two metaphors is a metaphor.
Relational (Kronecker) product is defined as expected,
\ensuremath{(\Varid{x},\Varid{y})\;(\Conid{R} \times \Conid{S})\;(\Varid{a},\Varid{b})~\Leftrightarrow~\Varid{x}\;\Conid{R}\;\Varid{a}\mathrel{\wedge}\Varid{y}\;\Conid{S}\;\Varid{b}}, which, in the case of functions,
becomes \ensuremath{(\Varid{f} \times \Varid{g})\;(\Varid{a},\Varid{b})\mathrel{=}(\Varid{f}\;\Varid{a},\Varid{g}\;\Varid{b})}. Both pairing and product can be written pointfree,
\begin{eqnarray}
	\ensuremath{\Conid{R} \times \Conid{S}\mathrel{=}{\Conid{R} \comp \p1}\kr{\Conid{S} \comp \p2}}
	\label{eq:960923c-rel}
\\
	\ensuremath{{\Conid{R}}\kr{\Conid{S}}\mathrel{=}{\conv{\p1} \comp \Conid{R}}\mathbin\cap{\conv{\p2} \comp \Conid{S}}}
	\label{eq:030418a}
\end{eqnarray}
where \ensuremath{\p1\;(\Varid{a},\Varid{b})\mathrel{=}\Varid{a}} and \ensuremath{\p2\;(\Varid{a},\Varid{b})\mathrel{=}\Varid{b}} are the standard projections.
The proof of (\ref{eq:160112a}) follows:
%|syd f g >< (syd h k)| = |kr((conv g . f . p1)) ((conv k . h. p2))| = |(conv((g><k))) . ((kr (f.p1)(h.p2)))| = |syd(f><h)(g><k)|
\begin{eqnarray*}
\start
	\ensuremath{\frac{\Varid{f}}{\Varid{g}} \times \frac{\Varid{h}}{\Varid{k}}}
\just={ (\ref{eq:960923c-rel}) ; (\ref{eq:160112b}) }
	\ensuremath{{\frac{\Varid{f} \comp \p1}{\Varid{g}}}\kr{\frac{\Varid{h} \comp \p2}{\Varid{k}}}}
\just={ (\ref{eq:030418a}) ; (\ref{eq:160112b}) }
	\ensuremath{{\frac{\Varid{f} \comp \p1}{\Varid{g} \comp \p1}}\mathbin\cap{\frac{\Varid{h} \comp \p2}{\Varid{k} \comp \p2}}}
\just={ (\ref{eq:160111a}) }
	\ensuremath{\frac{{\Varid{f} \comp \p1}\kr{\Varid{g} \comp \p1}}{{\Varid{h} \comp \p2}\kr{\Varid{k} \comp \p2}}}
	\eqnnewpage
\just={ (\ref{eq:960923c-rel}) twice }
	\ensuremath{\frac{\Varid{f} \times \Varid{g}}{\Varid{h} \times \Varid{k}}}
\qed
\end{eqnarray*}

\paragraph{Functorial metaphors}
Metaphor product rule (\ref{eq:160112a}) can be regarded as an instance of
a more general result: any relator \ensuremath{\fun F } \footnote{Recall from \secref{sec:170319a}
that a relator \ensuremath{\fun F } is an (endo)functor \ensuremath{\fun F } that preserves converses.}
distributes over a metaphor \ensuremath{\frac{\Varid{f}}{\Varid{g}}}:
\begin{eqnarray}
	\ensuremath{\fun F \;\frac{\Varid{f}}{\Varid{g}}\mathrel{=}\frac{\fun F \;\Varid{f}}{\fun F \;\Varid{g}}}
	\label{eq:160118d}
\end{eqnarray}
This result follows immediately from standard properties of relators, \ensuremath{\fun F \;(\Conid{R} \comp \Conid{S})\mathrel{=}(\fun F \;\Conid{R}) \comp (\fun F \;\Conid{S})} and
\ensuremath{\fun F \;(\conv{\Conid{R}})\mathrel{=}\conv{(\fun F \;\Conid{R})}}. Rule (\ref{eq:160112a}) corresponds to \ensuremath{\fun F \;(\Conid{R},\Conid{S})\mathrel{=}\Conid{R} \times \Conid{S}}, where
\ensuremath{\fun F } is binary. Thus
\begin{eqnarray}
	\ensuremath{\frac{\Varid{f}}{\Varid{g}}\mathbin{+}\frac{\Varid{h}}{\Varid{k}}} = \ensuremath{\frac{\Varid{f}\mathbin{+}\Varid{h}}{\Varid{g}\mathbin{+}\Varid{k}}}
	\label{eq:160118e}
\end{eqnarray}
also holds, where direct sum \ensuremath{\Conid{R}\mathbin{+}\Conid{S}} is the same as \ensuremath{\alt{i_1 \comp \Conid{R}}{i_2 \comp \Conid{S}}},
where \ensuremath{i_1} and \ensuremath{i_2} are the standard \emph{injections} associated to a datatype sum,
\(%begin{eqnarray*}
\myxym{
	\ensuremath{\Conid{A}}
		\ar[r]^-{\ensuremath{i_1}}
&
	\ensuremath{\Conid{A}\mathbin{+}\Conid{B}}
&
	\ensuremath{\Conid{B}}
		\ar[l]_-{\ensuremath{i_2}}
}
\) %end{eqnarray*}
and
\begin{eqnarray}
	\ensuremath{\larrow{\Conid{A}\mathbin{+}\Conid{B}}{\alt{\Conid{R}}{\Conid{S}}}{\Conid{C}}}
\end{eqnarray}
denotes the junction \ensuremath{{\Conid{R} \comp \conv{i_1}}\cup{\Conid{R} \comp \conv{i_2}}} of relations \ensuremath{\larrow{\Conid{A}}{\Conid{R}}{\Conid{C}}} and \ensuremath{\larrow{\Conid{B}}{\Conid{S}}{\Conid{C}}}.
By (\ref{eq:160118f}), one has
\begin{eqnarray}
	\ensuremath{\alt{\Varid{f}}{\Varid{g}}} \leq \ensuremath{\Varid{f}+\Varid{g}}
	\label{eq:150401a-fun}
\end{eqnarray}
since \ensuremath{\alt{\Varid{f}}{\Varid{g}}\mathrel{=}\alt{{id}}{{id}} \comp (\Varid{f}\mathbin{+}\Varid{g})} by coproduct laws.
Moreover,
\begin{eqnarray}
	\ensuremath{\Varid{f}+\Varid{g}} \leq \ensuremath{\Varid{h}+\Varid{k}} \wider\equiv \ensuremath{\Varid{f}\leq \Varid{h}} \land \ensuremath{\Varid{g}\leq \Varid{k}}
	\label{eq:150406j-fun}
\end{eqnarray}
holds by coproduct laws too, since
	\ensuremath{{\frac{\Varid{h}\mathbin{+}\Varid{k}}{\Varid{h}\mathbin{+}\Varid{k}}}\subseteq{\frac{\Varid{f}\mathbin{+}\Varid{g}}{\Varid{f}\mathbin{+}\Varid{g}}}} 
is equivalent to
	\ensuremath{{\frac{\Varid{h}}{\Varid{h}}\mathbin{+}\frac{\Varid{k}}{\Varid{k}}}\subseteq{\frac{\Varid{f}}{\Varid{f}}\mathbin{+}\frac{\Varid{g}}{\Varid{g}}}} by (\ref{eq:160118e}).

\paragraph{Difunctionality and uniformity}
A relation \ensuremath{\Conid{R}} is said to be \emph{difunctional} \cite{Ri48,Sc08} or
\emph{regular} \cite{JMBD91} wherever \ensuremath{\Conid{R} \comp \conv{\Conid{R}} \comp \Conid{R}\mathrel{=}\Conid{R}} holds, which
amounts to \ensuremath{\Conid{R} \comp \conv{\Conid{R}} \comp \Conid{R}\; \subseteq \;\Conid{R}} since the converse inclusion always
holds.

Metaphors are difunctional
because every symmetric division is so, as is easy to check by application
of laws (\ref{eq:151118d}) and (\ref{eq:151119a}).
%\begin{eqnarray*}
%\start
%	|sse (syd f g . (convsyd f g) . (syd f g)) (syd f g)|
%%
%\just\implied{ (\ref{eq:160112c}) ; (\ref{eq:160112d}) }
%%
%	|sse (syd f g . (syd f f)) (syd f g)|
%%
%\just\implied{ (\ref{eq:160112d}) }
%%
%	|sse (syd f g) (syd f g)|
%\qed
%\end{eqnarray*}
%For |g=id| above we get that any
The fact that every function \ensuremath{\Varid{f}} is difunctional can be expressed by \ensuremath{\Varid{f} \comp \frac{\Varid{f}}{\Varid{f}}\mathrel{=}\Varid{f}}.

A relation \ensuremath{\Conid{R}} is said to be \emph{uniform} \cite{JMBD91} if and only if \ensuremath{\Lambda{\Conid{R}}\leq \Conid{R}}, where
preorder (\ref{eq:160112f}) is extended to arbitrary relations
\begin{eqnarray}
	R \leq S & \wider\equiv & \ensuremath{\conv{\Conid{S}} \comp \Conid{S}} \subseteq \ensuremath{\conv{\Conid{R}} \comp \Conid{R}}
	\label{eq:041217a}
\end{eqnarray}
as in \cite{Ol14a}. 
Metaphors are uniform relations, because a relation is uniform iff it is
difunctional (regular), as the following calculation shows:
\begin{eqnarray*}
\start
	\mbox{\ensuremath{\Conid{R}} is uniform}
\just\equiv{ definition above }
	\ensuremath{\Lambda{\Conid{R}}\leq \Conid{R}}
\just\equiv{ (\ref{eq:041217a}) ; (\ref{eq:160117a}) }
	\ensuremath{{\conv{\Conid{R}} \comp \Conid{R}}\subseteq{\frac{\Lambda{\Conid{R}}}{\Lambda{\Conid{R}}}}}
\just\equiv{ \ensuremath{\Lambda } cancellation (\ref{eq:160108a}) }
	\ensuremath{{\conv{\Conid{R}} \comp \Conid{R}}\subseteq{\frac{\Conid{R}}{\Conid{R}}}}
\just\equiv{ universal property (\ref{eq:151118b}) of symmetric division }
	\ensuremath{{\Conid{R} \comp \conv{\Conid{R}} \comp \Conid{R}}\subseteq{\Conid{R}}}
\just\equiv{ definition above }
	\mbox{\ensuremath{\Conid{R}} is difunctional}
\qed
\end{eqnarray*}
Note how step
	\ensuremath{{\conv{\Conid{R}} \comp \Conid{R}}\subseteq{\frac{\Conid{R}}{\Conid{R}}}}
above captures the intuition about a regular (i.e.\ uniform, difunctional)
relation \ensuremath{\Conid{R}}, as given in the introduction, recall  (\ref{eq:170813a}):
\ensuremath{a_1 \;(\conv{\Conid{R}} \comp \Conid{R})\;a_2 } tells that \ensuremath{a_1 } and \ensuremath{a_2 } have \emph{some} common image;
\ensuremath{a_1 \;\frac{\Conid{R}}{\Conid{R}}\;a_2 } tells that they have \emph{exactly the same} image sets.

\paragraph{Functional metaphors}
When is a metaphor \ensuremath{\frac{\Varid{f}}{\Varid{g}}} a function? Shunting rules (\ref{eq:020617e},\ref{eq:020617f})
are equivalent to saying that \ensuremath{\Varid{f}} is total (or \emph{entire}) ---
\ensuremath{{{id}}\subseteq{\conv{\Varid{f}}\;\Varid{f}}}, which we have already seen in fraction notation (\ref{eq:160112e})
--- and deterministic (or \emph{simple}) --- \ensuremath{{\Varid{f} \comp \conv{\Varid{f}}}\subseteq{{id}}}.
We shall use notation \ensuremath{\ap{\rho}\Varid{f}\mathrel{=}\Varid{f} \comp \conv{\Varid{f}}} for the \emph{range} (of output values) of \ensuremath{\Varid{f}}.

By (\ref{eq:020617e},\ref{eq:020617f}), checking the totality of \ensuremath{\frac{\Varid{f}}{\Varid{g}}} ---
	\ensuremath{{{id}}\subseteq{\frac{\Varid{g}}{\Varid{f}}} \comp \frac{\Varid{f}}{\Varid{g}}}
--- amounts to \ensuremath{{\ap{\rho}\Varid{f}}\subseteq{\ap{\rho}\Varid{g}}}: the attribute value of any given vehicle is the
attribute value of some tenor. In case \ensuremath{\Varid{g}} is surjective (\ensuremath{\ap{\rho}\Varid{g}\mathrel{=}{id}}), \ensuremath{\frac{\Varid{f}}{\Varid{g}}} is total for any \ensuremath{\Varid{f}}.
For determinism to hold, \ensuremath{\ap{\rho}\frac{\Varid{f}}{\Varid{g}}\mathrel{=}\frac{\Varid{f}}{\Varid{g}} \comp \frac{\Varid{g}}{\Varid{f}}\; \subseteq \;{id}},
rule (\ref{eq:160112d}) offers the sufficient condition \ensuremath{\frac{\Varid{g}}{\Varid{g}}\mathrel{=}{id}}, that is, \ensuremath{\Varid{g}} injective
suffices.

For total metaphors, the inclusion \ensuremath{{\Varid{h}}\subseteq{\frac{\Varid{f}}{\Varid{g}}}} has at least a functional solution \ensuremath{\Varid{h}},
which can be calculated using the rule
\begin{eqnarray}
	\ensuremath{{\Varid{h}}\subseteq{\frac{\Varid{f}}{\Varid{g}}}~\Leftrightarrow~\Varid{g} \comp \Varid{h}\mathrel{=}\Varid{f}}
	\label{eq:160114a}
\end{eqnarray}
that relies on the useful law of \emph{function equality}
\begin{eqnarray}
	f \subseteq g \equiv f = g \equiv f \supseteq g
		\label{eq:020617g}
\end{eqnarray}
itself a follow-up of shunting rules (\ref{eq:020617e},\ref{eq:020617f}).

\section{Divide \& conquer metaphors} \label{sec:160125c}
\paragraph{``Shrinking" metaphors}
Thus far we have not taken into account the \emph{shrinking} part of (\ref{eq:150205e}),
which we now write using fraction notation:
\begin{eqnarray}
	\ensuremath{\Conid{M}\mathrel{=}{\frac{\Varid{f}}{\Varid{g}}}\shrunkby{\Conid{R}}}
	\label{eq:160123a}
\end{eqnarray}
By law (\ref{eq:fn-shrink-r}) one gets:
\begin{eqnarray*}
	\ensuremath{{\frac{\Varid{f}}{\Varid{g}}}\shrunkby{\Conid{R}}} = \ensuremath{({\frac{{id}}{\Varid{g}}}\shrunkby{\Conid{R}}) \comp \Varid{f}}
\end{eqnarray*}
Below we will show that this equality is an instance of a more general result
that underlies more elaborate metaphor transformations that prove useful
in the sequel. The main idea of such transformations is to split a \ensuremath{\larrow{\mathsf{V}}{}{\fun T }} metaphor in two parts mediated by an intermediate type, say \ensuremath{\fun W } in
\[ \xymatrix{\ensuremath{\fun T }&\ensuremath{\fun W }\ar[l]&\ensuremath{\mathsf{V}}\ar[l]} \] which is intended to gain control
of the ``pipeline''. This can be done in two ways. Suppose there is
a surjection \ensuremath{\Varid{h}} : \ensuremath{\fun W \to \fun T } onto the tenor side, that is,
\ensuremath{\ap{\rho}\Varid{h}\mathrel{=}\Varid{h} \comp \conv{\Varid{h}}\mathrel{=}{id}}. Then the splitting can be expressed as in the
following diagram
\begin{eqnarray}
\vcenter{\xymatrix@C=2.5ex@R=1ex{
	\ensuremath{\fun T }
&&&
	\ensuremath{\fun W }
		\ar[lll]_{\ensuremath{\Varid{h}}}
		\ar[rd]_{\ensuremath{\Varid{h}}}
&
&
&
	\ensuremath{\mathsf{V}}
		\ar[ldd]^{\ensuremath{\Varid{f}}}
		\ar[lll]_{\ensuremath{\Conid{X}}}
		\ar@{.>}@/_1.8pc/[llllll]_(.45){\ensuremath{{\frac{\Varid{f}}{\Varid{g}}}\shrunkby{\Conid{R}}}}
\\
&
&
&
&
	\ensuremath{\fun T }
		\ar[rd]_{\ensuremath{\Varid{g}}}
\\
&
&
&
&
&
	\ensuremath{\Conid{A}}
}}
	\label{eq:160124a}
\end{eqnarray}
provided one can find a relation \ensuremath{\Conid{X}} such that \ensuremath{\Varid{h} \comp \Conid{X}\mathrel{=}{\frac{\Varid{f}}{\Varid{g}}}\shrunkby{\Conid{R}}}.
Alternatively, we can imagine \emph{surjection} \ensuremath{\Varid{h}} onto the vehicle side, say \ensuremath{\Varid{h}} : \ensuremath{\fun W \to \mathsf{V}} in
\begin{eqnarray}
\vcenter{\xymatrix@C=2.3ex@R=1.4ex{
	\ensuremath{\fun T }
		\ar[rdd]_{\ensuremath{\Varid{g}}}
&
&
&
	\ensuremath{\fun W }
		\ar[lll]_{\ensuremath{\Conid{Y}}}
		\ar[ld]^{\ensuremath{\Varid{h}}}
&	
&	
	\ensuremath{\mathsf{V}}
		\ar[ll]_{\ensuremath{\conv{\Varid{h}}}}
		\ar@{.>}@/_1.8pc/[lllll]_(.45){\ensuremath{{\frac{\Varid{f}}{\Varid{g}}}\shrunkby{\Conid{R}}}}
\\
&
&
	\ensuremath{\mathsf{V}}
		\ar[ld]^{\ensuremath{\Varid{f}}}
\\
&
	\ensuremath{\Conid{A}}
&
}}
	\label{eq:160124b}
\end{eqnarray}
and try and find relation \ensuremath{\Conid{Y}} such that \ensuremath{\Conid{Y} \comp \conv{\Varid{h}}\mathrel{=}{\frac{\Varid{f}}{\Varid{g}}}\shrunkby{\Conid{R}}}.

Note how intermediate type \ensuremath{\fun W } is a \emph{representation} of \ensuremath{\fun T } or \ensuremath{\mathsf{V}}
in, respectively, (\ref{eq:160124a}) and (\ref{eq:160124b}), \ensuremath{\Varid{h}} acting as
a typical data refinement \emph{abstraction} function.\footnote{\label{fn:170517a}Following
the usual terminology \cite{Ol08a}, by an \emph{abstraction} we mean a \emph{simple}
(ie.\ functional) and \emph{surjective} relation. In this paper all abstractions are total (entire),
that is, they are functions. In symbols, \ensuremath{\alpha } is an abstraction \emph{function}
iff \ensuremath{{id}\; \subseteq \;\conv{\alpha } \comp \alpha } and \ensuremath{{id}\mathrel{=}\alpha  \comp \conv{\alpha }}.} Anticipating that the two-stage
schemas of (\ref{eq:160124a}) and (\ref{eq:160124b}) are intended to specify
\emph{divide \& conquer} implementations of the original metaphor, let us
calculate \emph{conquer} step \ensuremath{\Conid{Y}} in the first place:
\begin{eqnarray*}
\start
	\ensuremath{{\frac{\Varid{f}}{\Varid{g}}}\shrunkby{\Conid{R}}}
\just={ identity of composition }
	\ensuremath{({\frac{\Varid{f}}{\Varid{g}}}\shrunkby{\Conid{R}}) \comp {id}}
\just={ \ensuremath{\Varid{h}} assumed to be a surjection, \ensuremath{\ap{\rho}\Varid{h}\mathrel{=}\Varid{h} \comp \conv{\Varid{h}}\mathrel{=}{id}} }
	\ensuremath{({\frac{\Varid{f}}{\Varid{g}}}\shrunkby{\Conid{R}}) \comp \Varid{h} \comp \conv{\Varid{h}}}
\just={ law (\ref{eq:fn-shrink-r}) }
	\ensuremath{\underbrace{({\frac{\Varid{f} \comp \Varid{h}}{\Varid{g}}}\shrunkby{\Conid{R}})}_{\Conid{Y}} \comp \conv{\Varid{h}}}
\end{eqnarray*}
Clearly, in this refinement strategy, the optimization of the starting metaphor goes into
the \emph{conquer} stage, where it optimizes a richer metaphor between tenor \ensuremath{\fun T } and \ensuremath{\fun W },
the new vehicle. \emph{Divide} step \ensuremath{\conv{\Varid{h}}} is just a representation of the original vehicle \ensuremath{\mathsf{V}} into
the new vehicle \ensuremath{\fun W } (\ref{eq:160124b}). Altogether:
\begin{eqnarray}
	\ensuremath{{\frac{\Varid{f}}{\Varid{g}}}\shrunkby{\Conid{R}}} &=& \ensuremath{({\frac{\Varid{f} \comp \Varid{h}}{\Varid{g}}}\shrunkby{\Conid{R}}) \comp \conv{\Varid{h}}} ~~~~~~~~~ \mbox{for \ensuremath{\Varid{h}} surjective }
	\label{eq:160124f}
\end{eqnarray}
In a diagram, completing (\ref{eq:160124b}):
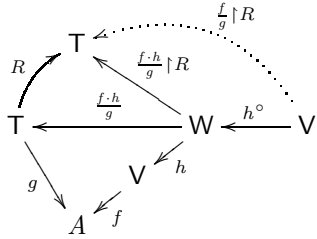
\begin{eqnarray*}
\xymatrix@C=2.3ex@R=1.4ex{
&
	\ensuremath{\fun T }
\\
\\
	\ensuremath{\fun T }
		\ar@/^0.5pc/[ruu]^(.5){\ensuremath{\Conid{R}}}
		\ar[rdd]_{\ensuremath{\Varid{g}}}
&
&
&
	\ensuremath{\fun W }
		\ar[lll]_{\ensuremath{\frac{\Varid{f} \comp \Varid{h}}{\Varid{g}}}}
		\ar[lluu]_{\ensuremath{{\frac{\Varid{f} \comp \Varid{h}}{\Varid{g}}}\shrunkby{\Conid{R}}}}
		\ar[ld]^{\ensuremath{\Varid{h}}}
&	
&	
	\ensuremath{\mathsf{V}}
		\ar[ll]_{\ensuremath{\conv{\Varid{h}}}}
		\ar@{.>}@/_1.8pc/[lllluu]_(.45){\ensuremath{{\frac{\Varid{f}}{\Varid{g}}}\shrunkby{\Conid{R}}}}
\\
&
&
	\ensuremath{\mathsf{V}}
		\ar[ld]^{\ensuremath{\Varid{f}}}
\\
&
	\ensuremath{\Conid{A}}
&
}
\end{eqnarray*}

Dually, it is to be expected that the derivation of \ensuremath{\Conid{X}} in (\ref{eq:160124a})
will yield an optimized \emph{divide} step where most of the work goes, running
\ensuremath{\Varid{h}} as \emph{conquer} step to abstract from \ensuremath{\fun W }, the new tenor, to \ensuremath{\fun T }, the
old tenor. Due to the asymmetry of \emph{shrinking}, the inference of \ensuremath{\Conid{X}}
is less immediate, calling for definition (\ref{eq:100211c}):
% The calculation of |X| is as follows:
\begin{eqnarray*}
\start
	\ensuremath{{\frac{\Varid{f}}{\Varid{g}}}\shrunkby{\Conid{R}}}
\just={ (\ref{eq:100211c}) ; converse of a metaphor (\ref{eq:160112c}) }
	\ensuremath{{\frac{\Varid{f}}{\Varid{g}}}\mathbin\cap{\Conid{R}\mathbin{/}\frac{\Varid{g}}{\Varid{f}}}}
\just={ \ensuremath{\Varid{h}} assumed to be a surjection, \ensuremath{\ap{\rho}\Varid{h}\mathrel{=}\Varid{h} \comp \conv{\Varid{h}}\mathrel{=}{id}} }
	\ensuremath{\Varid{h} \comp \conv{\Varid{h}} \comp ({\frac{\Varid{f}}{\Varid{g}}}\mathbin\cap{\Conid{R}\mathbin{/}\frac{\Varid{g}}{\Varid{f}}})}
\just={ injective \ensuremath{\conv{\Varid{h}}} distributes over \ensuremath{\cap } ; (\ref{eq:151118c}) }
	\ensuremath{\Varid{h} \comp ({\frac{\Varid{f}}{\Varid{g} \comp \Varid{h}}}\mathbin\cap{\conv{\Varid{h}} \comp \Conid{R}\mathbin{/}\frac{\Varid{g}}{\Varid{f}}})}
\just={ (\ref{eq:100707a}) ; shunting (\ref{eq:020617f}) }
	\ensuremath{\Varid{h} \comp \underbrace{({\frac{\Varid{f}}{\Varid{g} \comp \Varid{h}}}\mathbin\cap{\conv{\Varid{h}} \comp (\Conid{R}\mathbin{/}\Varid{g}) \comp \Varid{f}})}_{\Conid{X}}}
\end{eqnarray*}
Clearly, the choice of some intermediate \ensuremath{\Varid{w}} by \ensuremath{\Conid{X}} tells where the optimization
has moved to, as detailed below by rendering \ensuremath{\Conid{X}} in pointwise notation:
%|w X v <=> (g(h w) = f v) && (rcb forall t (f v = g t)((h w) R t))|
\begin{hscode}\SaveRestoreHook
\column{B}{@{}>{\hspre}l<{\hspost}@{}}%
\column{5}{@{}>{\hspre}l<{\hspost}@{}}%
\column{E}{@{}>{\hspre}l<{\hspost}@{}}%
\>[B]{}\Varid{w}\;\Conid{X}\;\Varid{v}~\Leftrightarrow~{}\<[E]%
\\
\>[B]{}\hsindent{5}{}\<[5]%
\>[5]{}\mathbf{let}\;\Varid{a}\mathrel{=}\Varid{f}\;\Varid{v}{}\<[E]%
\\
\>[B]{}\hsindent{5}{}\<[5]%
\>[5]{}\mathbf{in}\;(\Varid{g}\;(\Varid{h}\;\Varid{w})\mathrel{=}\Varid{a})\mathrel{\wedge}\rcb{\forall}{\Varid{t}}{\Varid{a}\mathrel{=}\Varid{g}\;\Varid{t}}{(\Varid{h}\;\Varid{w})\;\Conid{R}\;\Varid{t}}{}\<[E]%
\ColumnHook
\end{hscode}\resethooks
In words:
\begin{quote}\em
Given vehicle \ensuremath{\Varid{v}}, \ensuremath{\Conid{X}} will select those \ensuremath{\Varid{w}} that represent tenors (\ensuremath{\Varid{h}\;\Varid{w}})
with the same attribute (\ensuremath{\Varid{a}}) as vehicle \ensuremath{\Varid{v}}, and that are best among all
other tenors \ensuremath{\Varid{t}} exhibiting the same attribute \ensuremath{\Varid{a}}.
\end{quote}
Altogether:
\begin{eqnarray}
	\ensuremath{{\frac{\Varid{f}}{\Varid{g}}}\shrunkby{\Conid{R}}} &=&
	\ensuremath{\Varid{h} \comp ({\frac{\Varid{f}}{\Varid{g} \comp \Varid{h}}}\mathbin\cap{\conv{\Varid{h}} \comp (\Conid{R}\mathbin{/}\Varid{g}) \comp \Varid{f}})}
	~~~~~~~~~ \mbox{for \ensuremath{\Varid{h}} surjective }
	\label{eq:160124e}
\end{eqnarray}
A calculation similar to that showing \ensuremath{\frac{\Varid{f}}{\Varid{g}}} difunctional above,
will show that \ensuremath{\Conid{R}\mathbin{/}\Varid{g}} being difunctional is sufficient for factor
\ensuremath{\conv{\Varid{h}} \comp (\Conid{R}\mathbin{/}\Varid{g}) \comp \Varid{f}} in (\ref{eq:160124e}) to be so.

\paragraph{Post-conditioned metaphors}
Let us finally consider the following pattern of metaphor shrinking
\begin{eqnarray}
	\ensuremath{{\frac{\Varid{f}}{\Varid{g}}}\shrunkby{\frac{\Varid{true}}{\Varid{q}}}}
	\label{eq:160124c}
\end{eqnarray}
indicating that only the outputs satisfying \ensuremath{\Varid{q}} are regarded as good enough.
That is, \ensuremath{\Varid{q}} acts as a \emph{post-condition} on \ensuremath{\frac{\Varid{f}}{\Varid{g}}}. 
An example of (\ref{eq:160124c}) is the metaphor
\begin{eqnarray*}
	\ensuremath{\Conid{Sort}\mathrel{=}{\frac{\Varid{bag}}{\Varid{bag}}}\shrunkby{\frac{\Varid{true}}{\Varid{ordered}}}}
\end{eqnarray*}
where \ensuremath{\Varid{bag}} is the function that extracts the bag (multiset) of elements of a finite list
and \ensuremath{\Varid{ordered}} the predicate that checks whether a finite list is ordered according to some
predefined criterion. Clearly, \ensuremath{\mathsf{V}\mathrel{=}\fun T } in this example.

The laws developed above for metaphor shrinking can be instantiated for this pattern and reasoned about.
Alternatively, it can easily be shown that (\ref{eq:160124c}) reduces to
\begin{eqnarray}
	\ensuremath{{\frac{\Varid{f}}{\Varid{g}}}\mathbin\cap{\Varid{q}\hskip-1pt\mathit{?} \comp \top }}
	\label{eq:160124d}
\end{eqnarray}
provided \ensuremath{\frac{\Varid{f}}{\Varid{g}}} is entire (total), which is surely the case wherever \ensuremath{\Varid{f}\mathrel{=}\Varid{g}}, as we have seen.
This follows from this law of the shrinking operator proved in % Appendix
	\ref{sec:150329b}:
\begin{eqnarray}
	S \shrunkby (\ensuremath{\Varid{q}\hskip-1pt\mathit{?}} \comp \top) = \ensuremath{\Varid{q}\hskip-1pt\mathit{?}}\comp S ~~~ \implied ~~~~ \mbox{$S$ is entire }
	\label{eq:150214b}
\end{eqnarray}

Specifications of the form \ensuremath{{\frac{\Varid{f}}{\Varid{f}}}\mathbin\cap{\Varid{q}\hskip-1pt\mathit{?} \comp \top }}
are intersections of an \emph{equivalence relation with a rectangular relation},
a common specification pattern already identified by \citet{JMBD91}.
As intersections of rational relations are rational (regular) relations, pattern \ensuremath{{\frac{\Varid{f}}{\Varid{f}}}\mathbin\cap{\Varid{q}\hskip-1pt\mathit{?} \comp \top }} is
rational. By (\ref{eq:081025a}), (\ref{eq:160124d}) further reduces to
\begin{eqnarray*}
	\ensuremath{\Varid{q}\hskip-1pt\mathit{?} \comp \frac{\Varid{f}}{\Varid{g}}}
\end{eqnarray*}
a pattern to be referred to as a \emph{postconditioned metaphor}.
Sorting thus is one such metaphor,
\begin{eqnarray}
	\ensuremath{\Conid{Sort}\mathrel{=}\Varid{ordered}\hskip-1pt\mathit{?} \comp \Conid{Perm}} ~~~\ensuremath{\mathbf{where}} ~~~~ \ensuremath{\Conid{Perm}\mathrel{=}\frac{\Varid{bag}}{\Varid{bag}}}
	\label{eq:160118b}
\end{eqnarray}
where \ensuremath{\Varid{y}\;\Conid{Perm}\;\Varid{x}} means that \ensuremath{\Varid{y}} is a \emph{permutation} of \ensuremath{\Varid{x}}.

Understandably, the \emph{divide \& conquer} versions of a postconditioned
metaphor are easier to calculate than in the generic cases above,
because one can take advantage of \textsc{wp} laws such as e.g.\ (\ref{eq:150406c}).
Corresponding to (\ref{eq:160124e}), one gets
\begin{eqnarray}
	\ensuremath{\Varid{q}\hskip-1pt\mathit{?} \comp \frac{\Varid{f}}{\Varid{g}}\mathrel{=}\Varid{h} \comp \Varid{p}\hskip-1pt\mathit{?} \comp \frac{\Varid{f}}{\Varid{g} \comp \Varid{h}}}
	~~~~~~~~~ \mbox{for \ensuremath{\Varid{h}} surjective and \ensuremath{\Varid{p}\mathrel{=}\Varid{q} \comp \Varid{h}}}
	\label{eq:160125a}
\end{eqnarray}
since:
\begin{eqnarray*}
\start
	\ensuremath{\Varid{q}\hskip-1pt\mathit{?} \comp {id} \comp \frac{\Varid{f}}{\Varid{g}}}
\just={ \ensuremath{\Varid{h}} assumed surjective }
	\eqnnewpage
	\ensuremath{\Varid{q}\hskip-1pt\mathit{?} \comp \Varid{h} \comp \conv{\Varid{h}} \comp \frac{\Varid{f}}{\Varid{g}}}
\just={ switch to \textsc{wp} \ensuremath{\Varid{p}} (\ref{eq:150406c}), cf.\ \ensuremath{\Varid{q}\hskip-1pt\mathit{?} \comp \Varid{h}\mathrel{=}\Varid{h} \comp \Varid{p}\hskip-1pt\mathit{?}} }
	\ensuremath{\Varid{h} \comp \underbrace{\Varid{p}\hskip-1pt\mathit{?} \comp \frac{\Varid{f}}{\Varid{g} \comp \Varid{h}}}_{\Conid{X}}}
\end{eqnarray*}
The counterpart of (\ref{eq:160124f}) is even more immediate:
\begin{eqnarray}
	\ensuremath{\Varid{q}\hskip-1pt\mathit{?} \comp \frac{\Varid{f}}{\Varid{g}}\mathrel{=}\underbrace{\Varid{q}\hskip-1pt\mathit{?} \comp \frac{\Varid{f} \comp \Varid{h}}{\Varid{g}}}_{\Conid{Y}} \comp \conv{\Varid{h}}}
	~~~~~~~~~ \mbox{for \ensuremath{\Varid{h}} surjective}
	\label{eq:160125d}
\end{eqnarray}

\section{Metaphorisms} \label{sec:160125e}
%subsection{Representation changers as metaphorisms} \label{sec:160122c}
Thus far, types \ensuremath{\fun T }, \ensuremath{\mathsf{V}} and \ensuremath{\fun W } have been left uninterpreted.
We want now to address metaphors in which these 
are inductive (tree-like) types specified by initial algebras,
say
\ensuremath{\larrow{\fun F \;\fun T }{\mathsf{in}_{\fun F}}{\fun T }},
\ensuremath{\larrow{\fun G \;\fun W }{\mathsf{in}_{\fun G}}{\fun W }} and
\ensuremath{\larrow{\fun H \;\mathsf{V}}{\mathsf{in}_{\fun H}}{\mathsf{V}}},
assuming such algebras exist for functors \ensuremath{\fun F }, \ensuremath{\fun G } and \ensuremath{\fun H }, respectively.
Moreover, \ensuremath{\Varid{f}}, \ensuremath{\Varid{g}} and \ensuremath{\Varid{h}} become folds (catamorphisms) over such initial
types, recall \secref{sec:170319a}. We shall refer to such metaphors involving catamorphisms
over inductive types as \emph{metaphorisms} \cite{Ol15a}.

To facilitate linking each type with its functor, we shall adopt the familiar notation
\ensuremath{\muF } instead of \ensuremath{\fun T }, \ensuremath{\muG } instead of \ensuremath{\fun W } and \ensuremath{\muH } instead of \ensuremath{\mathsf{V}}.
The popular % fold (catamorphism)
notation \ensuremath{\mathopen{(\!|}\Conid{R}\mathclose{|\!)}} will be used to express % (relational, in general)
folds over such types, recall (\ref{eq:cataUniv-rel}). 
Also useful in the sequel is the fact that inductive predicates can be expressed by
folds too, in the form of partial identities:%
\footnote{\label{fn:170409a} %
Property (\ref{eq:cataUniv-rel}) establishes \ensuremath{\mathopen{(\!|}\Conid{R}\mathclose{|\!)}} as the unique fixpoint of the equation
\ensuremath{\Conid{X}\mathrel{=}\Conid{R} \comp (\fun F \;\Conid{X}) \comp \conv{\mathsf{in}_{\fun F}}}, and therefore the \emph{least prefix point} too:
	\ensuremath{{\mathopen{(\!|}\Conid{R}\mathclose{|\!)}}\subseteq{\Conid{X}}} \ensuremath{\Leftarrow} \ensuremath{{\Conid{R} \comp (\fun F \;\Conid{X}) \comp \conv{\mathsf{in}_{\fun F}}}\subseteq{\Conid{X}}} \cite{BM97}.
From this, % and (\ref{eq:150402b})
one can easily infer (\ref{eq:170409b}), that fold is a monotonic operator, etc. 
}
\begin{eqnarray}
	\ensuremath{\mathopen{(\!|}\Conid{R}\mathclose{|\!)}\; \subseteq \;{id}} & \implied & \ensuremath{\Conid{R}\; \subseteq \;\mathsf{in}_{\fun F}}
	\label{eq:170409b}
\end{eqnarray}

Our first example of metaphorism calculation by fusion (\ref{eq:150402a}) is the derivation of a simple
(functional) \emph{representation changer} \cite{HM93b}:
\begin{quote}\em
A representation changer is a function that converts a concrete representation
of an abstract value into a different concrete representation of that value.
\end{quote}
Metaphorisms of the form \ensuremath{\frac{\Varid{k} \comp \mathopen{(\!|}\Varid{y}\mathclose{|\!)}}{\mathopen{(\!|}\Varid{y}\mathclose{|\!)}}} are representation changers,
in which the change of representation consists in picking an attribute of the vehicle,
extracted by \ensuremath{\mathopen{(\!|}\Varid{y}\mathclose{|\!)}}, changing its value by applying \ensuremath{\Varid{k}} and then mapping the new
attribute value back to the tenor, which in this case is of the same type as the vehicle.

\begin{theorem}
Representation changer \ensuremath{\frac{\Varid{k} \comp \mathopen{(\!|}\Varid{y}\mathclose{|\!)}}{\mathopen{(\!|}\Varid{y}\mathclose{|\!)}}} is refined by a functional implementation \ensuremath{\mathopen{(\!|}\Varid{x}\mathclose{|\!)}} provided, for some \ensuremath{\Varid{z}}
such that \ensuremath{\Varid{k} \comp \Varid{y}\mathrel{=}\Varid{z} \comp \fun F \;\Varid{k}} holds, \ensuremath{\Varid{x}\; \subseteq \;\frac{\Varid{z} \comp \fun F \;\mathopen{(\!|}\Varid{y}\mathclose{|\!)}}{\mathopen{(\!|}\Varid{y}\mathclose{|\!)}}} also holds.
Proof: the proof relies on (double) fusion (\ref{eq:150402a}):
%, using abbreviation |(cata y)=cata y| where convenient:
\begin{eqnarray*}
\start
	\ensuremath{{\mathopen{(\!|}\Varid{x}\mathclose{|\!)}}\subseteq{\frac{\Varid{k} \comp \mathopen{(\!|}\Varid{y}\mathclose{|\!)}}{\mathopen{(\!|}\Varid{y}\mathclose{|\!)}}}}
\just\equiv{ (\ref{eq:160114a}) }
%
%	|(cata y) . (cata x) = k . (cata y)|
%
%
	\ensuremath{\mathopen{(\!|}\Varid{y}\mathclose{|\!)} \comp \mathopen{(\!|}\Varid{x}\mathclose{|\!)}\mathrel{=}\Varid{k} \comp \mathopen{(\!|}\Varid{y}\mathclose{|\!)}}
\just\equiv{ fuse \ensuremath{\Varid{k} \comp \mathopen{(\!|}\Varid{y}\mathclose{|\!)}} into \ensuremath{\mathopen{(\!|}\Varid{z}\mathclose{|\!)}} assuming \ensuremath{\Varid{k} \comp \Varid{y}\mathrel{=}\Varid{z} \comp \fun F \;\Varid{k}} (\ref{eq:150402a}) }
	\ensuremath{\mathopen{(\!|}\Varid{y}\mathclose{|\!)} \comp \mathopen{(\!|}\Varid{x}\mathclose{|\!)}\mathrel{=}\mathopen{(\!|}\Varid{z}\mathclose{|\!)}}
\just\implied{ fusion (\ref{eq:150402a}) again }
	\ensuremath{\mathopen{(\!|}\Varid{y}\mathclose{|\!)} \comp \Varid{x}\mathrel{=}\Varid{z} \comp \fun F \;\mathopen{(\!|}\Varid{y}\mathclose{|\!)}}
\just\equiv{ metaphors (\ref{eq:160114a}) }
	\ensuremath{\Varid{x}\; \subseteq \;\frac{\Varid{z} \comp \fun F \;\mathopen{(\!|}\Varid{y}\mathclose{|\!)}}{\mathopen{(\!|}\Varid{y}\mathclose{|\!)}}}
\end{eqnarray*}
\ensuremath{\ensuremath{\Box}}
\end{theorem}
Comparing the top and bottom lines of the calculation above we see that the
``banana brackets'' of \ensuremath{\mathopen{(\!|}\Varid{x}\mathclose{|\!)}} have disappeared.
This condition, together with the intermediate assumption \ensuremath{\Varid{k} \comp \Varid{y}\mathrel{=}\Varid{z} \comp \fun F \;\Varid{k}}, are sufficient
for the refinement to take place.

%Reference \cite{HM93b} gives an example of the situation above where |cata y|
%abstracts from the so-called \emph{carry-save} representation of natural
%numbers (|Nat0|) used in digital circuits
The example of application of this theorem given below is a quite simple one,
its purpose being mainly to illustrate the calculational style which will be followed
in the rest of the paper to derive programs from metaphorisms.
Let the initial algebra for finite lists be denoted by the familiar
\begin{eqnarray}
	\ensuremath{\mathsf{in}_{\fun F}\mathrel{=}\alt{\mathsf{nil}}{\mathsf{cons}}}
	\label{eq:160118a}
\end{eqnarray}
where \ensuremath{\mathsf{nil}\;\anonymous \mathrel{=}[\mskip1.5mu \mskip1.5mu]} is the constant function which yields the empty list and \ensuremath{\mathsf{cons}\;(\Varid{a},\Varid{s})\mathrel{=}\Varid{a}\mathbin{:}\Varid{s}}
adds \ensuremath{\Varid{a}} to the front of \ensuremath{\Varid{s}}. The underlying functor is \ensuremath{\fun F \;\Varid{f}\mathrel{=}{id}\mathbin{+}{id} \times \Varid{f}}, recall (\ref{eq:170518a}).
Let \ensuremath{\mathsf{add}\;(\Varid{x},\Varid{y})\mathrel{=}\Varid{x}\mathbin{+}\Varid{y}} denote natural number addition
and \ensuremath{\Varid{k}\mathrel{=}(\Varid{b}\mathbin{+})} be the unary function that adds \ensuremath{\Varid{b}} to its argument.
Define \ensuremath{\Varid{y}\mathrel{=}\alt{\mathsf{zero}}{\mathsf{add}}} where \ensuremath{\mathsf{zero}} is the everywhere-\ensuremath{\mathrm{0}} constant function.
So \ensuremath{\mathsf{sum}\mathrel{=}\mathopen{(\!|}\Varid{y}\mathclose{|\!)}} is the function which sums all elements of a list.

The intended change of representation between a vehicle \ensuremath{\Varid{v}} and tenor \ensuremath{\Varid{t}} is specified by
\ensuremath{\mathsf{sum}\;\Varid{t}\mathrel{=}\Varid{b}\mathbin{+}\mathsf{sum}\;\Varid{v}}.
% |zero _ = 0| yields |0| when the list is empty, otherwise |add(h,t)=h+t|
% adds the head |h| to the sum of tail |t|.  Let |k=(b+)| as in \cite{HM93b}.
Clearly, \ensuremath{(\Varid{b}\mathbin{+}) \comp \alt{\mathsf{zero}}{\mathsf{add}}\mathrel{=}\Varid{z} \comp ({id}\mathbin{+}{id} \times (\Varid{b}\mathbin{+}))}
has solution \ensuremath{\Varid{z}\mathrel{=}\alt{\kons{\Varid{b}}}{\mathsf{add}}}, since \ensuremath{\Varid{b}\mathbin{+}\mathrm{0}\mathrel{=}\Varid{b}} and \ensuremath{\Varid{b}\mathbin{+}(\Varid{h}\mathbin{+}\Varid{t})\mathrel{=}\Varid{h}\mathbin{+}(\Varid{b}\mathbin{+}\Varid{t})}.
Knowing \ensuremath{\Varid{z}}, our aim is to solve \ensuremath{\Varid{x}\; \subseteq \;\frac{\Varid{z} \comp \fun F \;\mathopen{(\!|}\Varid{y}\mathclose{|\!)}}{\mathopen{(\!|}\Varid{y}\mathclose{|\!)}}}
for \ensuremath{\Varid{x}\mathrel{=}\alt{x_1}{x_2}}, helped by the following law
\begin{eqnarray}
	\ensuremath{\alt{\Varid{x}}{\Varid{y}}\; \subseteq \;\frac{\alt{\Varid{g}}{\Varid{h}}}{\Varid{f}}} \ensuremath{~\Leftrightarrow~} \ensuremath{\Varid{x}\; \subseteq \;\frac{\Varid{g}}{\Varid{f}}\mathrel{\wedge}\Varid{y}\; \subseteq \;\frac{\Varid{h}}{\Varid{f}}}
	%label{eq:170701b}
\end{eqnarray}
easy to infer by coproduct and metaphor algebra.

Applied to our example, this yields
\ensuremath{x_1\; \subseteq \;\frac{\kons{\Varid{b}}}{\mathopen{(\!|}\Varid{y}\mathclose{|\!)}}} and \ensuremath{x_2\; \subseteq \;\frac{\mathsf{add} \comp ({id} \times \mathopen{(\!|}\Varid{y}\mathclose{|\!)})}{\mathopen{(\!|}\Varid{y}\mathclose{|\!)}}}, the latter equivalent to
\ensuremath{\mathopen{(\!|}\Varid{y}\mathclose{|\!)} \comp x_2\mathrel{=}\mathsf{add} \comp ({id} \times \mathopen{(\!|}\Varid{y}\mathclose{|\!)})}. From this  we get \ensuremath{x_2\mathrel{=}\mathsf{cons}} by cancellation (\ref{eq:150402b}).
%since |f=cata (either zero add)|;
%	|f(x2(h,t)) = h + (f t)|
On the other hand,
\ensuremath{x_1} is necessarily a constant function \ensuremath{\kons{\Varid{w}}} such that \ensuremath{\mathopen{(\!|}\Varid{y}\mathclose{|\!)}\;\Varid{w}\mathrel{=}\Varid{b}}.
The simplest choice for \ensuremath{\Varid{w}} is the singleton list
\ensuremath{[\mskip1.5mu \Varid{b}\mskip1.5mu]}. We therefore obtain the following functional solution for the given metaphor, unfolding \ensuremath{\Varid{r}\mathrel{=}\mathopen{(\!|}\Varid{x}\mathclose{|\!)}}
to pointwise notation:
\begin{eqnarray*}
\ensuremath{\Varid{r}\;[\mskip1.5mu \mskip1.5mu]} & = & \ensuremath{[\mskip1.5mu \Varid{b}\mskip1.5mu]}
\\
\ensuremath{\Varid{r}\;(\Varid{a}\mathbin{:}\Varid{t})} & = & \ensuremath{\Varid{a}\mathbin{+}\Varid{r}\;\Varid{t}}
\end{eqnarray*}

\section{Shrinking metaphorisms into hylomorphisms}\label{sec:150406g}
%section{More on metaphorism calculation} % Shrunken equivalence relations as metaphorisms
%label{sec:150406d}%
This section focusses on metaphorisms that  are equivalence relations over
inductive data types. Let \ensuremath{\larrow{\fun F \;\muF }{\mathsf{in}_{\fun F}}{\muF }}, and  let \ensuremath{\larrow{\fun F \;\Conid{A}}{\Varid{k}}{\Conid{A}}} be given,
so that \ensuremath{\larrow{\muF }{\mathopen{(\!|}\Varid{k}\mathclose{|\!)}}{\Conid{A}}}.
It turns out that not only is \ensuremath{\Conid{R}\mathrel{=}\frac{\mathopen{(\!|}\Varid{k}\mathclose{|\!)}}{\mathopen{(\!|}\Varid{k}\mathclose{|\!)}}} itself a relational fold
\begin{eqnarray*}
	\ensuremath{\Conid{R}\mathrel{=}\mathopen{(\!|}\Conid{R} \comp \mathsf{in}_{\fun F}\mathclose{|\!)}}
\end{eqnarray*}
of type \ensuremath{\larrow{\muF }{}{\muF }},
but also it is a \emph{congruence} for the algebra \ensuremath{\mathsf{in}_{\fun F}}.\footnote{The \ensuremath{\Conid{Perm}}
equivalence relation is an example of this, recall (\ref{eq:160118b}).}
This follows from the following theorem.

%paragraph{|fF|-congruences}
\begin{theorem}[\ensuremath{\fun F }-congruences]\label{th:150327a}
Let \ensuremath{\Conid{R}} be a congruence for an algebra \ensuremath{\Varid{h}\mathbin{:}\fun F \;\Conid{A}\to \Conid{A}} of functor \ensuremath{\fun F }, that is
\begin{eqnarray}
	\ensuremath{\Varid{h} \comp (\fun F \;\Conid{R})\; \subseteq \;\Conid{R} \comp \Varid{h}} & ~~ i.e. ~~ &
	 \ensuremath{\Varid{y}\;(\fun F \;\Conid{R})\;\Varid{x}\Rightarrow (\Varid{h}\;\Varid{y})\;\Conid{R}\;(\Varid{h}\;\Varid{x})}
	\label{eq:150326a}
\end{eqnarray}
hold and \ensuremath{\Conid{R}} is an equivalence relation.
Then (\ref{eq:150326a}) is equivalent to:
\begin{eqnarray}
	\ensuremath{\Conid{R} \comp \Varid{h}\mathrel{=}\Conid{R} \comp \Varid{h} \comp (\fun F \;\Conid{R})}
	\label{eq:150326b}
\end{eqnarray}
For the particular case \ensuremath{\Varid{h}\mathrel{=}\mathsf{in}_{\fun F}}, (\ref{eq:150326b}) is equivalent to:
\begin{eqnarray}
	\ensuremath{\Conid{R}\mathrel{=}\mathopen{(\!|}\Conid{R} \comp \mathsf{in}_{\fun F}\mathclose{|\!)}}
	\label{eq:160118c}
\end{eqnarray}
For \ensuremath{\Conid{R}} presented as a kernel metaphor \ensuremath{\Conid{R}\mathrel{=}\frac{\Varid{f}}{\Varid{f}}}, (\ref{eq:150326a}) is also equivalent to
\begin{eqnarray}
	\ensuremath{\Varid{f} \comp \Varid{h}\leq \fun F \;\Varid{f}}
	\label{eq:160120c}
\end{eqnarray}
where \ensuremath{\leq } is the injectivity preorder (\ref{eq:160112f}).
(Proof: see
	% Appendix
	\ref{sec:150329b}.)
\\$\Box$
\end{theorem}

A standard result in algebraic specification states that if a function \ensuremath{\Varid{f}}
defined on an initial algebra is a fold then \ensuremath{\frac{\Varid{f}}{\Varid{f}}} is a congruence
\cite{EM85,Gi16}. Although not strictly necessary, we give below a proof
that  frames this result in Theorem \ref{th:150327a} by making
\ensuremath{\Conid{R}\mathrel{=}\frac{\mathopen{(\!|}\Varid{k}\mathclose{|\!)}}{\mathopen{(\!|}\Varid{k}\mathclose{|\!)}}} in (\ref{eq:160118c}) and calculating:
\begin{eqnarray*}
\start
	\ensuremath{\frac{\mathopen{(\!|}\Varid{k}\mathclose{|\!)}}{\mathopen{(\!|}\Varid{k}\mathclose{|\!)}}\mathrel{=}\mathopen{(\!|}\frac{\mathopen{(\!|}\Varid{k}\mathclose{|\!)}}{\mathopen{(\!|}\Varid{k}\mathclose{|\!)}} \comp \mathsf{in}_{\fun F}\mathclose{|\!)}}
\just\equiv{ universal property (\ref{eq:cataUniv-rel}) ; metaphor algebra (\ref{eq:160118d}) }
	\ensuremath{\frac{\mathopen{(\!|}\Varid{k}\mathclose{|\!)} \comp \mathsf{in}_{\fun F}}{\mathopen{(\!|}\Varid{k}\mathclose{|\!)}}\mathrel{=}\frac{\mathopen{(\!|}\Varid{k}\mathclose{|\!)} \comp \mathsf{in}_{\fun F}}{\mathopen{(\!|}\Varid{k}\mathclose{|\!)}} \comp \frac{\fun F \;\mathopen{(\!|}\Varid{k}\mathclose{|\!)}}{\fun F \;\mathopen{(\!|}\Varid{k}\mathclose{|\!)}}}
\just\equiv{ cancellation (\ref{eq:150402b}) ; \ensuremath{\Varid{f} \comp \frac{\Varid{f}}{\Varid{f}}\mathrel{=}\Varid{f}}}
	\ensuremath{\frac{\mathopen{(\!|}\Varid{k}\mathclose{|\!)} \comp \mathsf{in}_{\fun F}}{\mathopen{(\!|}\Varid{k}\mathclose{|\!)}}\mathrel{=}\frac{\Varid{k} \comp \fun F \;\mathopen{(\!|}\Varid{k}\mathclose{|\!)}}{\mathopen{(\!|}\Varid{k}\mathclose{|\!)}}}
\just\implied{ Leibniz }
	\ensuremath{\mathopen{(\!|}\Varid{k}\mathclose{|\!)} \comp \mathsf{in}_{\fun F}\mathrel{=}\Varid{k} \comp \fun F \;\mathopen{(\!|}\Varid{k}\mathclose{|\!)}}
\just\equiv{ universal property (\ref{eq:cataUniv-rel}) }
	\ensuremath{\Varid{true}}
\qed
\end{eqnarray*}
For example, in the case \ensuremath{\Conid{R}\mathrel{=}\Conid{Perm}} (\ref{eq:160118b}), (\ref{eq:160118c}) instantiates to
\begin{eqnarray*}
	\ensuremath{\Conid{Perm} \comp \mathsf{in}_{\fun F}} = \ensuremath{\Conid{Perm} \comp \mathsf{in}_{\fun F} \comp (\ff \Conid{Perm})}
	%label{eq:150119b}
\end{eqnarray*}
whose useful part is
\begin{eqnarray*}
\ensuremath{\Conid{Perm} \comp \mathsf{cons}\mathrel{=}\Conid{Perm} \comp \mathsf{cons} \comp ({id} \times \Conid{Perm})}
	%label{eq:150328c}
\end{eqnarray*}
% recall the (Kronecker) \emph{product} (\ref{eq:960923c-rel}).
%|(b,d)(R >< S)(a,c)| holds iff both |b R a| and |d S c| hold.
% Thus (\ref{eq:150328c}) is the same as
%i.e.\
%\[
%	y\ Perm\ (a:x) = \rcb\exists z {z\ Perm\ x}{y\ Perm\ (a:z)}
%\]
%written pointwise.
In words, this means that permuting a sequence with at least one element is the same
as adding it to the front of a permutation of the tail and permuting again.

The main usefulness of (\ref{eq:150326b},\ref{eq:160118c}) is that the inductive
definition of a kernel equivalence relation generated by a fold is such
that the recursive branch (the \ensuremath{\fun F } term) can be added or removed where convenient,
as shown in the sequel.

To appreciate relational fold fusion (\ref{eq:150402a}) and Theorem \ref{th:150327a}
at work in metaphorism refinement let us consider metaphorisms of the postconditioned
form \ensuremath{\Conid{M}\mathrel{=}\Varid{q}\hskip-1pt\mathit{?} \comp \frac{\mathopen{(\!|}\Varid{k}\mathclose{|\!)}}{\mathopen{(\!|}\Varid{k}\mathclose{|\!)}}} instantiating diagram (\ref{eq:160124a})
for inductive types \ensuremath{\muF } and \ensuremath{\muG }:\footnote{Since there are folds over different
types in the diagram we tag each of them with the corresponding functor.}
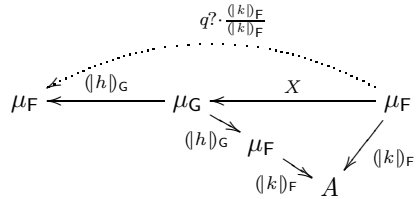
\begin{eqnarray*}
\vcenter{\xymatrix@C=2.5ex@R=1ex{
	\ensuremath{\muF }
&&&
	\ensuremath{\muG }
		\ar[lll]_{\ensuremath{\mathopen{(\!|}\Varid{h}\mathclose{|\!)_{\fun G}}}}
		\ar[rd]_{\ensuremath{\mathopen{(\!|}\Varid{h}\mathclose{|\!)_{\fun G}}}}
&
&
&
	\ensuremath{\muF }
		\ar[ldd]^{\ensuremath{\mathopen{(\!|}\Varid{k}\mathclose{|\!)_{\fun F}}}}
		\ar[lll]_{\ensuremath{\Conid{X}}}
		\ar@{.>}@/_1.8pc/[llllll]_(.45){\ensuremath{\Varid{q}\hskip-1pt\mathit{?} \comp \frac{\mathopen{(\!|}\Varid{k}\mathclose{|\!)_{\fun F}}}{\mathopen{(\!|}\Varid{k}\mathclose{|\!)_{\fun F}}}}}
\\
&
&
&
&
	\ensuremath{\muF }
	\ar[rd]_{\ensuremath{\mathopen{(\!|}\Varid{k}\mathclose{|\!)_{\fun F}}}}
\\
&
&
&
&
&
	\ensuremath{\Conid{A}}
}}
\end{eqnarray*}
As before, this assumes a (surjective) abstraction function \ensuremath{\mathopen{(\!|}\Varid{h}\mathclose{|\!)_{\fun G}}\mathbin{:}\muG \to \muF } 
ensuring that every inhabitant of \ensuremath{\muF } can be represented by at least one inhabitant
of the intermediate type \ensuremath{\muG }.
By direct application of (\ref{eq:160125a}) we obtain the equation
\begin{eqnarray}
	\ensuremath{\Varid{q}\hskip-1pt\mathit{?} \comp \frac{\mathopen{(\!|}\Varid{k}\mathclose{|\!)_{\fun F}}}{\mathopen{(\!|}\Varid{k}\mathclose{|\!)_{\fun F}}}\mathrel{=}\mathopen{(\!|}\Varid{h}\mathclose{|\!)_{\fun G}} \comp \underbrace{\Varid{p}\hskip-1pt\mathit{?} \comp \frac{\mathopen{(\!|}\Varid{k}\mathclose{|\!)_{\fun F}}}{\mathopen{(\!|}\Varid{k}\mathclose{|\!)_{\fun F}} \comp \mathopen{(\!|}\Varid{h}\mathclose{|\!)_{\fun G}}}}_{\Conid{X}}}
	\label{eq:151103a}
\end{eqnarray}
provided \ensuremath{\Varid{q}\hskip-1pt\mathit{?} \comp \mathopen{(\!|}\Varid{h}\mathclose{|\!)_{\fun G}}\mathrel{=}\mathopen{(\!|}\Varid{h}\mathclose{|\!)_{\fun G}} \comp \Varid{p}\hskip-1pt\mathit{?}} --- recall (\ref{eq:150406c}).
Our main goals are, therefore:
\begin{itemize}
\item
	to find \ensuremath{\Varid{p}} such that
\begin{eqnarray}
	\ensuremath{\mathopen{(\!|}\Varid{h}\mathclose{|\!)_{\fun G}} \comp \Varid{p}\hskip-1pt\mathit{?}\mathrel{=}\Varid{q}\hskip-1pt\mathit{?} \comp \mathopen{(\!|}\Varid{h}\mathclose{|\!)_{\fun G}}}
	\label{eq:150408a}
\end{eqnarray}
holds, where \ensuremath{\Varid{q}} is given;
\item
	to convert \ensuremath{\Conid{X}\mathrel{=}\Varid{p}\hskip-1pt\mathit{?} \comp \frac{\mathopen{(\!|}\Varid{k}\mathclose{|\!)_{\fun F}}}{\mathopen{(\!|}\Varid{k}\mathclose{|\!)_{\fun F}} \comp \mathopen{(\!|}\Varid{h}\mathclose{|\!)_{\fun G}}}},
	of type \ensuremath{\muG \leftarrow \muF } (\ref{eq:151103a}), into the converse of a fold,
	\ensuremath{\Conid{X}\mathrel{=}\conv{\mathopen{(\!|}\conv{\Conid{Z}}\mathclose{|\!)}}}, for some \ensuremath{\Conid{Z}\mathbin{:}\fun G \;\muF \leftarrow \muF }.
\end{itemize}
In general, we shall use notation \ensuremath{\mathopen{[\!(}\Conid{R}\mathclose{)\!]}} to abbreviate the expression
\ensuremath{\conv{\mathopen{(\!|}\conv{\Conid{R}}\mathclose{|\!)}}}.
% which we denote as usual by |ana X|, for some |X|, possibly a function |g|.\footnote{Converses of folds are usually termed \emph{unfolds} or anamorphisms. Notation |ana X| means |conv(cata (conv X))| \cite{BM97}.
In case \ensuremath{\Conid{Z}} above happens to be a function \ensuremath{\Varid{g}},
the original metaphorism (whose recursion is \ensuremath{\fun F }-shaped) will be converted into a so-called
\emph{hylomorphism} \cite{BM97}
\begin{quote}
	\ensuremath{\mathopen{(\!|}\Varid{h}\mathclose{|\!)_{\fun G}} \comp \mathopen{[\!(}\Varid{g}\mathclose{)\!]_{\fun G}}}
\end{quote}
whose recursion is \ensuremath{\fun G }-shaped, thus carrying
a ``change of virtual data-structure''.
%--- as in the following (hylomorphism) diagram in which the lace arrows are the partial
%identities of the predicates involved:
%\begin{eqnarray}
%\myxym{
%	|muF|
%		\ar@@(ul,l)[]_{|crflx q|}
%&
%	|fG muF|
%		\ar[l]_{|h|}
%\\
%	|muG|
%		\ar@@(ul,l)[]_{|crflx p|}
%		\ar[u]^{|cata h|}
%&
%	|fG muG|
%		\ar[u]_{|fG (cata h)|}
%		\ar[l]_{|inG|}
%\\
%	|muF|
%		\ar[u]^{|ana Z|}
%		\ar[r]_{|Z|}
%&
%	|fG muF|
%		\ar[u]_{|fG (ana Z)|}
%	\label{eq:160120b}
%\end{eqnarray}

%paragraph{\textsc{wp} calculus over inductive types}
\paragraph{Shifting the metaphor}
For the purposes of our calculations in this paper it is enough to consider the
partial identities (coreflexives) \ensuremath{\Varid{p}\hskip-1pt\mathit{?}} (resp.\ \ensuremath{\Varid{q}\hskip-1pt\mathit{?}}) in (\ref{eq:150408a})
on inductive type \ensuremath{\muG } (resp.\ \ensuremath{\muF }) generated by constraining the initial algebra
\ensuremath{\mathsf{in}_{\fun G}} (resp.\ \ensuremath{\mathsf{in}_{\fun F}}),
\begin{eqnarray*}
	\ensuremath{\Varid{p}\hskip-1pt\mathit{?}} &=& \scata{\xymatrix{\ensuremath{\muG } & \ensuremath{\fun G \;\muG }\ar[l]_{\ensuremath{\mathsf{in}_{\fun G}}} & \ensuremath{\fun G \;\muG }\ar[l]_{\ensuremath{\Varid{w}\hskip-1pt\mathit{?}}}}}
\\
	\ensuremath{\Varid{q}\hskip-1pt\mathit{?}} &=& \scata{\xymatrix{\ensuremath{\muF } & \ensuremath{\fun F \;\muF }\ar[l]_{\ensuremath{\mathsf{in}_{\fun F}}} & \ensuremath{\fun F \;\muF }\ar[l]_{\ensuremath{\Varid{t}\hskip-1pt\mathit{?}}}}}
\end{eqnarray*}
for suitable pre-conditions \ensuremath{\Varid{w}} and \ensuremath{\Varid{t}} --- recall (\ref{eq:170409b}).

The calculation of (\ref{eq:150408a}) proceeds by fusion (\ref{eq:150402a}),
aiming to reduce both \ensuremath{\mathopen{(\!|}\Varid{h}\mathclose{|\!)_{\fun G}} \comp \Varid{p}\hskip-1pt\mathit{?}} and \ensuremath{\Varid{q}\hskip-1pt\mathit{?} \comp \mathopen{(\!|}\Varid{h}\mathclose{|\!)_{\fun G}}} to some
relational fold \ensuremath{\mathopen{(\!|}\Conid{R}\mathclose{|\!)_{\fun G}}} over \ensuremath{\muG }. On the right hand side, fusion yields
\begin{eqnarray}
	\ensuremath{\Varid{q}\hskip-1pt\mathit{?} \comp \mathopen{(\!|}\Varid{h}\mathclose{|\!)_{\fun G}}\mathrel{=}\mathopen{(\!|}\Conid{R}\mathclose{|\!)_{\fun G}}} &\implied& \ensuremath{\Varid{q}\hskip-1pt\mathit{?} \comp \Varid{h}\mathrel{=}\Conid{R} \comp (\fun G \;\Varid{q}\hskip-1pt\mathit{?})}
	\label{eq:150329f}
\end{eqnarray}
On the other side:
\begin{eqnarray*}
\start
	\ensuremath{\mathopen{(\!|}\Varid{h}\mathclose{|\!)_{\fun G}} \comp \Varid{p}\hskip-1pt\mathit{?}\mathrel{=}\mathopen{(\!|}\Conid{R}\mathclose{|\!)_{\fun G}}}
\just\equiv{inline \ensuremath{\Varid{p}\hskip-1pt\mathit{?}\mathrel{=}\mathopen{(\!|}\mathsf{in}_{\fun G} \comp \Varid{w}\hskip-1pt\mathit{?}\mathclose{|\!)_{\fun G}}}}
	\eqnnewpage
	\ensuremath{\mathopen{(\!|}\Varid{h}\mathclose{|\!)_{\fun G}} \comp \mathopen{(\!|}\mathsf{in}_{\fun G} \comp \Varid{w}\hskip-1pt\mathit{?}\mathclose{|\!)_{\fun G}}\mathrel{=}\mathopen{(\!|}\Conid{R}\mathclose{|\!)_{\fun G}}}
\just\implied{fusion (\ref{eq:150402a})}
	\ensuremath{\mathopen{(\!|}\Varid{h}\mathclose{|\!)_{\fun G}} \comp \mathsf{in}_{\fun G} \comp \Varid{w}\hskip-1pt\mathit{?}\mathrel{=}\Conid{R} \comp \fun G \;\mathopen{(\!|}\Varid{h}\mathclose{|\!)_{\fun G}}}
\just\equiv{cancellation: \ensuremath{\mathopen{(\!|}\Varid{h}\mathclose{|\!)_{\fun G}} \comp \mathsf{in}_{\fun G}\mathrel{=}\Varid{h} \comp \fun G \;\mathopen{(\!|}\Varid{h}\mathclose{|\!)_{\fun G}}} (\ref{eq:150402b})}
	\ensuremath{\Varid{h} \comp \fun G \;\mathopen{(\!|}\Varid{h}\mathclose{|\!)_{\fun G}} \comp \Varid{w}\hskip-1pt\mathit{?}\mathrel{=}\Conid{R} \comp \fun G \;\mathopen{(\!|}\Varid{h}\mathclose{|\!)_{\fun G}}}
\just\equiv{ switch to \ensuremath{\Varid{r}\hskip-1pt\mathit{?}} such that \ensuremath{\fun G \;\mathopen{(\!|}\Varid{h}\mathclose{|\!)_{\fun G}} \comp \Varid{w}\hskip-1pt\mathit{?}\mathrel{=}\Varid{r}\hskip-1pt\mathit{?} \comp \fun G \;\mathopen{(\!|}\Varid{h}\mathclose{|\!)_{\fun G}}} holds (\ref{eq:150406c}) }
	\ensuremath{\Varid{h} \comp \Varid{r}\hskip-1pt\mathit{?} \comp \fun G \;\mathopen{(\!|}\Varid{h}\mathclose{|\!)_{\fun G}}\mathrel{=}\Conid{R} \comp \fun G \;\mathopen{(\!|}\Varid{h}\mathclose{|\!)_{\fun G}}}
\just\implied{Leibniz}
	\ensuremath{\Varid{h} \comp \Varid{r}\hskip-1pt\mathit{?}\mathrel{=}\Conid{R}}
\end{eqnarray*}
Thus \ensuremath{\Conid{R}\mathrel{=}\Varid{h} \comp \Varid{r}\hskip-1pt\mathit{?}} ensures proviso (\ref{eq:150408a}).
By replacing \ensuremath{\Conid{R}} in the other proviso --- the side condition of fusion step (\ref{eq:150329f}) ---
one obtains
\begin{eqnarray}
	\ensuremath{\Varid{q}\hskip-1pt\mathit{?} \comp \Varid{h}\mathrel{=}\Varid{h} \comp \Varid{r}\hskip-1pt\mathit{?} \comp \fun G \;\Varid{q}\hskip-1pt\mathit{?}}
&&
\myxym{
	\ensuremath{\muF }
		\ar[d]_{\ensuremath{\Varid{q}\hskip-1pt\mathit{?}}}
&
	~
&
	\ensuremath{\fun G \;\muF }
		\ar[d]^{\ensuremath{\fun G \;\Varid{q}\hskip-1pt\mathit{?}}}
		\ar[ll]_{\ensuremath{\Varid{h}}}
\\
	\ensuremath{\muF }
&
	\ensuremath{\fun G \;\muF }
		\ar[l]^{\ensuremath{\Varid{h}}}
&
	\ensuremath{\fun G \;\muF }
		\ar[l]^{\ensuremath{\Varid{r}\hskip-1pt\mathit{?}}}
}
	\label{eq:150328b}
\end{eqnarray}
that has to be ensured together with the other assumption above:
\begin{eqnarray}
	\ensuremath{\fun G \;\mathopen{(\!|}\Varid{h}\mathclose{|\!)_{\fun G}} \comp \Varid{w}\hskip-1pt\mathit{?}\mathrel{=}\Varid{r}\hskip-1pt\mathit{?} \comp \fun G \;\mathopen{(\!|}\Varid{h}\mathclose{|\!)_{\fun G}}}
&&
\myxym{
	\ensuremath{\fun G \;\muG }
		\ar[d]_{\ensuremath{\fun G \;\mathopen{(\!|}\Varid{h}\mathclose{|\!)_{\fun G}}}}
&
	\ensuremath{\fun G \;\muG }
		\ar[d]^{\ensuremath{\fun G \;\mathopen{(\!|}\Varid{h}\mathclose{|\!)_{\fun G}}}}
		\ar[l]_{\ensuremath{\Varid{w}\hskip-1pt\mathit{?}}}
\\
	\ensuremath{\fun G \;\muF }
&
	\ensuremath{\fun G \;\muF }
		\ar[l]^{\ensuremath{\Varid{r}\hskip-1pt\mathit{?}}}
}
	\label{eq:150330a}
\end{eqnarray}
Let us summarize these calculations in the form of a theorem.

\begin{theorem}\label{th:150327b}
Let \ensuremath{\rarrow{\muG }{\mathopen{(\!|}\Varid{h}\mathclose{|\!)_{\fun G}}}{\muF }} be an abstraction of inductive type
\ensuremath{\larrow{\fun F \;\muF }{\mathsf{in}_{\fun F}}{\muF }} by another inductive type
\ensuremath{\larrow{\fun G \;\muG }{\mathsf{in}_{\fun G}}{\muG }}, and
\ensuremath{\Varid{q}\hskip-1pt\mathit{?}\mathrel{=}\mathopen{(\!|}\mathsf{in}_{\fun F} \comp \Varid{t}\hskip-1pt\mathit{?}\mathclose{|\!)_{\fun F}}} be a partial identity
% and |(crflx p)=cata (inG.(crflx w))| be partial identities
representing an inductive predicate over \ensuremath{\muF }.

To calculate the weakest precondition \ensuremath{\Varid{p}} for \ensuremath{\mathopen{(\!|}\Varid{h}\mathclose{|\!)_{\fun G}}} to ensure \ensuremath{\Varid{q}} on its output,
say \ensuremath{\Varid{p}\hskip-1pt\mathit{?}\mathrel{=}\mathopen{(\!|}\mathsf{in}_{\fun G} \comp \Varid{w}\hskip-1pt\mathit{?}\mathclose{|\!)_{\fun G}}}, it suffices to find a predicate \ensuremath{\Varid{r}} on \ensuremath{\fun G \;\muF }
such that (\ref{eq:150328b}) and (\ref{eq:150330a}) hold.
\\$\Box$
\end{theorem}

Note how condition \ensuremath{\Varid{r}} on \ensuremath{\fun G \;\muF } in proviso (\ref{eq:150328b})
is the weakest precondition for algebra $h$ to maintain \ensuremath{\Varid{q}}, while
(\ref{eq:150330a}) establishes \ensuremath{\Varid{w}} as the {weakest precondition} for
the recursive branch \ensuremath{\fun G \;\mathopen{(\!|}\Varid{h}\mathclose{|\!)_{\fun G}}} to ensure \ensuremath{\Varid{r}} on its output.

\paragraph{Calculating the ``divide" step}
Armed with side conditions (\ref{eq:150328b}) and (\ref{eq:150330a}), our
final aim is to calculate \ensuremath{\Conid{X}\mathrel{=}\mathopen{[\!(}\Conid{Z}\mathclose{)\!]}} in (\ref{eq:151103a}):
\begin{eqnarray}
\start
	\ensuremath{\underbrace{\Varid{p}\hskip-1pt\mathit{?} \comp \frac{\mathopen{(\!|}\Varid{k}\mathclose{|\!)_{\fun F}}}{\mathopen{(\!|}\Varid{k}\mathclose{|\!)_{\fun F}} \comp \mathopen{(\!|}\Varid{h}\mathclose{|\!)_{\fun G}}}}_{\Conid{X}}\mathrel{=}\mathopen{[\!(}\Conid{Z}\mathclose{)\!]}}
	\nonumber
\just\equiv{ converses ; \ensuremath{\mathopen{[\!(}\Conid{Z}\mathclose{)\!]}\mathrel{=}\conv{\mathopen{(\!|}\conv{\Conid{Z}}\mathclose{|\!)}}} }
	\ensuremath{\frac{\mathopen{(\!|}\Varid{k}\mathclose{|\!)_{\fun F}} \comp \mathopen{(\!|}\Varid{h}\mathclose{|\!)_{\fun G}}}{\mathopen{(\!|}\Varid{k}\mathclose{|\!)_{\fun F}}} \comp \Varid{p}\hskip-1pt\mathit{?}\mathrel{=}\mathopen{(\!|}\conv{\Conid{Z}}\mathclose{|\!)}}
	\nonumber
\just\equiv{ \ensuremath{\mathopen{(\!|}\Varid{h}\mathclose{|\!)_{\fun G}} \comp \Varid{p}\hskip-1pt\mathit{?}\mathrel{=}\Varid{q}\hskip-1pt\mathit{?} \comp \mathopen{(\!|}\Varid{h}\mathclose{|\!)_{\fun G}}} assumed --- cf.\ (\ref{eq:150408a}) } %, Theorem \ref{th:150327b}}
	\ensuremath{\frac{\mathopen{(\!|}\Varid{k}\mathclose{|\!)_{\fun F}}}{\mathopen{(\!|}\Varid{k}\mathclose{|\!)_{\fun F}}} \comp \Varid{q}\hskip-1pt\mathit{?} \comp \mathopen{(\!|}\Varid{h}\mathclose{|\!)_{\fun G}}\mathrel{=}\mathopen{(\!|}\conv{\Conid{Z}}\mathclose{|\!)}}
	\nonumber
\just\implied{ fusion (\ref{eq:150402a}) ; functor \ensuremath{\fun G } }
	\ensuremath{\frac{\mathopen{(\!|}\Varid{k}\mathclose{|\!)_{\fun F}}}{\mathopen{(\!|}\Varid{k}\mathclose{|\!)_{\fun F}}} \comp \Varid{q}\hskip-1pt\mathit{?} \comp \Varid{h}\mathrel{=}\conv{\Conid{Z}} \comp \fun G \;\frac{\mathopen{(\!|}\Varid{k}\mathclose{|\!)_{\fun F}}}{\mathopen{(\!|}\Varid{k}\mathclose{|\!)_{\fun F}}} \comp \fun G \;\Varid{q}\hskip-1pt\mathit{?}}
	\nonumber
\just\equiv{ proviso (\ref{eq:150328b}): \ensuremath{\Varid{q}\hskip-1pt\mathit{?} \comp \Varid{h}\mathrel{=}\Varid{h} \comp \Varid{r}\hskip-1pt\mathit{?} \comp \fun G \;\Varid{q}\hskip-1pt\mathit{?}} }
	\ensuremath{\frac{\mathopen{(\!|}\Varid{k}\mathclose{|\!)_{\fun F}}}{\mathopen{(\!|}\Varid{k}\mathclose{|\!)_{\fun F}}} \comp \Varid{h} \comp \Varid{r}\hskip-1pt\mathit{?} \comp \fun G \;\Varid{q}\hskip-1pt\mathit{?}\mathrel{=}\conv{\Conid{Z}} \comp \fun G \;\frac{\mathopen{(\!|}\Varid{k}\mathclose{|\!)_{\fun F}}}{\mathopen{(\!|}\Varid{k}\mathclose{|\!)_{\fun F}}} \comp \fun G \;\Varid{q}\hskip-1pt\mathit{?}}
	\nonumber
\just\implied{Leibniz}
	\ensuremath{\frac{\mathopen{(\!|}\Varid{k}\mathclose{|\!)_{\fun F}}}{\mathopen{(\!|}\Varid{k}\mathclose{|\!)_{\fun F}}} \comp \Varid{h} \comp \Varid{r}\hskip-1pt\mathit{?}\mathrel{=}\conv{\Conid{Z}} \comp \fun G \;\frac{\mathopen{(\!|}\Varid{k}\mathclose{|\!)_{\fun F}}}{\mathopen{(\!|}\Varid{k}\mathclose{|\!)_{\fun F}}}}
	\label{eq:160125b}
\end{eqnarray}

We are still far from having a closed formula for \ensuremath{\Conid{Z}}. Can we get rid of term
\ensuremath{\fun G \;\frac{\mathopen{(\!|}\Varid{k}\mathclose{|\!)_{\fun F}}}{\mathopen{(\!|}\Varid{k}\mathclose{|\!)_{\fun F}}}} from the right hand side?
This is where Theorem \ref{th:150327a} plays a role, enabling such a cancellation 
provided we ensure that equivalence \ensuremath{\frac{\mathopen{(\!|}\Varid{k}\mathclose{|\!)_{\fun F}}}{\mathopen{(\!|}\Varid{k}\mathclose{|\!)_{\fun F}}}} is a \emph{congruence} for algebra $h$,
which (by virtue of Theorem \ref{th:150327a}) amounts to ensuring
	\ensuremath{\mathopen{(\!|}\Varid{k}\mathclose{|\!)_{\fun F}} \comp \Varid{h}\leq \fun G \;\mathopen{(\!|}\Varid{k}\mathclose{|\!)_{\fun F}}}.
In words: \ensuremath{\mathopen{(\!|}\Varid{k}\mathclose{|\!)_{\fun F}} \comp \Varid{h}} should be \emph{no more injective} (\ref{eq:041217a})
than the recursive branch \ensuremath{\fun G \;\mathopen{(\!|}\Varid{k}\mathclose{|\!)_{\fun F}}}.
It turns out that we shall need yet another similar injectivity clause involving
\ensuremath{\Varid{r}} in the sequel. Altogether:
\begin{eqnarray}
	\ensuremath{\mathopen{(\!|}\Varid{k}\mathclose{|\!)_{\fun F}} \comp \Varid{h}} &\leq& \ensuremath{\fun G \;\mathopen{(\!|}\Varid{k}\mathclose{|\!)_{\fun F}}}
	\label{eq:150331a}
\\
	\ensuremath{\Varid{r}} &\leq& \ensuremath{\fun G \;\mathopen{(\!|}\Varid{k}\mathclose{|\!)_{\fun F}}}
	\label{eq:170625a}
\end{eqnarray}
Below we resume the calculation of (\ref{eq:160125b}) assuming (\ref{eq:150331a}) and (\ref{eq:170625a}):
\begin{eqnarray}
\start
	\ensuremath{\frac{\mathopen{(\!|}\Varid{k}\mathclose{|\!)_{\fun F}}}{\mathopen{(\!|}\Varid{k}\mathclose{|\!)_{\fun F}}} \comp \Varid{h} \comp \Varid{r}\hskip-1pt\mathit{?}\mathrel{=}\conv{\Conid{Z}} \comp \fun G \;\frac{\mathopen{(\!|}\Varid{k}\mathclose{|\!)_{\fun F}}}{\mathopen{(\!|}\Varid{k}\mathclose{|\!)_{\fun F}}}}
	\nonumber
\just\equiv{ (\ref{eq:150326b}) }
	\ensuremath{\frac{\mathopen{(\!|}\Varid{k}\mathclose{|\!)_{\fun F}}}{\mathopen{(\!|}\Varid{k}\mathclose{|\!)_{\fun F}}} \comp \Varid{h} \comp \fun G \;\frac{\mathopen{(\!|}\Varid{k}\mathclose{|\!)_{\fun F}}}{\mathopen{(\!|}\Varid{k}\mathclose{|\!)_{\fun F}}} \comp \Varid{r}\hskip-1pt\mathit{?}\mathrel{=}\conv{\Conid{Z}} \comp \fun G \;\frac{\mathopen{(\!|}\Varid{k}\mathclose{|\!)_{\fun F}}}{\mathopen{(\!|}\Varid{k}\mathclose{|\!)_{\fun F}}}}
	\nonumber
	\eqnnewpagex
\just\equiv{ (\ref{eq:160118d}) first, then (\ref{eq:150407a-modified}) thanks to (\ref{eq:170625a}) } % (\ref{eq:150407a}), assuming: |rcb exists u () (r = u . fG(cataF k)|
	\ensuremath{\frac{\mathopen{(\!|}\Varid{k}\mathclose{|\!)_{\fun F}}}{\mathopen{(\!|}\Varid{k}\mathclose{|\!)_{\fun F}}} \comp \Varid{h} \comp \Varid{r}\hskip-1pt\mathit{?} \comp \frac{\fun G \;\mathopen{(\!|}\Varid{k}\mathclose{|\!)_{\fun F}}}{\fun G \;\mathopen{(\!|}\Varid{k}\mathclose{|\!)_{\fun F}}}\mathrel{=}\conv{\Conid{Z}} \comp \frac{\fun G \;\mathopen{(\!|}\Varid{k}\mathclose{|\!)_{\fun F}}}{\fun G \;\mathopen{(\!|}\Varid{k}\mathclose{|\!)_{\fun F}}}}
	%label{eq:160121a}
	\nonumber
\just\implied{ drop \ensuremath{\frac{\fun G \;\mathopen{(\!|}\Varid{k}\mathclose{|\!)_{\fun F}}}{\fun G \;\mathopen{(\!|}\Varid{k}\mathclose{|\!)_{\fun F}}}} (Leibniz) }
	\ensuremath{\frac{\mathopen{(\!|}\Varid{k}\mathclose{|\!)_{\fun F}}}{\mathopen{(\!|}\Varid{k}\mathclose{|\!)_{\fun F}}} \comp \Varid{h} \comp \Varid{r}\hskip-1pt\mathit{?}\mathrel{=}\conv{\Conid{Z}}}
	\label{eq:160119a}
\end{eqnarray}
Taking converses, we get
\begin{eqnarray}
	\ensuremath{\Conid{Z}\mathrel{=}\Varid{r}\hskip-1pt\mathit{?} \comp \frac{\mathopen{(\!|}\Varid{k}\mathclose{|\!)_{\fun F}}}{\mathopen{(\!|}\Varid{k}\mathclose{|\!)_{\fun F}} \comp \Varid{h}}}
	\label{eq:150709a}
\end{eqnarray}
from (\ref{eq:160119a}) --- another metaphorism, of the expected type \ensuremath{\larrow{\muF }{}{\fun G \;\muF }}.
% Consider, however, the requirement that |r| must be a \textsc{wp} wrt.\ |fG(cataF k)| --- cf.\ step (\ref{eq:160121a}).

Summing up, note how the original metaphorism \ensuremath{\Varid{q}\hskip-1pt\mathit{?} \comp \frac{\mathopen{(\!|}\Varid{k}\mathclose{|\!)_{\fun F}}}{\mathopen{(\!|}\Varid{k}\mathclose{|\!)_{\fun F}}}}
gets converted into a hylomorphism whose \emph{divide} step is another
metaphorism (\ref{eq:150709a}).
%The following diagram decorates diagram (\ref{eq:160120b})
%with further laces, representing all the conditions involved:
%\begin{eqnarray*}
%\myxym{
%	|muF|
%		\ar@@(ul,l)[]_{|crflx q|}
%&
%	|fG muF|
%		\ar[l]_{|h|}
%		\ar@@(ur,r)[]^{|crflx r|}
%		\ar@@(u,ul)[]_{|crflx (fG q)|}
%\\
%	|muG|
%		\ar[u]^{|cataG h|}
%		\ar@@(ul,l)[]_{|crflx p|}
%&
%	|fG muG|
%		\ar@@(ur,r)[]^{|crflx w|}
%		\ar[u]_{|fG (cataG h)|}
%		\ar[l]_{|inG|}
%\\
%	|muF|
%		\ar[u]^{|ana Z|}
%		\ar[r]_{|Z|}
%&
%	|fG muF|
%		\ar[u]_{|fG (ana Z)|}
%		\ar@@(dr,r)[]_{|crflx r|}
%\end{eqnarray*}
Recall that \ensuremath{\Varid{r}} acts as \textsc{wp} for algebra $h$ to maintain \ensuremath{\Varid{q}}
and that \ensuremath{\Varid{w}} is the \textsc{wp} for
the recursive branch \ensuremath{\fun G \;\mathopen{(\!|}\Varid{h}\mathclose{|\!)_{\fun G}}} to ensure \ensuremath{\Varid{r}}.

Altogether, the ``outer'' metaphor which we started from (involving only \ensuremath{\muF })
disappears and gives place to an ``inner'' metaphor between inductive types \ensuremath{\muG } and \ensuremath{\muF }
(the divide step), whereby the optimization is internalized.
This ``inner'' metaphor is more interesting, as we can see by looking at an
example of all this reasoning. Before this, we close this section with
the checklist of all provisos that have to be verified for \ensuremath{\Conid{Z}} (\ref{eq:150709a}) to exist:
\begin{itemize}
\item	
	(\ref{eq:150328b}) --- establishes \ensuremath{\Varid{r}} as the weakest precondition for \ensuremath{\fun G }-algebra
	\ensuremath{\Varid{h}} to \emph{maintain} \ensuremath{\Varid{q}} as an invariant.
\item
	(\ref{eq:150330a}) --- establishes \ensuremath{\Varid{w}} as the {weakest precondition} for
	the recursive branch \ensuremath{\fun G \;\mathopen{(\!|}\Varid{h}\mathclose{|\!)_{\fun G}}} to ensure \ensuremath{\Varid{r}} as post-condition.
\item
	(\ref{eq:150331a}) --- \ensuremath{\mathopen{(\!|}\Varid{k}\mathclose{|\!)_{\fun F}} \comp \Varid{h}} should be {no more injective} than the recursive branch \ensuremath{\fun G \;\mathopen{(\!|}\Varid{k}\mathclose{|\!)_{\fun F}}}.
\item
	(\ref{eq:170625a}) --- inputs undistinguishable by \ensuremath{\fun G \;\mathopen{(\!|}\Varid{k}\mathclose{|\!)_{\fun F}}} should also be undistinguishable by predicate \ensuremath{\Varid{r}}.
\end{itemize}

\section{Example: Quicksort} \label{sec:150406i}

This section shows how the derivation of \emph{quicksort} as given e.g.\ by
\citet{BM97} corresponds to the implementation strategy for metaphorisms
given above, under the following instantiations:
\begin{itemize}
\item	The starting metaphorism is (\ref{eq:160118b}) where \ensuremath{\Conid{Perm}} is the list permutation
	relationship.
\item	\ensuremath{\muF } is the usual finite list datatype with constructors (say) \ensuremath{\mathsf{nil}} and \ensuremath{\mathsf{cons}}, that is,
	\ensuremath{\mathsf{in}_{\fun F}\mathrel{=}\alt{\mathsf{nil}}{\mathsf{cons}}} with base \ensuremath{\fun F \;\Varid{f}\mathrel{=}{id}\mathbin{+}{id} \times \Varid{f}} (\ref{eq:170518a}).
\item	\ensuremath{\muG } is the binary tree data type whose base is \ensuremath{\fun G \;\Varid{f}\mathrel{=}{id}\mathbin{+}{id} \times {\Varid{f}}^{\mathrm{2}}} and whose initial
	algebra is (say) \ensuremath{\mathsf{in}_{\fun G}\mathrel{=}\alt{\mathsf{empty}}{\mathsf{node}}}. (We use abbreviation \ensuremath{{\Varid{f}}^{\mathrm{2}}} for \ensuremath{\Varid{f} \times \Varid{f}}.)
\item	\ensuremath{\mathopen{(\!|}\Varid{k}\mathclose{|\!)_{\fun F}}\mathrel{=}\Varid{bag}}, the function which converts a list into the bag (multiset) of its elements.
\item	\ensuremath{\mathopen{(\!|}\Varid{h}\mathclose{|\!)_{\fun G}}\mathrel{=}\Varid{flatten}}, for \ensuremath{\Varid{h}\mathrel{=}\alt{\mathsf{nil}}{\Varid{inord}}} where \ensuremath{\Varid{inord}\;(\Varid{a},(\Varid{x},\Varid{y}))\mathrel{=}\Varid{x}\mathbin{+\!\!+}[\mskip1.5mu \Varid{a}\mskip1.5mu]\mathbin{+\!\!+}\Varid{y}}; that is,
	\ensuremath{\Varid{flatten}} is the binary tree into finite list (inorder) traversal surjection.
\item	\ensuremath{\Varid{q}\mathrel{=}\Varid{ordered}} (\ref{eq:160118b}), that is, \ensuremath{\Varid{q}\hskip-1pt\mathit{?}\mathrel{=}\mathopen{(\!|}\alt{\mathsf{nil}}{\mathsf{cons}} \comp ({id}\mathbin{+}\Varid{mn}\hskip-1pt\mathit{?})\mathclose{|\!)}},
	for \ensuremath{\Varid{mn}\;(\Varid{x},\Varid{xs})\mathrel{=}\rcb{\forall}{\Varid{x'}}{\Varid{x'}\;\epsilon_{\muF}\;\Varid{xs}}{\Varid{x'}\geq \Varid{x}}}
	where \ensuremath{\epsilon_{\muF}} denotes list membership;
	that is, predicate \ensuremath{\Varid{mn}\;(\Varid{x},\Varid{xs})} ensures that list \ensuremath{\Varid{x}\mathbin{:}\Varid{xs}} is such that \ensuremath{\Varid{x}}
	is at most the minimum of \ensuremath{\Varid{xs}}, if it exists.
\end{itemize}
As seen in Sect.~\ref{sec:150406g}, we first have to search for some predicate \ensuremath{\Varid{r}} that,
following (\ref{eq:150328b}), should be the weakest precondition for \ensuremath{\alt{\mathsf{nil}}{\Varid{inord}}}
to preserve ordered lists (\ensuremath{\Varid{q}\hskip-1pt\mathit{?}}). We calculate:
\begin{eqnarray*}
\start
	\ensuremath{\Varid{q}\hskip-1pt\mathit{?} \comp \alt{\mathsf{nil}}{\Varid{inord}}\mathrel{=}\alt{\mathsf{nil}}{\Varid{inord}} \comp \Varid{r}\hskip-1pt\mathit{?} \comp ({id}\mathbin{+}{id} \times (\Varid{q}\hskip-1pt\mathit{?} \times \Varid{q}\hskip-1pt\mathit{?}))}
\just\equiv{ switch to \ensuremath{\Varid{s}} such that \ensuremath{\Varid{r}\hskip-1pt\mathit{?}\mathrel{=}{id}\mathbin{+}\Varid{s}\hskip-1pt\mathit{?}}; coproducts }
	\ensuremath{\alt{\Varid{q}\hskip-1pt\mathit{?} \comp \mathsf{nil}}{\Varid{q}\hskip-1pt\mathit{?} \comp \Varid{inord}}\mathrel{=}\alt{\mathsf{nil}}{\Varid{inord} \comp \Varid{s}\hskip-1pt\mathit{?} \comp ({id} \times (\Varid{q}\hskip-1pt\mathit{?} \times \Varid{q}\hskip-1pt\mathit{?}))}}
\just\equiv{the empty list is trivially ordered}
	\ensuremath{\Varid{q}\hskip-1pt\mathit{?} \comp \Varid{inord}\mathrel{=}\Varid{inord} \comp \Varid{s}\hskip-1pt\mathit{?} \comp ({id} \times (\Varid{q}\hskip-1pt\mathit{?} \times \Varid{q}\hskip-1pt\mathit{?}))}
\just\equiv{ universal property (\ref{eq:150406c}) }
	\ensuremath{\Varid{s}\hskip-1pt\mathit{?} \comp ({id} \times (\Varid{q}\hskip-1pt\mathit{?} \times \Varid{q}\hskip-1pt\mathit{?}))} = \ensuremath{(\Varid{q} \comp \Varid{inord})\hskip-1pt\mathit{?}}
\end{eqnarray*}
Knowing the definitions of \ensuremath{\Varid{q}} and \ensuremath{\Varid{inord}}, we easily infer \ensuremath{\Varid{s}} by going pointwise:
\begin{eqnarray}
\start
	\ensuremath{\Varid{q}\;(\Varid{x}\mathbin{+\!\!+}[\mskip1.5mu \Varid{a}\mskip1.5mu]\mathbin{+\!\!+}\Varid{y})}
	\nonumber
\just\equiv{ pointwise definition of ordered lists }
	\ensuremath{\begin{lcbr}(\Varid{q}\;\Varid{x})\mathrel{\wedge}(\Varid{q}\;\Varid{y})\\\underbrace{\rcb{\forall}{\Varid{b}}{\Varid{b}\;\epsilon_{\muF}\;\Varid{x}}{\Varid{b}\leq \Varid{a}}\mathrel{\wedge}\rcb{\forall}{\Varid{b}}{\Varid{b}\;\epsilon_{\muF}\;\Varid{y}}{\Varid{a}\leq \Varid{b}}}_{\Varid{s}\;(\Varid{a},(\Varid{x},\Varid{y}))}\end{lcbr}}
	\label{eq:160120a}
\end{eqnarray}
%Thus |(a',(x',y'))vUpsilon(a,(x,y))| holds wherever |x'=x && a'=a && y'=y && s(x,a,y)| --- a partial identity.
Knowing \ensuremath{\Varid{s}} and thus \ensuremath{\Varid{r}}, we go back to (\ref{eq:150709a}) to calculate the
(relational) {coalgebra} that shall control the \emph{divide} part,
still letting \ensuremath{\Varid{r}\hskip-1pt\mathit{?}\mathrel{=}{id}\mathbin{+}\Varid{s}\hskip-1pt\mathit{?}},
\begin{eqnarray}
\arrayin{
&&\ensuremath{\Conid{Z}\mathbin{:}\mathrm{1}\mathbin{+}\muF  \times (\muF  \times \muF )\leftarrow \muF }
\\
&&\ensuremath{\Conid{Z}\mathrel{=}({id}\mathbin{+}\Varid{s}\hskip-1pt\mathit{?}) \comp \frac{\Varid{bag}}{\Varid{bag} \comp \alt{\mathsf{nil}}{\Varid{inord}}}}
}
	\label{eq:150408b}
\end{eqnarray}
as follows:
\begin{eqnarray*}
\start
	\ensuremath{\Conid{Z}\mathrel{=}({id}\mathbin{+}\Varid{s}\hskip-1pt\mathit{?}) \comp \frac{\Varid{bag}}{\Varid{bag} \comp \alt{\mathsf{nil}}{\Varid{inord}}}}
\just\equiv{ let \ensuremath{\conv{\Conid{Z}}\mathrel{=}\alt{\conv{Z_1}}{\conv{Z_2}}}}
	\ensuremath{\conv{\alt{\conv{Z_1}}{\conv{Z_2}}}\mathrel{=}({id}\mathbin{+}\Varid{s}\hskip-1pt\mathit{?}) \comp \frac{\Varid{bag}}{\Varid{bag} \comp \alt{\mathsf{nil}}{\Varid{inord}}}}
\just\equiv{ take converses }
	\ensuremath{\alt{\conv{Z_1}}{\conv{Z_2}}} = \ensuremath{\frac{\Varid{bag} \comp \alt{\mathsf{nil}}{\Varid{inord}}}{\Varid{bag}} \comp ({id}\mathbin{+}\Varid{s}\hskip-1pt\mathit{?})}
\just\equiv{ \ensuremath{\Conid{Perm}} (\ref{eq:160118b}) ; coproducts }
	\ensuremath{\alt{\conv{Z_1}}{\conv{Z_2}}} = \ensuremath{\alt{\Conid{Perm} \comp \mathsf{nil}}{\Conid{Perm} \comp \Varid{inord} \comp \Varid{s}\hskip-1pt\mathit{?}}}
\just\equiv{ coproducts; \ensuremath{\Conid{Perm} \comp \mathsf{nil}\mathrel{=}\mathsf{nil}}; converses }
	\ensuremath{\begin{lcbr}Z_1\mathrel{=}\conv{\mathsf{nil}}\\Z_2\mathrel{=}\Varid{s}\hskip-1pt\mathit{?} \comp \conv{\Varid{inord}} \comp \Conid{Perm}\end{lcbr}}
\just\equiv{go pointwise}
\begin{lcbr}
	\ensuremath{\anonymous \;Z_1\;\Varid{x}} \equiv \ensuremath{\Varid{x}\mathrel{=}[\mskip1.5mu \mskip1.5mu]}
\\
	\ensuremath{(\Varid{a},(\Varid{y},\Varid{z}))\;Z_2\;\Varid{x}} \equiv \ensuremath{(\Varid{a},(\Varid{y},\Varid{z}))\;(\Varid{s}\hskip-1pt\mathit{?} \comp \conv{\Varid{inord}} \comp \Conid{Perm})\;\Varid{x}}
\end{lcbr}
\end{eqnarray*}
The second clause of the bottom line just above unfolds to:
\begin{eqnarray*}
\ensuremath{(\Varid{a},(\Varid{y},\Varid{z}))\;Z_2\;\Varid{x}} \equiv \ensuremath{\Varid{s}\;(\Varid{a},(\Varid{y},\Varid{z}))\mathrel{\wedge}(\Varid{y}\mathbin{+\!\!+}[\mskip1.5mu \Varid{a}\mskip1.5mu]\mathbin{+\!\!+}\Varid{z})\;\Conid{Perm}\;\Varid{x}}
\end{eqnarray*}

% From this we get the following relational definition of the \emph{divide} step (\ref{eq:150709a})
% of the implementation,
% \begin{eqnarray}
% \arrayin{
% &&|R : [A] -> 1 + A >< ([A]><[A])|
% \\
% &&|R=(id+crflx s) . (conv((bag.(either nil inord)))) . bag|
% }
% 	\label{eq:150408b}
% \end{eqnarray}
% which we unfold as follows, by letting |conv R = (either (conv R1)(conv R2))| and using the
% converse of (\ref{eq:150408b}):
% \begin{eqnarray*}
% \start
% %	|conv(either (conv R1)(conv R2))| = |(id+crflx s) . (conv((bag.(either nil inord)))) . bag|
% %
% %just\equiv{ converses }
% %
% 	|either (conv R1)(conv R2)| = |conv bag . (bag.(either nil inord)). (id+crflx s)|
% %
% \just\equiv{ |conv bag.bag=Perm|; |Perm.nil=nil|; converses }
% %
% 	|lcbr (R1=conv nil)(R2=crflx s.(conv inord).Perm)|
% %
% \end{eqnarray*}
\noindent
In words, \ensuremath{\Varid{y}\;\Conid{Z}\;\Varid{x}} has the following meaning: either \ensuremath{\Varid{x}\mathrel{=}[\mskip1.5mu \mskip1.5mu]} and \ensuremath{\Conid{Z}} yields
the unique inhabitant of singleton type \ensuremath{\mathrm{1}} (cf.\ \ensuremath{Z_1}) or \ensuremath{\Varid{x}} is non-empty
and \ensuremath{\Conid{Z}} splits a permutation of \ensuremath{\Varid{x}} into two halves \ensuremath{\Varid{y}} and \ensuremath{\Varid{z}}
separated by a ``pivot" \ensuremath{\Varid{a}}, all subject to \ensuremath{\Varid{s}} calculated above (\ref{eq:160120a}).
Note the free choice of ``pivot'' \ensuremath{\Varid{a}} provided \ensuremath{\Varid{s}} holds.
% Pivot |a| can be taken from any position in the list.
In the standard version, \ensuremath{\Varid{a}} is the head of \ensuremath{\Varid{x}}
and \ensuremath{Z_2} is rendered deterministic as follows (Haskell notation):
\begin{hscode}\SaveRestoreHook
\column{B}{@{}>{\hspre}l<{\hspost}@{}}%
\column{9}{@{}>{\hspre}l<{\hspost}@{}}%
\column{E}{@{}>{\hspre}l<{\hspost}@{}}%
\>[9]{}x_2\;(\Varid{h}\mathbin{:}\Varid{t})\mathrel{=}(\Varid{h},([\mskip1.5mu \Varid{a}\mid \Varid{a}\leftarrow \Varid{t},\Varid{a}\leq \Varid{h}\mskip1.5mu],[\mskip1.5mu \Varid{a}\mid \Varid{a}\leftarrow \Varid{t},\Varid{a}\mathbin{>}\Varid{h}\mskip1.5mu])){}\<[E]%
\ColumnHook
\end{hscode}\resethooks
It is easy to show that the particular partition chosen in this standard
version meets predicate \ensuremath{\Varid{s}}. But there is, still, a check-list of proofs to discharge.

\paragraph{Ensuring bi-ordered (virtual) intermediate trees}
This corresponds to (\ref{eq:150330a}) of the check-list which, instantiated for this exercise, is:
\begin{eqnarray*}
	\ensuremath{\fun G \;\Varid{flatten} \comp ({id}\mathbin{+}\Varid{x}\hskip-1pt\mathit{?})\mathrel{=}({id}\mathbin{+}\Varid{s}\hskip-1pt\mathit{?}) \comp \fun G \;\Varid{flatten}}
\end{eqnarray*}
Letting \ensuremath{\Varid{w}\hskip-1pt\mathit{?}\mathrel{=}{id}\mathbin{+}\Varid{x}\hskip-1pt\mathit{?}}, the goal is to find weakest precondition \ensuremath{\Varid{x}} that is basically
\ensuremath{\Varid{s}} ``passed along" \ensuremath{\fun G \;\Varid{flatten}} from lists to trees:%
\footnote{As before, we abbreviate \ensuremath{\Varid{flatten} \times \Varid{flatten}} by \ensuremath{{\Varid{flatten}}^{\mathrm{2}}} for economy of notation.}
\begin{eqnarray*}
\start
	\ensuremath{({id} \times {\Varid{flatten}}^{\mathrm{2}}) \comp \Varid{x}\hskip-1pt\mathit{?}\mathrel{=}\Varid{s}\hskip-1pt\mathit{?} \comp ({id} \times {\Varid{flatten}}^{\mathrm{2}})}
\just\equiv{ (\ref{eq:150406c}) }
	\ensuremath{\Varid{x}\mathrel{=}\Varid{s} \comp ({id} \times {\Varid{flatten}}^{\mathrm{2}})}
\just\equiv{ go pointwise }
	\ensuremath{\Varid{x}\;(\Varid{a},(t_1,t_2))\mathrel{=}\Varid{s}\;(\Varid{a},(\Varid{flatten}\;t_1,\Varid{flatten}\;t_2))}
\just\equiv{ definition of \ensuremath{\Varid{s}} }
	\ensuremath{\Varid{x}\;(\Varid{a},(t_1,t_2))\mathrel{=}\begin{lcbr}\rcb{\forall}{\Varid{b}}{\Varid{b}\;\epsilon_{\muF}\;(\Varid{flatten}\;t_1)}{\Varid{b}\leq \Varid{a}}\\\rcb{\forall}{\Varid{b}}{\Varid{b}\;\epsilon_{\muF}\;(\Varid{flatten}\;t_2)}{\Varid{a}\leq \Varid{b}}\end{lcbr}}
\just\equiv{ define \ensuremath{\epsilon_{\muG}\mathrel{=}\epsilon_{\muF} \comp \Varid{flatten}} }
	\ensuremath{\Varid{x}\;(\Varid{a},(t_1,t_2))\mathrel{=}\rcb{\forall}{\Varid{b}}{\Varid{b}\;\epsilon_{\muG}\;t_1}{\Varid{b}\leq \Varid{a}}\mathrel{\wedge}\rcb{\forall}{\Varid{b}}{\Varid{b}\;\epsilon_{\muG}\;t_2}{\Varid{a}\leq \Varid{b}})}
\end{eqnarray*}
%Recall that |(crflx w) = id + (crflx x)|.
In words, \ensuremath{\Varid{x}} in \ensuremath{\Varid{p}\hskip-1pt\mathit{?}\mathrel{=}\mathopen{(\!|}\mathsf{in}_{\fun G} \comp \Varid{w}\hskip-1pt\mathit{?}\mathclose{|\!)_{\fun G}}} = \ensuremath{\mathopen{(\!|}\mathsf{in}_{\fun G} \comp ({id}\mathbin{+}\Varid{x}\hskip-1pt\mathit{?})\mathclose{|\!)_{\fun G}}}
ensures that the first part of the implementation, controlled by the
\emph{divide step} coalgebra \ensuremath{\Conid{Z}} calculated above (\ref{eq:150408b})
% --- cf.\ arrow |M| in diagram (\ref{eq:150329e}) ---
yields trees which are \emph{bi-ordered}.
Trees with this property are known as \emph{binary search trees} \cite{Kn97}.

\paragraph{Preserving the metaphor}
Next we consider side condition (\ref{eq:150331a}) of the check-list, which instantiates to:
\begin{eqnarray*}
\start
	\ensuremath{\Varid{bag} \comp \alt{\mathsf{nil}}{\Varid{inord}}} \leq \ensuremath{{id}\mathbin{+}{id} \times {\Varid{bag}}^{\mathrm{2}}}
\just\implied{ coproducts; (\ref{eq:150401a-fun}) }
	\ensuremath{\Varid{bag} \comp \mathsf{nil}\mathbin{+}\Varid{bag} \comp \Varid{inord}} \leq \ensuremath{{id}\mathbin{+}{id} \times {\Varid{bag}}^{\mathrm{2}}}
\just\equiv{ (\ref{eq:150406j-fun}) ; any \ensuremath{\Varid{f}\leq {id}} \cite{Ol14a} } % ; let |bag' = bag . inord|
	\ensuremath{\Varid{bag} \comp \Varid{inord}} \leq \ensuremath{{id} \times {\Varid{bag}}^{\mathrm{2}}}
\just\equiv{ (\ref{eq:160118f}) }
	\ensuremath{\rcb{\exists }{\Varid{k}}{}{\Varid{bag} \comp \Varid{inord}\mathrel{=}\Varid{k} \comp ({id} \times {\Varid{bag}}^{\mathrm{2}})}}
\end{eqnarray*}
That \ensuremath{\Varid{k}} exists arises from the fact that \ensuremath{\Varid{bag}} is a homomorphism between
the monoid of lists and that of bags: algebra \ensuremath{\Varid{k}} will join two bags and
a singleton bag in the same way as \ensuremath{\Varid{inord}\;(\Varid{a},(\Varid{x},\Varid{y}))} yields \ensuremath{\Varid{x}\mathbin{+\!\!+}[\mskip1.5mu \Varid{a}\mskip1.5mu]\mathbin{+\!\!+}\Varid{y}}, at
list level.

\paragraph{Down to the multiset level} \label{pg:170625b}
Finally, we have to check the last assumption (\ref{eq:170625a}) of the ckeck-list.
By (\ref{eq:160118f}) and (\ref{eq:150406c}),
this amounts to finding \ensuremath{\Varid{u}} such that \ensuremath{\fun G \;\Varid{bag} \comp \Varid{r}\hskip-1pt\mathit{?}\mathrel{=}\Varid{u}\hskip-1pt\mathit{?} \comp \fun G \;\Varid{bag}}:
\begin{eqnarray*}
\start
	\ensuremath{\fun G \;\Varid{bag} \comp \Varid{r}\hskip-1pt\mathit{?}\mathrel{=}\Varid{u}\hskip-1pt\mathit{?} \comp \fun G \;\Varid{bag}}
\just\equiv{\ensuremath{\fun G \;\Conid{R}\mathrel{=}{id}\mathbin{+}{id} \times {\Conid{R}}^{\mathrm{2}}} ; \ensuremath{\Varid{r}\hskip-1pt\mathit{?}\mathrel{=}{id}\mathbin{+}\Varid{s}\hskip-1pt\mathit{?}}; let \ensuremath{\Varid{u}\hskip-1pt\mathit{?}\mathrel{=}{id}\mathbin{+}\Varid{v}\hskip-1pt\mathit{?}}}
	\ensuremath{({id}\mathbin{+}{id} \times {\Varid{bag}}^{\mathrm{2}}) \comp ({id}\mathbin{+}\Varid{s}\hskip-1pt\mathit{?})\mathrel{=}({id}\mathbin{+}\Varid{v}\hskip-1pt\mathit{?}) \comp ({id}\mathbin{+}{id} \times {\Varid{bag}}^{\mathrm{2}})}
\just\equiv{ coproducts }
	\ensuremath{({id} \times {\Varid{bag}}^{\mathrm{2}}) \comp \Varid{s}\hskip-1pt\mathit{?}\mathrel{=}\Varid{v}\hskip-1pt\mathit{?} \comp ({id} \times {\Varid{bag}}^{\mathrm{2}})}
\just\equiv{ (\ref{eq:150406c}) }
	\ensuremath{\Varid{s}\mathrel{=}\Varid{v} \comp ({id} \times {\Varid{bag}}^{\mathrm{2}})}
\just\equiv{ go pointwise }
	\ensuremath{\Varid{s}\;(\Varid{a},(\Varid{x},\Varid{y}))\mathrel{=}\Varid{v}\;(\Varid{a},(\Varid{bag}\;\Varid{x},\Varid{bag}\;\Varid{y}))}
\just\equiv{ unfold \ensuremath{\Varid{s}} }
	\ensuremath{\Varid{v}\;(\Varid{a},(\Varid{bag}\;\Varid{x},\Varid{bag}\;\Varid{y}))\mathrel{=}\begin{lcbr}\rcb{\forall}{\Varid{b}}{\Varid{b}\;\epsilon_{\muF}\;\Varid{x}}{\Varid{b}\leq \Varid{a}}\\\rcb{\forall}{\Varid{b}}{\Varid{b}\;\epsilon_{\muF}\;\Varid{y}}{\Varid{a}\leq \Varid{b}}\end{lcbr}}
\just\equiv{ assume \ensuremath{\epsilon_{\fun B}} such that \ensuremath{\epsilon_{\muF}\mathrel{=}\epsilon_{\fun B} \comp \Varid{bag}}}
	\ensuremath{\Varid{v}\;(\Varid{a},(\Varid{bag}\;\Varid{x},\Varid{bag}\;\Varid{y}))\mathrel{=}\begin{lcbr}\rcb{\forall}{\Varid{b}}{\Varid{b}\;\epsilon_{\fun B}\;(\Varid{bag}\;\Varid{x})}{\Varid{b}\leq \Varid{a}}\\\rcb{\forall}{\Varid{b}}{\Varid{b}\;\epsilon_{\fun B}\;(\Varid{bag}\;\Varid{y})}{\Varid{a}\leq \Varid{b}}\end{lcbr}}
\just\implied{ substitution }
	\ensuremath{\Varid{v}\;(\Varid{a},(b_1,b_2))\mathrel{=}\begin{lcbr}\rcb{\forall}{\Varid{b}}{\Varid{b}\;\epsilon_{\fun B}\;b_1}{\Varid{b}\leq \Varid{a}}\\\rcb{\forall}{\Varid{b}}{\Varid{b}\;\epsilon_{\fun B}\;b_2}{\Varid{a}\leq \Varid{b}}\end{lcbr}}
%qed
\end{eqnarray*}
Thus we have found post-condition \ensuremath{\Varid{u}} ensured by \ensuremath{{id} \times {\Varid{bag}}^{\mathrm{2}}} with \ensuremath{\Varid{s}} as weakest-precondition.

Finally, multiset membership \ensuremath{\epsilon_{\fun B}\mathrel{=}{\in} \comp \Varid{support}} can be obtained by taking
multiset \emph{supports}, whereby we land in standard set membership (\ensuremath{{\in}}).
Altogether, we have relied on a chain of memberships, from sets, to multisets, to finite
lists and finally to binary (search) trees.

Note how this last proof of the check-list goes down to the very
essence of \emph{sorting as a metaphorism}: the attribute of a finite list which
any sorting function is bound to preserve is the multiset (bag) of its elements
--- the \emph{invariant} part of the sorting metaphor.

\section{Example: Mergesort} \label{sec:160121b}

In a landmark paper on algorithm classification and synthesis \cite{Da78},
Darlington carries out a derivation of sorting algorithms that
places \emph{quicksort} and \emph{mergesort} in different branches of a derivation tree.
In this section we give a calculation of mergesort which shows precisely where they
differ, given that both are \emph{divide \& conquer} algorithms.

The fact that mergesort relies on a different kind of tree, called \emph{leaf
tree} and based on a different base functor, say \ensuremath{\fun K \;\Varid{f}\mathrel{=}{id}\mathbin{+}{\Varid{f}}^{\mathrm{2}}},
is not the main difference. This resides chiefly in the division of work of
\emph{mergesort} which, contrary to \emph{quicksort}, does almost everything
in the \emph{conquer} step. In our setting,
\begin{quote}
while \emph{quicksort} follows generic metaphorism refinement plan (\ref{eq:160124a}),
\emph{mergesort} follows plan (\ref{eq:160124b}), recall \secref{sec:160125c}.
\end{quote}
With no further detours we go back to (\ref{eq:160125d}), the instance of (\ref{eq:160124b})
which fits the sorting metaphorism, to obtain
\begin{eqnarray*}
	\ensuremath{\Varid{q}\hskip-1pt\mathit{?} \comp \frac{\Varid{bag}}{\Varid{bag}}\mathrel{=}\underbrace{\Varid{q}\hskip-1pt\mathit{?} \comp \frac{\Varid{bag} \comp \Varid{tips}}{\Varid{bag}}}_{\mathopen{(\!|}\Conid{X}\mathclose{|\!)_{\fun K}}} \comp \conv{\Varid{tips}}}
\end{eqnarray*}
where \ensuremath{\Varid{tips}\mathrel{=}\mathopen{(\!|}\Varid{t}\mathclose{|\!)_{\fun K}}} is the fold which converts a leaf tree into a sequence of leaves,
in the obvious way: \ensuremath{\Varid{t}\mathrel{=}\alt{\Varid{singl}}{\mathit{conc}}}, where \ensuremath{\Varid{singl}\;\Varid{a}\mathrel{=}[\mskip1.5mu \Varid{a}\mskip1.5mu]} and \ensuremath{\mathit{conc}\;(\Varid{x},\Varid{y})\mathrel{=}\Varid{x}\mathbin{+\!\!+}\Varid{y}}.%
\footnote{Note that the trivial case of empty list is treated separately from this scheme.
}

\emph{Divide} step \ensuremath{\conv{\Varid{tips}}} can be refined into a function using
standard ``converse of a function" theorems, see e.g.\ \cite{BGM02,MuB04}. % pag 149 de AoP
Our aim is to calculate \ensuremath{\Conid{X}}, the \ensuremath{\fun K }-algebra that shall control the \emph{conquer} step.
We reason:
\begin{eqnarray*}
\start
	\ensuremath{\mathopen{(\!|}\Conid{X}\mathclose{|\!)_{\fun K}}\mathrel{=}\Varid{q}\hskip-1pt\mathit{?} \comp \frac{\Varid{bag}}{\Varid{bag}} \comp \mathopen{(\!|}\Varid{t}\mathclose{|\!)_{\fun K}}}
\just\implied{ fusion (\ref{eq:150402a}) ; functor \ensuremath{\fun K } }
	\ensuremath{\Varid{q}\hskip-1pt\mathit{?} \comp \frac{\Varid{bag}}{\Varid{bag}} \comp \Varid{t}\mathrel{=}\Conid{X} \comp (\fun K \;\Varid{q}\hskip-1pt\mathit{?}) \comp \fun K \;\frac{\Varid{bag}}{\Varid{bag}}}
\just\implied{ (\ref{eq:150326b}) assuming \ensuremath{\frac{\Varid{bag}}{\Varid{bag}}} is a \ensuremath{\fun K }-congruence for algebra \ensuremath{\Varid{t}} ; Leibniz }
	\ensuremath{\Varid{q}\hskip-1pt\mathit{?} \comp \frac{\Varid{bag}}{\Varid{bag}} \comp \Varid{t}\mathrel{=}\Conid{X} \comp \fun K \;\Varid{q}\hskip-1pt\mathit{?}}
\end{eqnarray*}
Next, we head for a functional implementation \ensuremath{{\Varid{x}}\subseteq{\Conid{X}}}:
\begin{eqnarray*}
\start
	\ensuremath{{\Varid{x} \comp \fun K \;\Varid{q}\hskip-1pt\mathit{?}}\subseteq{\Varid{q}\hskip-1pt\mathit{?} \comp \frac{\Varid{bag}}{\Varid{bag}} \comp \Varid{t}}}
\just\implied{ cancel \ensuremath{\Varid{q}\hskip-1pt\mathit{?}} assuming \ensuremath{\Varid{x} \comp \fun K \;\Varid{q}\hskip-1pt\mathit{?}\mathrel{=}\Varid{q}\hskip-1pt\mathit{?} \comp \Varid{x}} (\ref{eq:150406c}) }
	\ensuremath{\Varid{x}\; \subseteq \;\frac{\Varid{bag} \comp \Varid{t}}{\Varid{bag}}}
\end{eqnarray*}

Again we obtain solution \ensuremath{\Varid{x}\mathbin{:}\fun K \;\muF \to \muF } as a metaphor implementation, essentially requiring
that \ensuremath{\Varid{x}} preserves the bag of elements of the lists involved, as in \emph{quicksort}.
Note that the surjectivity of \ensuremath{\Varid{bag}} allows for a total solution \ensuremath{\Varid{x}}, whose
standard implementation is the well-known \emph{list merge} function
that merges two ordered lists into an ordered list.
This behaviour is in fact required by the last assumption above,
\ensuremath{\Varid{x} \comp \fun K \;\Varid{q}\hskip-1pt\mathit{?}\mathrel{=}\Varid{q}\hskip-1pt\mathit{?} \comp \Varid{x}}.
The other assumption, that \ensuremath{\frac{\Varid{bag}}{\Varid{bag}}} is a congruence for algebra \ensuremath{\Varid{t}},
amounts to (recall Theorem \ref{th:150327a}):
\begin{eqnarray*}
\start
	\ensuremath{\Varid{bag} \comp \Varid{t}} \leq \ensuremath{\fun K \;\Varid{bag}}
\just\implied{ \ensuremath{\Varid{t}\mathrel{=}\alt{\Varid{singl}}{\mathit{conc}}}; coproduct injectivity (\ref{eq:150401a-fun},\ref{eq:150406j-fun}) ; \ensuremath{\fun K \;\Varid{f}\mathrel{=}{id}\mathbin{+}{\Varid{f}}^{\mathrm{2}}} }
	\ensuremath{\begin{lcbr}\Varid{bag} \comp \Varid{singl}\leq {id}\\\Varid{bag} \comp \mathit{conc}\leq {\Varid{bag}}^{\mathrm{2}}\end{lcbr}}
\just\equiv{ any \ensuremath{\Varid{f}\leq {id}} }
	\ensuremath{\Varid{bag} \comp \mathit{conc}\leq {\Varid{bag}}^{\mathrm{2}}}
\just\equiv{ (\ref{eq:160118f}) }
	\ensuremath{\rcb{\exists }{\Varid{k}}{}{\Varid{bag}\;(\Varid{x}\mathbin{+\!\!+}\Varid{y})\mathrel{=}\Varid{k}\;(\Varid{bag}\;\Varid{x},\Varid{bag}\;\Varid{y})}}
\just\equiv{ same argument as in quicksort }
	\ensuremath{\Varid{true}}
\qed
\end{eqnarray*}

Summing up, the workload inversion in \emph{mergesort},
compared to \emph{quicksort},
can be felt right at the start of the derivation, by grafting the (range
of the) virtual tree representation at the front rather than at the rear of the pipeline.

\section{Example: minimum height trees} \label{sec:170322a}
Our last example addresses a metaphorism \ensuremath{{\frac{\mathopen{(\!|}\Varid{f}\mathclose{|\!)}}{\mathopen{(\!|}\Varid{g}\mathclose{|\!)}}}\shrunkby{\Conid{R}}} in which \ensuremath{\Conid{R}} is an optimization preorder.
% and an intermediate data structure is introduced not for algorithmic control but rather for efficiency purposes.
It is adapted from \cite{BGM02} where it is labelled \emph{tree with minimum
height}. Rephrased in our setting, the problem to be addressed is that of
\emph{reshaping a binary tree so as to minimize its height}:\footnote{That is to say, leaves are
numbers representing heights of subtrees, and the problem is to assemble the subtrees into a single tree of minimum height.}
\begin{eqnarray}
	\ensuremath{{\frac{\Varid{tips}}{\Varid{tips}}}\shrunkby{\leq _{\Varid{height}}}}
	\label{eq:170322b}
\end{eqnarray}
Recall that \ensuremath{\Varid{tips}\mathrel{=}\mathopen{(\!|}\alt{\Varid{singl}}{\mathit{conc}}\mathclose{|\!)_{\fun K}}} converts a tree into the sequence
of its leaves.\footnote{Note that \ensuremath{\Varid{tips}} is called \ensuremath{\Varid{flatten}} in \cite{BGM02}. Recall
\ensuremath{\fun K \;\Varid{f}\mathrel{=}{id}\mathbin{+}{\Varid{f}}^{\mathrm{2}}} and \ensuremath{\mathsf{in}_{\fun K}\mathrel{=}\alt{\mathsf{leaf}}{\mathsf{fork}}} from \secref{sec:160121b}.}
Heights of trees are calculated by function \ensuremath{\Varid{height}\mathrel{=}\mathopen{(\!|}\alt{{id}}{\Varid{ht}}\mathclose{|\!)_{\fun K}}} where
\ensuremath{\Varid{ht}\;(\Varid{a},\Varid{b})\mathrel{=}(\Varid{a}\sqcup \Varid{b})\mathbin{+}\mathrm{1}}, for \ensuremath{\Varid{a}\sqcup \Varid{b}\mathrel{=}\Varid{b}~\Leftrightarrow~\Varid{a}\leq \Varid{b}}. Finally, \ensuremath{\leq _{\Varid{height}}} is
a shorthand for \ensuremath{\conv{\Varid{height}} \comp (\leq ) \comp \Varid{height}}, the preorder that ranks trees
according to their height.

By rule (\ref{eq:fn-shrink-r}), (\ref{eq:170322b}) is the same as
	\ensuremath{({\conv{\Varid{tips}}}\shrunkby{\leq _{\Varid{height}}}) \comp \Varid{tips}}.
\citet{BGM02} show the advantage of handling \ensuremath{\conv{\Varid{tips}}} using
a different format for trees known as \emph{left spine}.\footnote{%
In essence, left spines offer a bottom-up access to trees, in this case more
convenient than the standard top-down traversal.} Let \ensuremath{\larrow{\Conid{A}\mathbin{+}{\muK }^{\mathrm{2}}}{\mathsf{in}_{\fun K}}{\muK }}
be our datatype of trees with leaves of type \ensuremath{\Conid{A}}. The corresponding left spine datatype is
\ensuremath{\Conid{S}\mathrel{=}\Conid{A} \times {\muK }^{*}} and it is isomorphic to \ensuremath{\muK }. This isomorphism, termed
\ensuremath{\Varid{roll}} as in \cite{BGM02}, is depicted below in the form of a diagram, where
\ensuremath{\alpha } is obvious and \ensuremath{\mathsf{in}_{sn}\mathrel{=}\alt{\mathsf{nil}}{\mathsf{snoc}}} is the ``snoc" variant of initial
algebra of lists:\footnote{Recall \ensuremath{\mathsf{snoc}\;(\Varid{x},\Varid{xs})\mathrel{=}\Varid{xs}\mathbin{+\!\!+}[\mskip1.5mu \Varid{x}\mskip1.5mu]} from \cite{BM97}.}
\begin{eqnarray*}
\myxym{
	\ensuremath{\Conid{A} \times {\muK }^{*}}
		\ar[d]_{\ensuremath{\Varid{roll}}}
&
	\ensuremath{\Conid{A} \times (\mathrm{1}\mathbin{+}\muK  \times {\muK }^{*})}
		\ar[l]_-{\ensuremath{{id} \times \mathsf{in}_{sn}}}
&
	\ensuremath{\Conid{A}\mathbin{+}(\Conid{A} \times {\muK }^{*}) \times \muK }
		\ar[l]_-{\ensuremath{\alpha }}
		\ar[d]^{\ensuremath{{id}\mathbin{+}\Varid{roll} \times {id}}}
\\
	\ensuremath{\muK }
&
&
	\ensuremath{\Conid{A}\mathbin{+}\muK  \times \muK }
		\ar[ll]^{\ensuremath{\mathsf{in}_{\fun K}}}
}
\end{eqnarray*}
To obtain \ensuremath{\conv{\Varid{roll}}} one just has to reverse all arrows in the
diagram, since they are all isomorphisms.

The left spine representation is introduced as in the previous examples:
\begin{eqnarray*}
\start
	\ensuremath{({\conv{\Varid{tips}}}\shrunkby{\leq _{\Varid{height}}}) \comp \Varid{tips}}
\just={\ensuremath{\Varid{roll} \comp \conv{\Varid{roll}}\mathrel{=}{id}}}
	\ensuremath{({(\Varid{roll} \comp \conv{\Varid{roll}} \comp \conv{\Varid{tips}})}\shrunkby{\leq _{\Varid{height}}}) \comp \Varid{tips}}
\just={ by (\ref{eq:fn-shrink-l}) abbreviating \ensuremath{{\preceq }^\prime\mathrel{=}(\leq )_{\Varid{height} \comp \Varid{roll}}} and \ensuremath{\Varid{troll}\mathrel{=}\Varid{tips} \comp \Varid{roll}}}
	\ensuremath{\Varid{roll} \comp ({\conv{\Varid{troll}}}\shrunkby{{\preceq }^\prime}) \comp \Varid{tips}}
\end{eqnarray*}
Altogether, we are lead to a metaphorism between binary trees and left spines, post processed by \ensuremath{\Varid{roll}}:
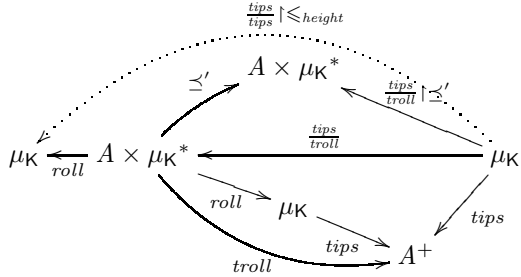
\begin{eqnarray*}
\xymatrix@C=3.5ex@R=1.4ex{
&
&
	\ensuremath{\Conid{A} \times {\muK }^{*}}
\\
\\
	\ensuremath{\muK }
&
	\ensuremath{\Conid{A} \times {\muK }^{*}}
		\ar[l]^-{\ensuremath{\Varid{roll}}}
		\ar[rd]_-{\ensuremath{\Varid{roll}}}
		\ar@/^0.5pc/[ruu]^(.5){\ensuremath{{\preceq }^\prime}}
		\ar@/_1.5pc/[ddrr]_(.5){\ensuremath{\Varid{troll}}}
&
&
&
	\ensuremath{\muK }
		\ar@{.>}@/_3.8pc/[llll]_(.45){\ensuremath{{\frac{\Varid{tips}}{\Varid{tips}}}\shrunkby{\leq _{\Varid{height}}}}}
		\ar[ldd]^-{\ensuremath{\Varid{tips}}}
		\ar[lll]_{\ensuremath{\frac{\Varid{tips}}{\Varid{troll}}}}
		\ar[lluu]_{\ensuremath{{\frac{\Varid{tips}}{\Varid{troll}}}\shrunkby{{\preceq }^\prime}}}
&	
&	
\\
&
&
	\ensuremath{\muK }
		\ar[rd]_-{\ensuremath{\Varid{tips}}}
&
\\
&
&
&
	\ensuremath{{\Conid{A}}^{+}}
}
\end{eqnarray*}
The hard bit above is \ensuremath{{\conv{\Varid{troll}}}\shrunkby{{\preceq }^\prime}}, to be addressed in two steps:
first, we convert \ensuremath{\conv{\Varid{troll}}\mathrel{=}\mathopen{(\!|}\Conid{S}\mathclose{|\!)}} for some \ensuremath{\Conid{S}} using the
\emph{converse of a function} theorem by \citet{BM97}. % Theorem 6.4, pag 149 de AoP
Then we use the \emph{greedy theorem} of shrinking \cite{MO12a} to refine \ensuremath{{\mathopen{(\!|}\Conid{S}\mathclose{|\!)}}\shrunkby{{\preceq }^\prime}} into
\ensuremath{\mathopen{(\!|}{\Conid{S}}\shrunkby{{\preceq }^\prime}\mathclose{|\!)}}. For easy reference, we quote both theorems below from their sources. 

\begin{theorem}[Converse of a function]\label{th:170322c}
Let \ensuremath{\rarrow{\fun F \;\fun T }{\mathsf{in}_{\fun T}}{\fun T }} and \ensuremath{\rarrow{\Conid{A}}{\Varid{f}}{\fun T }} be given. Then \ensuremath{\conv{\Varid{f}}\mathrel{=}\mathopen{(\!|}\Conid{R}\mathclose{|\!)}} provided \ensuremath{\Conid{R}\mathbin{:}\Conid{A}\leftarrow \fun F \;\Conid{A}} is surjective and such that \ensuremath{\Varid{f} \comp \Conid{R}\; \subseteq \;\mathsf{in}_{\fun T} \comp \fun F \;\Varid{f}}.
(Proof: see theorem 6.4 in \cite{BM97}.)
\\\ensuremath{\ensuremath{\Box}}
\end{theorem}

\begin{theorem}[Greedy shrinking]\label{th:170322d}
\ensuremath{{\mathopen{(\!|}{\Conid{S}}\shrunkby{\Conid{R}}\mathclose{|\!)}}\subseteq{{\mathopen{(\!|}\Conid{S}\mathclose{|\!)}}\shrunkby{\Conid{R}}}} provided \ensuremath{\Conid{R}} is transitive and \ensuremath{\Conid{S}} is monotonic with respect to \ensuremath{\conv{\Conid{R}}}, that is,
\ensuremath{\Conid{S} \comp (\fun F \;\conv{\Conid{R}})\; \subseteq \;\conv{\Conid{R}} \comp \Conid{S}}.
(Proof: see theorem 1 in \cite{MO12a}.)
\\\ensuremath{\ensuremath{\Box}}
\end{theorem}

In our case, \ensuremath{\mathsf{in}_{\fun T}\mathbin{:}\fun F \;{\Conid{A}}^{+}\to {\Conid{A}}^{+}} (non-empty lists)
where \ensuremath{\fun F \;\Conid{X}\mathrel{=}\Conid{A}\mathbin{+}\Conid{A} \times \Conid{X}}, assuming \ensuremath{\Conid{A}} fixed. (\ensuremath{\Conid{A}\mathrel{=}\mathbb{Z}} in \cite{BGM02}.)
We aim at \ensuremath{\conv{\Varid{troll}}\mathrel{=}\mathopen{(\!|}\alt{\Conid{R}}{\Conid{Q}}\mathclose{|\!)}}, where \ensuremath{\Conid{R}\mathbin{:}\Conid{A} \times {\muK }^{*}\leftarrow \Conid{A}}
and \ensuremath{\Conid{Q}\mathbin{:}\Conid{A} \times {\muK }^{*}\leftarrow \Conid{A} \times (\Conid{A} \times {\muK }^{*})}.
While \ensuremath{\Conid{R}\mathrel{=}{{id}}\kr{\mathsf{nil}}} is immediate, \ensuremath{\Conid{Q}} is constrained by theorem \ref{th:170322c}
\begin{eqnarray}
%|tips.roll . S atmost cons . (id >< tips . roll)|
	\ensuremath{\Conid{Q}\; \subseteq \;\frac{\mathsf{cons} \comp ({id} \times \Varid{troll})}{\Varid{troll}}}
	\label{eq:170329a}
\end{eqnarray}
and by theorem \ref{th:170322d}:
\begin{eqnarray}
\ensuremath{\Conid{Q} \comp ({id} \times \conv{({\preceq }^\prime)})\; \subseteq \;\conv{({\preceq }^\prime)} \comp \Conid{Q}}
	\label{eq:170326a}
\end{eqnarray}
Solutions will be of the form
%begin{quote}
\ensuremath{\mathopen{(\!|}\alt{{\mathsf{one}}\kr{\mathsf{nil}}}{{\Conid{Q}}\shrunkby{{\preceq }^\prime}}\mathclose{|\!)}} 
%end{quote}
by the following properties of shrinking \cite{MO12a}: 
\begin{eqnarray*}
\start
	\junc{S}{T}\shrunkby R = \junc{S\shrunkby R}{T\shrunkby R}
	%label{eq:shrink-dist-join}
\more
	\ensuremath{{\Varid{f}}\shrunkby{\Conid{R}}\mathrel{=}\Varid{f}} ~ \implied ~ \mbox{\ensuremath{\Conid{R}} is reflexive}
\end{eqnarray*}
The following property of \ensuremath{\Varid{troll}}
\begin{eqnarray*}
	\ensuremath{\Varid{troll} \comp ({id} \times \mathsf{cons} \comp (\mathsf{leaf} \times {id}))\mathrel{=}\mathsf{cons} \comp ({id} \times \Varid{troll})}
\end{eqnarray*}
naively suggests solution \ensuremath{\Conid{Q}\mathrel{=}{id} \times \mathsf{cons} \comp (\mathsf{leaf} \times {id})} which,
however, does not work: for input tree 
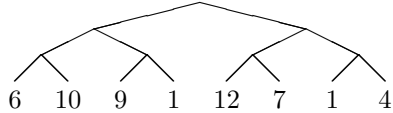
\begin{quote}
%|t = Fork (Fork (Fork (Leaf 6,Leaf 10),Fork (Leaf 9,Leaf 1)),Fork (Fork (Leaf 12,Leaf 7),Fork (Leaf 1,Leaf 4)))|
\unitlength=1.0em
\begin{picture}(15,5)(0,-2)
\put(0,-0.7){\makebox(0,0)[c]{6}}
\put(2,-0.7){\makebox(0,0)[c]{10}}
\put(1,1){\line(-1,-1){1}}
\put(1,1){\line(1,-1){1}}
\put(4,-0.7){\makebox(0,0)[c]{9}}
\put(6,-0.7){\makebox(0,0)[c]{1}}
\put(5,1){\line(-1,-1){1}}
\put(5,1){\line(1,-1){1}}
\put(3,2){\line(-2,-1){2}}
\put(3,2){\line(2,-1){2}}
\put(8,-0.7){\makebox(0,0)[c]{12}}
\put(10,-0.7){\makebox(0,0)[c]{7}}
\put(9,1){\line(-1,-1){1}}
\put(9,1){\line(1,-1){1}}
\put(12,-0.7){\makebox(0,0)[c]{1}}
\put(14,-0.7){\makebox(0,0)[c]{4}}
\put(13,1){\line(-1,-1){1}}
\put(13,1){\line(1,-1){1}}
\put(11,2){\line(-2,-1){2}}
\put(11,2){\line(2,-1){2}}
\put(7,3){\line(-4,-1){4}}
\put(7,3){\line(4,-1){4}}
\end{picture}
\end{quote}
with height \ensuremath{\mathrm{15}}, \ensuremath{\mathopen{(\!|}{id} \times \mathsf{cons} \comp (\mathsf{leaf} \times {id})\mathclose{|\!)} \comp \Varid{tips}} would generate output tree
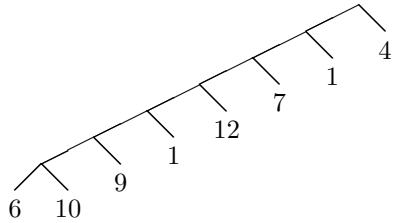
\begin{quote}
%Fork (Fork (Fork (Fork (Fork (Fork (Fork (Leaf 6,Leaf 10),Leaf 9),Leaf 1),Leaf 12),Leaf 7),Leaf 1),Leaf 4)
\unitlength=1.0em
\begin{picture}(15,9)(0,-2)
\put(0,-0.7){\makebox(0,0)[c]{6}}
\put(2,-0.7){\makebox(0,0)[c]{10}}
\put(1,1){\line(-1,-1){1}}
\put(1,1){\line(1,-1){1}}
\put(4,0.3){\makebox(0,0)[c]{9}}
\put(3,2){\line(-2,-1){2}}
\put(3,2){\line(1,-1){1}}
\put(6,1.3){\makebox(0,0)[c]{1}}
\put(5,3){\line(-2,-1){2}}
\put(5,3){\line(1,-1){1}}
\put(8,2.3){\makebox(0,0)[c]{12}}
\put(7,4){\line(-2,-1){2}}
\put(7,4){\line(1,-1){1}}
\put(10,3.3){\makebox(0,0)[c]{7}}
\put(9,5){\line(-2,-1){2}}
\put(9,5){\line(1,-1){1}}
\put(12,4.3){\makebox(0,0)[c]{1}}
\put(11,6){\line(-2,-1){2}}
\put(11,6){\line(1,-1){1}}
\put(14,5.3){\makebox(0,0)[c]{4}}
\put(13,7){\line(-2,-1){2}}
\put(13,7){\line(1,-1){1}}
\end{picture}
\end{quote}
with height \ensuremath{\mathrm{17}}, worsening the input rather than improving it. Still, because
\begin{eqnarray*}
	\ensuremath{\mathsf{head} \comp \Varid{troll}\mathrel{=}\p1}
\end{eqnarray*}
one may stick to the pattern \ensuremath{\Conid{Q}\mathrel{=}{\p1}\kr{\Conid{U} \comp ({id} \times (\mathsf{leaf} \times {id}))}}, 
for some \ensuremath{\Conid{U}\mathbin{:}{\muK }^{*}\leftarrow \Conid{A} \times (\muK  \times {\muK }^{*})}.

It turns out that finding \ensuremath{\Conid{U}} such that \ensuremath{\Conid{Q}} satisfies (\ref{eq:170329a},\ref{eq:170326a})
is not easy \cite{BGM02}.
The solution is a strategy allowed by the monotonicity of shrinking on the optimization relation:
if \ensuremath{\Conid{P}\; \subseteq \;\Conid{R}} then \ensuremath{{{\Conid{S}}\shrunkby{\Conid{P}}}\subseteq{{\Conid{S}}\shrunkby{\Conid{R}}}} and therefore
\ensuremath{{\mathopen{(\!|}{\Conid{S}}\shrunkby{\Conid{P}}\mathclose{|\!)}}\subseteq{\mathopen{(\!|}{\Conid{S}}\shrunkby{\Conid{R}}\mathclose{|\!)}}}.%
\footnote{The \emph{refined greedy theorem} by \citet{BGM02} corresponds to this use of theorem \ref{th:170322d}.}
At this point we can pick the refinement \ensuremath{\Conid{U}\mathrel{=}\Varid{minsplit}} given in \cite{BGM02}:
\begin{hscode}\SaveRestoreHook
\column{B}{@{}>{\hspre}l<{\hspost}@{}}%
\column{3}{@{}>{\hspre}l<{\hspost}@{}}%
\column{E}{@{}>{\hspre}l<{\hspost}@{}}%
\>[B]{}\Varid{minsplit}\;(\Varid{a},(\Varid{x},[\mskip1.5mu \mskip1.5mu]))\mathrel{=}[\mskip1.5mu \Varid{x}\mskip1.5mu]{}\<[E]%
\\
\>[B]{}\Varid{minsplit}\;(\Varid{a},(\Varid{x},\Varid{y}\mathbin{:}\Varid{xs})){}\<[E]%
\\
\>[B]{}\hsindent{3}{}\<[3]%
\>[3]{}\mid (\Varid{a}\sqcup \Varid{height}\;\Varid{x})\mathbin{<}\Varid{height}\;\Varid{y}\mathrel{=}\Varid{x}\mathbin{:}\Varid{y}\mathbin{:}\Varid{xs}{}\<[E]%
\\
\>[B]{}\hsindent{3}{}\<[3]%
\>[3]{}\mid \mathsf{otherwise}\mathrel{=}\Varid{minsplit}\;(\Varid{a},(\mathsf{fork}\;(\Varid{x},\Varid{y}),\Varid{xs})){}\<[E]%
\ColumnHook
\end{hscode}\resethooks
For the input tree given above, this refinement will yield
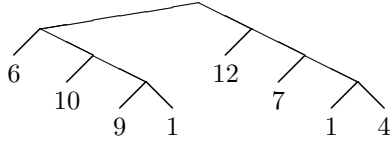
\begin{quote}
% Fork (Fork (Leaf 6,Fork (Leaf 10,Fork (Leaf 9,Leaf 1))),Fork (Leaf 12,Fork (Leaf 7,Fork (Leaf 1,Leaf 4))))
\unitlength=1.0em
\begin{picture}(15,5)(0,-3)
\put(0,-0.7){\makebox(0,0)[c]{6}}
\put(2,-1.7){\makebox(0,0)[c]{10}}
\put(4,-2.7){\makebox(0,0)[c]{9}}
\put(6,-2.7){\makebox(0,0)[c]{1}}
\put(5,-1){\line(-1,-1){1}}
\put(5,-1){\line(1,-1){1}}
\put(3,0){\line(-1,-1){1}}
\put(3,0){\line(2,-1){2}}
\put(1,1){\line(-1,-1){1}}
\put(1,1){\line(2,-1){2}}
\put(8,-0.7){\makebox(0,0)[c]{12}}
\put(10,-1.7){\makebox(0,0)[c]{7}}
\put(12,-2.7){\makebox(0,0)[c]{1}}
\put(14,-2.7){\makebox(0,0)[c]{4}}
\put(13,-1){\line(-1,-1){1}}
\put(13,-1){\line(1,-1){1}}
\put(11,0){\line(-1,-1){1}}
\put(11,0){\line(2,-1){2}}
\put(9,1){\line(-1,-1){1}}
\put(9,1){\line(2,-1){2}}
\put(7,2){\line(-6,-1){6}}
\put(7,2){\line(2,-1){2}}
\end{picture}
\end{quote}
with (minimum) height \ensuremath{\mathrm{14}}.

To follow the reasoning by \citet{BGM02} that leads to the above solution
note that, although shrinking is not used there, it is in a sense implicit.
Take \ensuremath{{\Conid{S}}\shrunkby{\Conid{R}}} and apply the power-transpose to \ensuremath{\Conid{S}}, obtaining \ensuremath{{({\in} \comp \Lambda{\Conid{S}})}\shrunkby{\Conid{R}}}. Since \ensuremath{\Lambda{\Conid{S}}} is a function we can use (\ref{eq:fn-shrink-r})
to get \ensuremath{({{\in}}\shrunkby{\Conid{R}}) \comp \Lambda{\Conid{S}}}. The relation \ensuremath{{{\in}}\shrunkby{\Conid{R}}\mathbin{:}\Conid{A}\leftarrow \fun P \;\Conid{A}}, which picks minimal elements of a set according to criterion \ensuremath{\Conid{R}} is written
\ensuremath{\Varid{min}\;\Conid{R}} in \cite{BGM02}. So expressions of the form (\ensuremath{\Varid{min}\;\Conid{R} \comp \Lambda{\Conid{S}}}) in that
paper express the same as \ensuremath{{\Conid{S}}\shrunkby{\Conid{R}}} in the current paper.

The ordering on left spines found in \cite{BGM02} to ensure the monotonicity of \ensuremath{\Conid{Q}} is
\ensuremath{\sqsubseteq _{\Varid{lspinecosts}}}
where \ensuremath{\Varid{lspinecosts}\;(\Varid{a},\Varid{ts})\mathrel{=}[\mskip1.5mu (\Varid{height} \comp \Varid{roll})\;(\Varid{a},\Varid{x})\mid \Varid{x}\leftarrow \Varid{pref}\;\Varid{ts}\mskip1.5mu]},
\ensuremath{\Varid{pref}\;\Varid{ts}} lists the prefixes of \ensuremath{\Varid{ts}} in length-decreasing order, and
\ensuremath{[\mskip1.5mu \Varid{a}_{\mathrm{1}}\mathinner{\ldotp\ldotp}\Varid{a}_{\Varid{m}}\mskip1.5mu]\sqsubseteq [\mskip1.5mu \Varid{b}_{\mathrm{1}}\mathinner{\ldotp\ldotp}\Varid{b}_{\Varid{n}}\mskip1.5mu]~\Leftrightarrow~\Varid{m}\leq \Varid{n}\mathrel{\wedge}\rcb{\forall}{\Varid{i}}{\Varid{i}\leq \Varid{m}}{\Varid{a}_{\Varid{i}}\leq \Varid{b}_{\Varid{i}}}}. As examples, let \ensuremath{\Varid{s}} be the left spine of the first,
balanced tree given above as example, and \ensuremath{\Varid{s'}} be that of the last, minimal-height
one. We have \ensuremath{\Varid{lspinecosts}\;\Varid{s}\mathrel{=}[\mskip1.5mu \mathrm{15},\mathrm{12},\mathrm{11},\mathrm{6}\mskip1.5mu]} while \ensuremath{\Varid{lspinecosts}\;\Varid{s'}\mathrel{=}[\mskip1.5mu \mathrm{14},\mathrm{12},\mathrm{6}\mskip1.5mu]}, meaning
\ensuremath{\Varid{s'}\;\sqsubseteq _{\Varid{lspinecosts}}\;\Varid{s}}. That \ensuremath{\sqsubseteq _{\Varid{lspinecosts}}\; \subseteq \;{\preceq }^\prime\mathrel{=}\leq _{\Varid{height} \comp \Varid{roll}}} holds follows immediately from \ensuremath{\mathsf{head} \comp \Varid{lspinecosts}\mathrel{=}\Varid{height} \comp \Varid{roll}}.

\section{Conclusions} \label{sec:150406h}

This paper identifies a pattern of relational specification, termed \emph{metaphorism},
in which some kernel information of the input is preserved at the same time some
form of optimization takes place towards the output of an algorithmic process.
Text processing, sorting and representation changers are given as examples
of metaphorisms.

Metaphorisms expose the \emph{variant}/\emph{invariant} duality essential to program correctness
in their own way: there are two main attributes in
the game, one is to be preserved (the essence of the metaphor,
cf.\ \emph{invariant}) while the other is to be mini(maxi)mized
(the essence of the optimization, cf.\ \emph{variant}).

At the heart of relational specifications of this kind the paper identifies
the occurrence of \emph{metaphors} characterized as \emph{symmetric divisions}
\cite{FS90,SS93} of functions. This makes it possible to regard them as \emph{rational}
(regular) relations \cite{JMBD91} and develop an algebra of metaphors that  contains
much of what is needed for refining {metaphorisms}
into recursive programs.

In particular, the kind of metaphorism refinement studied in the paper is known as \emph{changing
the virtual data structure}, whereby algorithms are re-structured in a \emph{divide \& conquer} fashion.
The paper gives sufficient conditions for such implementations
to be calculated in general and gives the derivation of \emph{quicksort}
and \emph{mergesort} as examples. The former can be regarded as a generalization
of the reasoning about the same algorithm given by \citet{BM97}.

Altogether, the paper shows how such \emph{divide \& conquer} refinement
strategies consist of replacing the ``outer metaphor'' of the starting specification
(metaphorism) by a more implicit but more interesting ``inner metaphor'',
which governs the implementation. Where exactly this inner metaphor
is located depends on the overall refinement plan.

The \emph{quicksort} example shows how the outer metaphor,
relating lists which permute each other, gives place to an
inner metaphor located in the \emph{divide} step that
relates lists to binary search trees.
This provides technical evidence for \emph{quicksort} being usually classified
as a ``\emph{Hard Split, Easy Join}'' \cite{Ho94} algorithm: indeed, the
``metaphor shift'' calculated in \secref{sec:150406i} shows the workload
passing along the conquer layer towards the \emph{divide} one, eventually landing into
the \emph{coalgebra} which governs the \emph{``hard'' divide} process. Conversely,
the inner metaphor in the case of \emph{mergesort} goes into the \emph{algebra}
of the \emph{conquer} step, explaining why this is regarded as a ``\emph{Easy
Split, Hard Join}'' algorithm by \citet{Ho94}. As seen in the paper, this
has to do with \emph{where} the virtual data structure is placed, either at the
front or at the rear of the starting metaphorism.

From the linguistics perspective, metaphorisms are \emph{formal} metaphors
and not exactly \emph{cognitive} metaphors. But computer science is full of these as well,
as its terminology (e.g.\ ``stack'', ``pipe", ``memory", ``driver") amply shows.
If a picture is worth a thousand words, perhaps a good metaphor(ism) is worth a thousand
axioms?

\section{Future work} \label{sec:160122b}

The research reported in this paper falls into the area of investigating
how to manage or refine specification vagueness (non-determinism) by means
of the ``shrinking" combinator proposed elsewhere  \cite{MO12a,OF13}.
The interplay between this combinator and metaphors (as symmetric divisions)
has room for further research. Can \ensuremath{\frac{\Varid{f}}{\Varid{g}}} be generalized to some \ensuremath{\frac{\Conid{R}}{\Conid{S}}}
and still retain metaphors' ability to \emph{equate objects of incompatible orders} \cite{Be81}? 
Facts (\ref{eq:170413a}), (\ref{eq:170413b}) and (\ref{eq:160108a}) point
towards such a generalization.
% Vagueness, yes, is an ingredient of formal specification, \emph{ma non troppo}.
This relates to another direction for possible genericity:
metaphorisms as given in this paper call for a \emph{division allegory} \cite{FS90}
such as that of binary relations. Can this be generalized? \citet{Gi16}
asks a similar question and suggests \emph{regular} categories as the right
abstraction. It will be interesting to generalize metaphorisms in a similar,
categorial way.

Another direction for future research is to generalize \emph{shrinking} in
metaphorisms to \emph{thinning} \cite{BM97}. A notation similar to \ensuremath{{\Conid{S}}\shrunkby{\Conid{R}}}
can be adopted for thinning,
\begin{eqnarray*}
	\ensuremath{{\Conid{S}}\thinnedby{\Conid{R}}\mathrel{=}{{\in}\setminus \Conid{S}}\mathbin\cap{(\conv{{\in}} \comp \Conid{R})\mathbin{/}\conv{\Conid{S}}}}
\end{eqnarray*}
where \ensuremath{{\Conid{S}}\thinnedby{\Conid{R}}} is a set-valued relation: \ensuremath{\Varid{x}\;({\Conid{S}}\thinnedby{\Conid{R}})\;\Varid{a}} holds for exactly those \ensuremath{\Varid{x}} such that
\ensuremath{\Varid{x}\; \subseteq \;\Lambda{\Conid{S}}\;\Varid{a}} and \ensuremath{\Varid{x}} is lower-bounded with respect to \ensuremath{\Conid{R}}.
\ensuremath{{\Conid{S}}\shrunkby{\Conid{R}}} corresponds to that part of \ensuremath{{\Conid{S}}\thinnedby{\Conid{R}}} whose outputs are singletons containing minima,
\ensuremath{\eta  \comp ({\Conid{S}}\shrunkby{\Conid{R}})\; \subseteq \;{\Conid{S}}\thinnedby{\Conid{R}}} where \ensuremath{\eta \;\Varid{b}\mathrel{=}\{\mskip1.5mu \Varid{b}\mskip1.5mu\}}.
For \ensuremath{\Conid{R}} a preorder one has:
\begin{quote}
	\ensuremath{{\Conid{S}}\shrunkby{\Conid{R}}\mathrel{=}({{\in}}\shrunkby{\Conid{R}}) \comp ({\Conid{S}}\thinnedby{\Conid{R}})}
\end{quote}
So preorder shrinking can be  expressed by thinning. Not surprisingly, shrinking and thinning share similar laws,
namely (\ref{eq:fn-shrink-r}), cf. \ensuremath{{(\Conid{S} \comp \Varid{f})}\thinnedby{\Conid{R}}\mathrel{=}({\Conid{S}}\thinnedby{\Conid{R}}) \comp \Varid{f}}.
Thus, refining metaphorisms under thinning can also follow the \emph{converse of a function} strategy
enabled by theorem \ref{th:170322c}. Moreover, the thinning counterpart to
the greedy theorem,
% Theorem 8.1:
\ensuremath{\mathopen{(\!|}{(\Conid{S} \comp \fun F \;{\in})}\thinnedby{\Conid{R}}\mathclose{|\!)}\; \subseteq \;{\mathopen{(\!|}\Conid{S}\mathclose{|\!)}}\thinnedby{\Conid{R}}}
suggests similar refinement processes. Whether thinning offers genuinely new opportunities for metaphorism
reasoning as compared to shrinking is a subject for future research.

% This paper is intended as basis for future work in exploiting the metaphorism concept in program derivation. Candidate case studies in program refactoring or text processing already pose significant challenges when compared to the sorting examples given in the current paper. Comparative work is also welcome, in particular checking what benefits can be expected from regarding arbitrary representation changers \cite{HM93b} from the metaphorism perspective, or (back to sorting) checking how the ideas of this paper combine with the work on parametric permutation functions by Henglein \cite{He09}.

\section*{Acknowledgements}
The origin of this paper is not computer science. When reading 
Leonard Bernstein's \emph{The Unanswered Question} \cite{Be81}
the author's attention was driven by the \emph{triangular formations}
mentioned by the maestro in his metaphorical analysis of music:
\begin{quote}\em % page 127
(...) in music as in poetry, the A and B of a metaphor must both relate to some X-factor (...) such as rhythm or
(...) harmonic progressions. You see there still that triangular formation of A, B and X to be reckoned with.
\end{quote}
Bernstein's ``triangles'' inspired the ``cospans'' of \cite{Ol15a} and of this paper (\ref{eq:150328a}),
ventured earlier in \cite{Ol13c}. Music has to do with sequences of sound events
and several of its stylistic features can be described by metaphorisms.
It was the generalization of these to arbitrary finite sequences that suggested the sorting
examples and then the theory behind this paper.

The author is indebted to his colleague and linguist {\'Alvaro Iriarte}
for inviting him to contribute to the 2013 \emph{Humanities and Sciences} colloquium where \cite{Ol13c} was presented.
This was followed by several interesting coffee-time conversations in which \'Alvaro eventually pointed to
Lakoff's work \cite{LJ80}. Reading this classic changed the author's perception of natural language for ever.

On the technical side, comments and suggestions by Ali Jaoua and Ali Mili
are gratefully acknowledged. Last but not least, the author thanks the detailed
suggestions and comments made by the anonymous referees which helped to improve the original manuscript.

% The author wishes to thank the anonymous referees for their comments and suggestions.
This work is funded by ERDF - European Regional Development Fund through
the COMPETE Programme (operational programme for competitiveness) and by
National Funds through the \emph{FCT - Funda\c{c}\~ao para a Ci\^encia e a Tecnologia}
(Portuguese Foundation for Science and Technology) within project \textsc{Trust}
(POCI-01-0145-FEDER-016826).

%bibliography{/Users/jno/share/texinputs/jno}

\appendix

\section{Auxiliary lemmas and proofs}
\label{sec:150329b}
%label{sec:150329a}

\begin{lemma}
Given predicate \ensuremath{\Varid{q}} and function \ensuremath{\Varid{f}},
\begin{eqnarray}
	\ensuremath{(\Varid{q} \comp \Varid{f})\hskip-1pt\mathit{?}} = \ensuremath{\ap{\delta}(\Varid{q}\hskip-1pt\mathit{?} \comp \Varid{f})}
	\label{eq:150924a}
\end{eqnarray}
holds, where
\begin{eqnarray}
	\ensuremath{\ap{\delta}\Conid{R}\mathrel{=}{{id}}\mathbin\cap{\conv{\Conid{R}} \comp \Conid{R}}}
	\label{eq:020624j}
\end{eqnarray}
is the \emph{domain} of \ensuremath{\Conid{R}}.
\\ Proof:
\begin{eqnarray*}
\start
	\ensuremath{(\Varid{q} \comp \Varid{f})\hskip-1pt\mathit{?}}
\just={ (\ref{eq:160121c}) }
	\ensuremath{{{id}}\mathbin\cap{\frac{\Varid{true}}{\Varid{q} \comp \Varid{f}}}}
\just={ since \ensuremath{\frac{\Varid{f}}{\Varid{f}}} is reflexive (\ref{eq:160112e}) }
	\ensuremath{{{id}}\mathbin\cap{{\frac{\Varid{f}}{\Varid{f}}}\mathbin\cap{\frac{\Varid{true} \comp \Varid{f}}{\Varid{q} \comp \Varid{f}}}}}
\just={ (\ref{eq:160111a}) ; products }
	\ensuremath{{{id}}\mathbin\cap{\frac{({{id}}\kr{\Varid{true}}) \comp \Varid{f}}{({{id}}\kr{\Varid{q}}) \comp \Varid{f}}}}
\just={ (\ref{eq:160117a}) ; (\ref{eq:160111a}) }
	\ensuremath{{{id}}\mathbin\cap{\conv{\Varid{f}} \comp ({{id}}\mathbin\cap{\frac{\Varid{true}}{\Varid{q}}}) \comp \Varid{f}}}
\just={ (\ref{eq:160121c}) }
	\ensuremath{{{id}}\mathbin\cap{\conv{\Varid{f}} \comp \Varid{q}\hskip-1pt\mathit{?} \comp \Varid{f}}}
\just={ (\ref{eq:020624j}) }
	\ensuremath{\ap{\delta}(\Varid{q}\hskip-1pt\mathit{?} \comp \Varid{f})}
\qed
\end{eqnarray*}
\end{lemma}
The rest of this appendix provides proofs of results left pending in the main text. % \label{sec:150329b}

\paragraph{Proof of property (\ref{eq:160115b})}
\textbf{Part (\ensuremath{\Rightarrow})} --- \ensuremath{\frac{\Conid{R}}{\Conid{R}}} is always an equivalence relation, recall \secref{sec:170319a}.
\textbf{Part (\ensuremath{\Leftarrow})} --- assume that \ensuremath{\Conid{R}} is an equivalence relation. Then:
% ex 4.49 of page 108 \cite{BM97}
\begin{eqnarray*}
\start
	\ensuremath{\Conid{R}\mathrel{=}\Conid{R}\setminus \Conid{R}}
\just\equiv{ since \ensuremath{{\Conid{R}}\subseteq{\Conid{R}\setminus \Conid{R}}} just states that \ensuremath{\Conid{R}} is transitive (\ref{eq:020614b}) }
	\ensuremath{{\Conid{R}\setminus \Conid{R}}\subseteq{\Conid{R}}}
\just\implied{ since \ensuremath{{\Conid{R} \comp (\Conid{R}\setminus \Conid{R})}\subseteq{\Conid{R}}} by (\ref{eq:020614b}) }
	\ensuremath{{\Conid{R}\setminus \Conid{R}}\subseteq{\Conid{R} \comp (\Conid{R}\setminus \Conid{R})}}
	\eqnnewpage
\just\implied{ composition is monotone }
	\ensuremath{{{id}}\subseteq{\Conid{R}}}
\just\equiv{ \ensuremath{\Conid{R}} is reflexive }
	\ensuremath{\Varid{true}}
\end{eqnarray*}
Then:
\begin{eqnarray*}
\start
	\ensuremath{\Conid{R}\mathrel{=}\frac{\Conid{R}}{\Conid{R}}}
\just\equiv{ (\ref{eq:160122a}) ; \ensuremath{\conv{\Conid{R}}\mathbin{/}\conv{\Conid{R}}\mathrel{=}\conv{(\Conid{R}\setminus \Conid{R})}}}
	\ensuremath{\Conid{R}\mathrel{=}{\Conid{R}\rdiv \Conid{R}}\mathbin\cap{\conv{(\Conid{R}\setminus \Conid{R})}}}
\just\equiv{ \ensuremath{\Conid{R}\mathrel{=}\Conid{R}\setminus \Conid{R}} above }
	\ensuremath{\Conid{R}\mathrel{=}{\Conid{R}}\mathbin\cap{\conv{\Conid{R}}}}
\just\equiv{ since \ensuremath{\Conid{R}} is symmetric: \ensuremath{\Conid{R}\mathrel{=}\conv{\Conid{R}}} }
	\ensuremath{\Varid{true}}
\qed
\end{eqnarray*}

\paragraph{Proof of property (\ref{eq:150406c})}
 % in theorem \ref{th:150406b}:
\textbf{Part ($\implies$)} --- show that \ensuremath{\Varid{p}\mathrel{=}\Varid{q} \comp \Varid{f}} follows from \ensuremath{\Varid{f} \comp \Varid{p}\hskip-1pt\mathit{?}\mathrel{=}\Varid{q}\hskip-1pt\mathit{?} \comp \Varid{f}}:
\begin{eqnarray*}
\start
	\ensuremath{\Varid{p}\mathrel{=}\Varid{q} \comp \Varid{f}}
\just\equiv{ bijection between predicates and partial identities }
	\ensuremath{\Varid{p}\hskip-1pt\mathit{?}\mathrel{=}(\Varid{q} \comp \Varid{f})\hskip-1pt\mathit{?}}
\just\equiv{ (\ref{eq:150924a}) ; \ensuremath{\Varid{f} \comp \Varid{p}\hskip-1pt\mathit{?}\mathrel{=}\Varid{q}\hskip-1pt\mathit{?} \comp \Varid{f}} assumed }
	\ensuremath{\Varid{p}\hskip-1pt\mathit{?}\mathrel{=}\ap{\delta}(\Varid{f} \comp \Varid{p}\hskip-1pt\mathit{?})}
\just\equiv{ \ensuremath{\ap{\delta}(\Conid{R} \comp \Conid{S})\mathrel{=}\ap{\delta}(\ap{\delta}\Conid{R} \comp \Conid{S})}}
	\ensuremath{\Varid{p}\hskip-1pt\mathit{?}\mathrel{=}\ap{\delta}(\ap{\delta}\Varid{f} \comp \Varid{p}\hskip-1pt\mathit{?})}
\just\equiv{ \ensuremath{\ap{\delta}\Varid{f}\mathrel{=}{id}}}
	\ensuremath{\Varid{p}\hskip-1pt\mathit{?}\mathrel{=}\ap{\delta}(\Varid{p}\hskip-1pt\mathit{?})}
\just\equiv{ domain of a partial identity is itself }
	\ensuremath{\Varid{true}}
\qed
\end{eqnarray*}
\textbf{Part ($\implied$)} --- show that \ensuremath{\Varid{f} \comp \Varid{p}\hskip-1pt\mathit{?}\mathrel{=}\Varid{q}\hskip-1pt\mathit{?} \comp \Varid{f}} holds assuming \ensuremath{\Varid{p}\mathrel{=}\Varid{q} \comp \Varid{f}}:
\begin{eqnarray*}
\start
	\ensuremath{\Varid{f} \comp \Varid{p}\hskip-1pt\mathit{?}} = \ensuremath{\Varid{q}\hskip-1pt\mathit{?} \comp \Varid{f}}
\just\equiv{ substitution \ensuremath{\Varid{p}\mathbin{:=}\Varid{q} \comp \Varid{f}}; (\ref{eq:150924a}) }
	\ensuremath{\Varid{f} \comp \ap{\delta}(\Varid{q}\hskip-1pt\mathit{?} \comp \Varid{f})} = \ensuremath{\Varid{q}\hskip-1pt\mathit{?} \comp \Varid{f}}
\just\implied{ \ensuremath{ \subseteq }-antisymmetry, since \ensuremath{{\ap{\delta}(\Varid{q}\hskip-1pt\mathit{?} \comp \Varid{f})}\subseteq{\conv{\Varid{f}} \comp \Varid{q}\hskip-1pt\mathit{?} \comp \Varid{f}}} and \ensuremath{\Varid{f} \comp \conv{\Varid{f}}\; \subseteq \;{id}}}
	\ensuremath{{\Varid{q}\hskip-1pt\mathit{?} \comp \Varid{f}}\subseteq{\Varid{f} \comp \ap{\delta}(\Varid{q}\hskip-1pt\mathit{?} \comp \Varid{f})}}
\just\equiv{ \ensuremath{\Conid{R}\mathrel{=}\Conid{R} \comp \ap{\delta}\Conid{R}} }
	\ensuremath{{\Varid{q}\hskip-1pt\mathit{?} \comp \Varid{f} \comp \ap{\delta}(\Varid{q}\hskip-1pt\mathit{?} \comp \Varid{f})}\subseteq{\Varid{f} \comp \ap{\delta}(\Varid{q}\hskip-1pt\mathit{?} \comp \Varid{f})}}
\just\implied{ monotonicity of composition }
	\ensuremath{\Varid{q}\hskip-1pt\mathit{?}} \wider\subseteq \ensuremath{{id}}
\just\equiv{\ensuremath{\Varid{q}\hskip-1pt\mathit{?}} is a partial identity}
	\ensuremath{\Varid{true}}
\qed
\end{eqnarray*}

\paragraph{Proof of property (\ref{eq:150407a-modified})}
By (\ref{eq:160112f}), \ensuremath{\Varid{p}\;{\leq_{\cdot }}\;\Varid{f}} is equivalent to the 
existence of some \ensuremath{\Varid{q}} such that \ensuremath{\Varid{p}\mathrel{=}\Varid{q} \comp \Varid{f}} holds, which in turn is equivalent to
\ensuremath{\Varid{f} \comp \Varid{p}\hskip-1pt\mathit{?}} = \ensuremath{\Varid{q}\hskip-1pt\mathit{?} \comp \Varid{f}} by (\ref{eq:150406c}). Then:
\begin{eqnarray*}
\start
	\ensuremath{\frac{\Varid{f}}{\Varid{f}} \comp \Varid{p}\hskip-1pt\mathit{?}}
\just={ metaphors (\ref{eq:160117a}) ; (\ref{eq:150406c}) }
	\ensuremath{\conv{\Varid{f}} \comp \Varid{q}\hskip-1pt\mathit{?} \comp \Varid{f}}
\just={ converses ; partial identities }
	\ensuremath{\conv{(\Varid{q}\hskip-1pt\mathit{?} \comp \Varid{f})} \comp \Varid{f}}
\just={ again (\ref{eq:150406c}) and (\ref{eq:160117a}) }
	\ensuremath{\Varid{p}\hskip-1pt\mathit{?} \comp \frac{\Varid{f}}{\Varid{f}}}
\qed
\end{eqnarray*}
%----------------------------------------------

\paragraph{Proof of property (\ref{eq:150214b})}
Our strategy is indirect equality carried over the universal property of the shrinking operator (\ref{eq:100116d}):
\begin{eqnarray*}
\start
	X \subseteq S \shrunkby (\ensuremath{\Varid{q}\hskip-1pt\mathit{?}} \comp \top)
\just\equiv{ (\ref{eq:100116d}) ; (\ref{eq:060124a}) }
	X \subseteq S \land X\comp\conv S \subseteq \ensuremath{\Varid{q}\hskip-1pt\mathit{?}} \comp \ensuremath{\frac{\mathop{!}}{\mathop{!}}}
	\eqnnewpage
\just\equiv{ shunting (\ref{eq:020617f}) ; converses }
	X \subseteq S \land X\comp\conv{(\bang\comp S)} \subseteq \ensuremath{\Varid{q}\hskip-1pt\mathit{?}} \comp \conv\bang
\just\equiv{ assume \ensuremath{\Conid{S}} entire, so \ensuremath{\mathop{!} \comp \Conid{S}\mathrel{=}\mathop{!}} }
	X \subseteq S \land X\comp\conv\bang \subseteq \ensuremath{\Varid{q}\hskip-1pt\mathit{?}} \comp \conv\bang
\just\equiv{ shunting (\ref{eq:020617f}) ;  (\ref{eq:060124a}) }
	X \subseteq S \land X \subseteq \ensuremath{\Varid{q}\hskip-1pt\mathit{?} \comp \top }
\just\equiv{ (\ref{eq:071215a}) below }
%
%	X \subseteq S \cap |(crflx q)| \comp \top
%
	X \subseteq \ensuremath{\Varid{q}\hskip-1pt\mathit{?}}\comp S
\just{::}{ indirect equality}
	\ensuremath{{\Conid{S}}\shrunkby{(\Varid{q}\hskip-1pt\mathit{?} \comp \top )}\mathrel{=}\Varid{q}\hskip-1pt\mathit{?} \comp \Conid{S}} 
\qed
\end{eqnarray*}
The proof relies on a well-known property of partial identities,
given below together with its converse version:
\begin{eqnarray}
R\comp \ensuremath{\Varid{p}\hskip-1pt\mathit{?}} & = & R \cap \top\comp \ensuremath{\Varid{p}\hskip-1pt\mathit{?}}
\label{eq:081025a}
\\
\ensuremath{\Varid{q}\hskip-1pt\mathit{?}}\comp R & = & R \cap \ensuremath{\Varid{q}\hskip-1pt\mathit{?}}\comp \top
\label{eq:071215a}
\end{eqnarray}
see e.g.\ \cite{Ol08b}.
%-----------------------------------

\paragraph{Proof of Theorem \ref{th:150327a}}
Equality (\ref{eq:160118c}) follows immediately from (\ref{eq:150326b})
by fold-cancellation (\ref{eq:150402b}). Next we show the equivalence
between (\ref{eq:150326b}) and (\ref{eq:150326a}):
\begin{eqnarray*}
\start
	\ensuremath{\Conid{R} \comp \Varid{h}\mathrel{=}\Conid{R} \comp \Varid{h} \comp (\fun F \;\Conid{R})}
\just\equiv{\ensuremath{\Conid{R} \comp \Varid{h}\; \subseteq \;\Conid{R} \comp \Varid{h} \comp (\fun F \;\Conid{R})} holds by \ensuremath{{{id}}\subseteq{\fun F \;\Conid{R}}}, since \ensuremath{{{id}}\subseteq{\Conid{R}}}}
	\ensuremath{\Conid{R} \comp \Varid{h} \comp (\fun F \;\Conid{R})\; \subseteq \;\Conid{R} \comp \Varid{h}}
\just\equiv{ \ensuremath{(\Conid{R} \comp )} is a {closure} operation, see (\ref{eq:060613c-modified})  below}
	\ensuremath{\Varid{h} \comp (\fun F \;\Conid{R})\; \subseteq \;\Conid{R} \comp \Varid{h}}
\qed
\end{eqnarray*}
The last step relies on the fact that composition with equivalence relations is a \emph{closure}
operation:
\begin{eqnarray}
	 R \comp S \subseteq R \comp Q & \equiv & S \subseteq R \comp Q 
	\label{eq:060613c-modified}
\end{eqnarray}
This fact is used elsewhere \cite{Ol11} to reason about functional dependencies in databases.
Below we rephrase its proof using the power transpose \ensuremath{\Lambda{\Conid{R}}} which maps
objects to their \ensuremath{\Conid{R}}-equivalence classes (\ref{eq:160115a}):
\begin{eqnarray*}
\start
	\ensuremath{\Conid{R} \comp \Conid{S}\; \subseteq \;\Conid{R} \comp \Conid{Q}}
\just\equiv{ \ensuremath{\Conid{R}\mathrel{=}\frac{\Lambda{\Conid{R}}}{\Lambda{\Conid{R}}}} (\ref{eq:160115a}) }
	\ensuremath{\frac{\Lambda{\Conid{R}}}{\Lambda{\Conid{R}}} \comp \Conid{S}\; \subseteq \;\frac{\Lambda{\Conid{R}}}{\Lambda{\Conid{R}}} \comp \Conid{Q}}
\just\equiv{ \ensuremath{\frac{\Lambda{\Conid{R}}}{\Lambda{\Conid{R}}}\mathrel{=}\conv{\Lambda{\Conid{R}}} \comp \Lambda{\Conid{R}}} (\ref{eq:160117a}) ; shunting (\ref{eq:020617e}) }
	\ensuremath{\Lambda{\Conid{R}} \comp \conv{\Lambda{\Conid{R}}} \comp \Lambda{\Conid{R}} \comp \Conid{S}\; \subseteq \;\Lambda{\Conid{R}} \comp \Conid{Q}}
\just\equiv{\ensuremath{\Varid{f} \comp \conv{\Varid{f}} \comp \Varid{f}\mathrel{=}\Varid{f}} (difunctionality)}
	\ensuremath{\Lambda{\Conid{R}} \comp \Conid{S}\; \subseteq \;\Lambda{\Conid{R}} \comp \Conid{Q}}
\just\equiv{ shunting (\ref{eq:020617e}) ; \ensuremath{\conv{\Lambda{\Conid{R}}} \comp \Lambda{\Conid{R}}\mathrel{=}\frac{\Lambda{\Conid{R}}}{\Lambda{\Conid{R}}}\mathrel{=}\Conid{R}} (\ref{eq:160115a})}
	\ensuremath{\Conid{S}\; \subseteq \;\Conid{R} \comp \Conid{Q}}
\qed
\end{eqnarray*}
Finally, the proof that (\ref{eq:160120c}) is equivalent to (\ref{eq:150326a}) for the special case \ensuremath{\Conid{R}\mathrel{=}\frac{\Varid{f}}{\Varid{f}}}:
\begin{eqnarray*}
\start
	\ensuremath{\Varid{h} \comp (\fun F \;\frac{\Varid{f}}{\Varid{f}})\; \subseteq \;\frac{\Varid{f}}{\Varid{f}} \comp \Varid{h}}	
\just\equiv{ metaphor algebra: (\ref{eq:160117a}) etc }	
	\ensuremath{\fun F \;\frac{\Varid{f}}{\Varid{f}}\; \subseteq \;\frac{\Varid{f} \comp \Varid{h}}{\Varid{f} \comp \Varid{h}}}	
\just\equiv{ injectivity preorder (\ref{eq:160112f}) ; relator \ensuremath{\fun F } (\ref{eq:160118d}) }	
	\ensuremath{\Varid{f} \comp \Varid{h}\leq \fun F \;\Varid{f}}	
\qed
\end{eqnarray*}
%-----------------------------------

\end{document}